\documentclass[twocolumn,showpacs,preprintnumbers,amsmath,amssymb]{revtex4}
\usepackage{graphicx}
\usepackage{dcolumn}
\usepackage{bm}
\newcommand\ed{\tilde{\delta}} 

\begin{document}

\title{Double Distribution of Dark Matter Halos \\
with respect to Mass and Local Overdensity}
\author{Vasiliki Pavlidou} 
\author{Brian D. Fields}
\affiliation{Center for Theoretical Astrophysics, 
Department of Astronomy \\
University of Illinois, Urbana, IL 61801}

\date{\today}

\begin{abstract}

We present a double distribution function of dark matter halos,
with respect to both object mass and local over- (or under-) density. 
This analytical tool provides a statistical treatment of the 
properties of matter {\em surrounding} collapsed objects, and can be
used to study environmental effects on hierarchical structure
formation. 
The size of the ``local environment'' of a collapsed object is defined 
to depend on the mass of the object. 
The Press-Schechter mass function is recovered by integration 
of our double distribution over the density contrast. 
We also present a detailed treatment of the evolution of overdensities 
{\em and underdensities} in
$\Omega_{\rm m}+\Omega_\Lambda=1$ and $\Omega_{\rm m}=1$ universes according to
the spherical evolution model. We explicitly distinguish
between true and linearly extrapolated overdensities and provide 
conversion relations between the two quantities.
\end{abstract}

\maketitle
\section{Introduction}\label{intro}

Cosmological distributions have long 
been used with great success as an analytical tool complementary to 
numerical simulations. They have been used to constrain the cosmological 
parameters; interpret results of cosmological simulations; study
regions of the parameter space which cannot be approached by simulations 
due to prohibitive computational
cost; exploring the effects of various physical 
processes in an efficient if approximate way.
The analytical tool used most widely in cosmology is
the mass function of dark matter halos (distribution of the number
density of halos with respect to halo mass). Analytical descriptions
of dark matter halos are usually based on the Press-Schechter
formalism (\cite{ps},\cite{b91}) 
and its extensions (e.g.,\cite{lc93}, \cite{ph90}, \cite{j95},
\cite{bow91}). The Press-Schechter mass function has
is in good agreement with the results of N-body
simulations (e.g. \cite{wef}, \cite{lc94},  \cite{ecf}).
More sophisticated approaches taking into
account deviations from spherical symmetry (e.g., \cite{ls98}, 
\cite{smt01},\cite{st02}) 
have improved this agreement even further.

To derive the Press-Schechter mass function, regions in
space are smoothed on successively smaller scales.
The mass of a collapsed object is then taken to be the 
largest smoothing mass scale
for which the average linear overdensity exceeds some threshold. 
In this way, matter in the universe is
distributed among collapsed structures of different masses, which all
share the same value of average overdensity (the threshold value). 
Information about the local environment of the
collapsed objects (whether they live in underdensities or
overdensities) is thus erased. 

For this reason, and despite its wide applicability, 
the expression for the mass function cannot
be used to address environment-related questions: Does the mass function of
structures in superclusters differ from the mass function 
inside  voids? Are structures of some particular mass more likely to reside
inside underdense or overdense regions in space? How does such a 
preference evolve with redshift, and  how sensitively does it depend
on the cosmological parameters? How does the state of the material 
surrounding and accreted by a collapsed object depend on the
mass of the object and the cosmic epoch? 

To address such questions, we seek a {\em double distribution} of the number 
density of structures with 
respect to mass but also to local overdensity (or underdensity). In
order to extract information about the surroundings of collapsed
structures, we use the same random walk formalism  which rigorously yields 
the Press-Schechter mass function (\cite{b91},
\cite{lc93}). Integration of this distribution over density contrast should
return the Press-Schechter mass function so that the successes of the
Press-Schechter formalism be retained. 

Two complications arise in the effort to expand the Press-Schechter
mass function to incorporate a description of the local overdensity.
First, the concept of the ``local environment''
is somewhat vague and needs to be defined in a more rigorous way. The size
of the ``local environment'' cannot be the same for all structures. If
this was the case, very small structures would represent only a tiny
fraction of the ``environment'', while very large structures could
even exceed the size of the ``environment'', which would be an
unphysical situation. This problem is not exclusive to analytical
tools, but also needs to be addressed when analyzing the results of 
numerical simulations. 
Second, the Press-Schechter treatment of the density field uses linear 
theory, and ways of converting this information to a more physical 
non-linear result need to be determined. 

We address the first problem 
by introducing a {\em clustering scale parameter}, $\beta$, 
which allows us to define
the size of the ``environment'' of each structure as a function of its
mass. We address the second concern by calculating conversion
relations between the linear-theory overdensities (or underdensities)
and those predicted by the spherical-evolution model. 

Environment-related questions in cosmological structure
formation have also beed addressed using analytical models for the
clustering properties of dark halos which evaluate quantities such as  
the cross-correlation function between dark halos and matter, and the
biasing factor (e.g. \cite{mw96}, \cite{kns97},
\cite{s98}, \cite{tp98}, \cite{lk99}, \cite{st99}, \cite{sj00}, also
see review by \cite{cs02} and references therein). These analyses are
based on  the same random walk formalism which we use here to derive our double
distribution (also see \cite{j98} for fitting formulae from N-body
simulation results). However, the information content of the double distribution, 
which treats the ``environment'' in a mass-dependent fashion, is
complementary to that of correlation functions, which 
describe the clustering properties of the dark halo
population at some {\em fixed spatial scale}. 
The double distribution is ideally fitted for population studies of
cosmological objects. If the properties of a single object can be
parametrized as a function of its mass and its environment, then the
double distribution can be used to predict the statistical properties
of such objects, as well as their evolution with time, for any
cosmological model.

Our paper is organized as follows. In \S \ref{form} we briefly review 
the ``random walk'' formalism used in the derivation of the
Press-Schechter mass function by \cite{b91} and \cite{lc93}. We then use 
the same formalism to derive the double distribution for dark matter
halos. We also discuss converting between linear and
spherical-evolution density contrasts, and we present some 
interesting derivative quantities of the double distribution. 
In \S \ref{res} we explore the information content of our
double-distribution by plotting the distribution itself as well as its
derivative quantities for concordance ($\Omega_{\rm
  m}+\Omega_\Lambda =1$) and Einstein-deSitter universes.
Finally, we
conclude and discuss our findings in \S \ref{disc}.

\section{Formalism}\label{form}

\subsection{Labeling of Cosmic Epochs}

Throughout this paper we 
use the value of the
dimensionless scale factor of the universe, $a$, to label different
cosmic epochs. We normalize it so that the
scale factor of the present is $a_0=1$. Then, a given value of
$a$ corresponds to a redshift 
\begin{equation} z = a^{-1}-1 \,,\end{equation}
independently of the cosmological model used. On the other hand, the
conversion between $a$ and time $t$ does depend on the assumed cosmology.
For a flat
($\Omega_{\rm tot} = \rho_{\rm tot}/\rho_{\rm c} = 1$)
universe containing only matter and vacuum energy in the form of a
cosmological constant (with present-day density parameters
$\Omega_{\rm m} = \rho_{\rm m,0}/\rho_{\rm c, 0}$ and $\Omega_{\rm
  \Lambda} 
= \rho_{\rm \Lambda,0}/\rho_{\rm c,
  0}$ respectively), the conversion equation is
\begin{equation} \label{timeL}
t = \frac{2}{3}H_0^{-1}\Omega_{\rm \Lambda}^{-1/2} \sinh ^{-1} 
\left(\sqrt{a^3\Omega_{\rm \Lambda}/\Omega_{\rm m}}\right)
\,,
\end{equation}
where $H_0$ is the present-day value of the Hubble parameter. In the
limiting case when $\Omega_{\rm \Lambda} =0$, $\Omega_{\rm m} =1$,
equation (\ref{timeL}) becomes
\begin{equation}
t = \frac{2}{3}H_0^{-1} a^{3/2}\,.
\end{equation}

We choose to cast our results in terms of $a$ because it has 
a trivial and cosmology-independent relation with the directly
observable redshift. In addition, it increases with time, 
which allows for a more intuitive interpretation of the evolution of the 
quantities we consider here. 

\subsection{Random Walks and the Press-Schechter Mass Function}

The ``random walk'' formalism was introduced for the derivation of
cosmological mass functions by \cite{b91} and by \cite{lc93}. Here, we
briefly review the basic concepts of this formalism, before 
extending it to the case of the double distribution in the next section.

The Press-Schechter mass function 
of collapsed structures is the
comoving number density of virialized objects per differential mass interval,
$dn/dm$, for every cosmic epoch $a$ \footnote{The {\em proper} mass function, i.e. the number of
collapsed structures per unit proper volume per differential mass
interval is simply related to the comoving mass function via 
$dn/dm|_{\rm proper} = a^{-3} dn/dm|_{\rm comoving}$}.
A related quantity is the mass fraction, $P(>m, a)$, which is 
the fraction of matter in the universe belonging to
collapsed structures with mass $>m$. If $P(>m,a)$ is known, then
$dn/dm$ can be calculated from 
\begin{equation}\label{funcfrac}
\frac{dn}{dm}(m,a)= \frac{\rho_{\rm m,0}}{m}
\left|\frac{d}{dm}
P(>m,a)
\right|\,,
\end{equation}
where $\rho_{\rm m,0}$ is the present-day matter density of the
universe.

$P(>m,a)$ is in turn calculated by assigning, at each epoch $a$, every
infinitesimal element $dm$ in the universe to a
collapsed structure of some mass $m$.
A structure is considered ``collapsed'' if its
mean overdensity 
\begin{equation}\label{contrastdef}
\langle\delta\rangle = 
\frac{\langle\rho_{\rm structure}\rangle-\rho_{{\rm m}, a}}{\rho_{{\rm m}, a}}
\end{equation}
exceeds a certain critical value, $\delta_{\rm c}(a)$.
In equation (\ref{contrastdef}), $\rho_{{\rm m}, a}$ 
is the mean matter density of the universe
at epoch $a$. The critical oversendity  $\delta_{\rm c}(a)$
is the mean overdensity predicted by the
spherical evolution model for a
structure virializing at epoch $a$. 
For each point in space, one calculates the
mean local overdensity by smoothing the
overdensity field $\delta(\vec{x},a)$ with a spherically symmetric
filter function of varying mass scale, starting
from $m \rightarrow \infty$, where 
one averages over the whole universe and finds
identically $\langle \delta \rangle = 0$, and proceeding to
successively smaller scales. When a mass scale is found for which the
mean overdensity becomes equal to $\delta_{\rm c}(a)$, it is taken
to be the mass $m$ of the parent object of the infinitesimal mass
at the point under consideration. 
This way of assigning object masses circumvents
the structure-in-structure problem, since
the mass of the parent object is always
the {\em largest possible mass} satisfying the criterion for
collapse. All information on substructure within
collapsed structures is thus erased from the resulting mass function. 

The way the average overdensity $\langle \delta \rangle$ 
changes when the smoothing mass scale is varied
resembles, under certain conditions, a 1D random walk
\cite{b91}. For all
``particles''(points in space in our case), the walk begins 
at the ``spatial origin'' ($\langle \delta \rangle =0$), at ``time zero''
($m \rightarrow \infty$). As ``time progresses'' ($m$ decreases), each
``particle'' may move either to the ``left''
 (negative $\langle \delta \rangle$) or to the ``right'' (positive 
 $\langle \delta \rangle$). An ``absorbing wall'' exists at $\delta_{\rm c}(a)$.
If this ``wall'' is reached,
the ``particle'' is ``removed'' from the walk (the point is
assigned its parent object mass and
removed from further consideration at smaller values of $m$). $P(>m,a)$
is then the fraction of ``particles'' which
have been ``lost'' by  ``time''
$m$, and it can be calculated using random-walk
theory. 

However, we must first ensure that simple random-walk theory is indeed applicable. 
First, each ``step'' of the ``walk'' should
be completely independent from the previous step. This requires that
the $k$-modes producing an increase $\Delta \langle \delta \rangle$
in the space-like variable not appear in any of the previous steps in
$\langle \delta \rangle$. A smoothing window function sharp in
$k-$space, 
\begin{equation}\label{what}
\hat{W}_m(k)=\left\{
\begin{array}{ll}
1 & k<k_{\rm c}(m)\\
0 &k>k_{\rm c}(m)
\end{array}
\right. \,.
\end{equation}
(see \cite{b91} and \cite{lc93} for more extended discussions on the
consequences of such a choice) enforces this condition, since
\begin{equation}
\langle \delta \rangle _{m,\vec{x}_0} =
\int W_m\left(|\vec{x}_0-\vec{x}|\right)\delta(\vec{x})d^3\vec{x}
= \int_{k \le k_{\rm c}(m)} 
\!\!\!\!\! \!\!\!\!\!
\!\!\!\!\!\!\!\!
\delta_k {\rm e}^{i\vec{x}_0\cdot \vec{k}} d^3\vec{k}
\end{equation}
and
\begin{equation}
\Delta \langle \delta \rangle _{\vec{x}_0} =
\int_{_{k_{\rm c}(m) \le k \le k_{\rm c}(m-dm)}} \!\!\!\!\!\!\!\!\!\!\!\!\!\!\!\!\!\!\!\! 
\!\!\!\!\!\!\!\!\!\!
\delta_k {\rm e}^{i\vec{x}_0\cdot \vec{k}} d^3\vec{k}
\end{equation}
which only involves new $k$-modes corresponding to scales from
$m$ to $m-dm$.

Second, there must be an
equal probability for the system to ``move'' towards any one of the
two available ``directions''. A Gaussian overdensity field 
(which is the usual assumption for deriving analytic mass functions
and which we adopt in this paper) guarantees that
this condition is satisfied.

Finally, the
appropriate ``time-like'' variable (which should depend on $m$)
needs to be selected, given that 
the ``space-like'' variable is $\langle \delta \rangle$.
By direct analogy to the 1D 
random walk theory result $\langle x^2\rangle = 2Dt$, and from the
definition of the variance of the overdensity field $S(m)$, 
\begin{equation}\label{sdef}
S(m) =\sigma^2(m) = \langle|\delta (m,\vec{x})|^2\rangle = 
\int_{k=0}^{k(m)}\!\!\!k^2dk |\delta_k|^2
\end{equation}
we can immediately identify $Dt \rightarrow S(m)/2$.

Three further complications need to be addressed. 
First, our knowledge of $\delta_k$ and subsequently $S(m)$ 
is limited at late times. 
In the early universe, right after matter-radiation equality, 
$\langle|\delta_k|^2\rangle$ can be simply described in terms of a 
power-law in $k$ modified by a transfer function,
$\langle|\delta_k|^2\rangle \propto T^2(k)k^n$. 
While all $\delta$ are still in their linear regime, 
they simply grow by the linear growth factor
(independent of $k$). However, at later times, when certain structures start
departing from the linear regime, 
we cannot use our simple early-universe expressions for $\delta_k$.  
Second, $\langle \delta \rangle$ is limited 
to be $\ge -1$, which introduces a second,
reflecting ``wall'' at a value of $\langle \delta \rangle = -1$,
further complicating the random-walk calculations. 
 Finally, the true overdensity field
loses its Gaussianity as it evolves past the linear regime.

To circumvent these problems, we define the 
{\em linearly extrapolated overdensity field},
$\ed(\vec{x},a)$, as the overdensity field that would result if all
structures continued to grow according to the linear theory until time
$a$. Now $\ed(\vec{x},a)$ 
is not limited to be $\ge -1$, since it does not 
represent real overdensities. In addition, we can always
calculate $S(m)$ for  $\ed(\vec{x},a)$, since
$\ed_k$ is modified from its simple early-universe expression 
only by the linear growth factor. Finally, the extrapolated field
remains Gaussian at all times.

The linearly extrapolated overdensity $\ed(\vec{x},a)$
and the associated variance $S(m)$,
are time-varying, but  
the time dependence is well-known (see appendices \ref{ap1} and
\ref{ap2}), and the same for both $S$ and $\ed^2$ \footnote{ as seen by
equation (\ref{sdef}) re-written for the extrapolated rather than the true
overdensity field}. Thus, the time dependence drops out of ratios 
$\ed/\sqrt{S}$ which appear in the mass function. For this reason we may,  without loss
of generality, choose any single epoch to evaluate these quantities, 
with the stipulation
that $\ed$ and $S(m)$ must refer to the same epoch.
Given that $\sigma(m)$ is often normalized to the present
value of $\sigma_8$, a convenient choice of epoch
is the present.  Then, 
equation (\ref{sdef}) gives for $S(m)$ 
\begin{equation}
S(m) = \sigma_8^2 
\frac{\int_{k=0}^{k(m)} T^2(k)k^{n+2}dk}
{\int_{k=0}^{k(m_8)} T^2(k)k^{n+2}dk} \,.
\end{equation}

Thus we only consider $\ed(\vec{x},a_0)$ (the overdensity field
linearly extrapolated to the present epoch), which we use instead of 
the true field $\delta(\vec{x},a)$ in our
random walk formalism \footnote{Physically, the substitution of the true field by the extrapolated
field in the ``random walk'' corresponds to
smoothing the extrapolated overdensity field, and then mapping the mean
extrapolated overdensity value to a true overdensity value. That true
overdensity value is then assumed to accurately represent the result of
a smoothing of the true field, which implies
$
\delta\left(\left\langle\ed\right\rangle\right) = 
\left\langle\delta\left(\ed\right)\right\rangle$.
This would be exactly true only if $\delta(\ed)$ was linear in $\ed$,
which is not the case (see appendices \ref{ap1} and \ref{ap2}). 
This assumption introduces an inaccuracy 
inherent to all calculations which employ it, including the
Press-Schechter mass function as well as the 
double distribution. }. 
To find the mass function at a particular cosmic epoch
$a$, we calculate the location of the ``absorbing wall'', $\ed_{\rm
  c}(a)$.
If a structure is predicted to 
collapse at epoch $a$ according to the spherical evolution model,
then $\ed_{\rm 0,c}(a)$ is the overdensity this same structure would have had if, 
instead of turning around and collapsing, 
it had continued its linear evolution until the present. 
This $\ed_{\rm 0,c}(a)$ is then our ``absorbing wall''.

We can now derive the mass fraction and mass function 
using random walk theory. If a particle executes a one-dimensional
random walk with an absorbing boundary at a point $x_1$, then its 
probability ${\cal W}(x,t) $ to be 
between $x$ and
$x+dx$ at time $t$ is \cite{ch}
\begin{equation}\label{rweq}
{\cal W}(x,t,x_1)dx = \frac{
\exp\left[-\frac{x^2}{4Dt}\right]
-\exp \left[-\frac{(2x_1-x)^2}{4Dt}\right]
}{2\sqrt{\pi Dt}}dx\, ,
\end{equation}
where $x \le x_1$.
In our case, the probability that a
point in space will be assigned an average extrapolated overdensity 
between $\ed$ and $\ed+d\ed$ when filtered at a scale $m$
corresponding to a variance of $S(m)$ is
\begin{equation}\label{wforPS}
{\cal W}(\ed,S,\ed_{\rm 0,c})d\ed = \frac{
\exp\left[-\frac{\ed^2}{2S}\right]
-\exp \left[-\frac{(2\ed_{\rm 0,c}-\ed)^2}{2S}\right]
}{\sqrt{2\pi S}}d\ed\, ,
\end{equation}
with $\ed \le \ed_{\rm 0,c}$.
The mass fraction $P(>m,a)$ is then the fraction of points 
already ``lost'' from the walk when filtering at higher mass scales,
which is one minus the fraction of points remaining in the walk,
\begin{eqnarray}
P(>m,a) &=& P(>\ed_{\rm 0,c}) 
= 1-\int_{-\infty}^{\ed_{\rm 0, c}} {\cal W}(\ed,S, \ed_{\rm 0,c})d\ed 
\nonumber \\
&=& {\rm erfc} \left(\frac{\ed_{\rm 0,c}(a)}{\sqrt{2S(m)}}\right)\,.
\end{eqnarray}
Then, 
\begin{equation}\label{dPdm}
\frac{dP(>m,a)}{dm} = 
\frac{1}{\sqrt{2\pi}} \frac{\ed_{\rm 0,c}(a)}{S(m)^{3/2}}\frac{dS}{dm}
\exp \left[-\frac{\ed_{\rm 0,c}(a)^2}{2S(m)}\right]\,,
\end{equation}
and the Press-Schechter mass function can be found using equation
(\ref{funcfrac}), 
\begin{equation}
\frac{dn}{dm}(m,a) = \sqrt{\frac{2}{\pi}}\frac{\rho_{\rm m,0}}{m^2}
\frac{\ed_{\rm 0,c}(a)}{\sqrt{S(m)}} \left|\frac{d\ln \sqrt{S}}{d\ln m}\right|
\exp \left[-\frac{\ed_{\rm 0,c}(a)^2}{2S(m)}\right]\,.
\end{equation}

\subsection{Derivation of the Double Distribution}

We now use the random walk formalism described in the previous
section to derive the double distribution of the comoving number
density of collapsed structures with respect to object mass $m$ and local
environment overdensity $\delta_{\ell}$, $dn/(dm \, d\delta_{\ell})$.

For the reasons described in the previous section, 
we replace the true overdensity
field, $\delta(\vec{x},a)$, with its linear extrapolation to the
present time, $\ed(\vec{x},a_0)$. Thus, we derive the double
distribution of comoving $n$ with respect to object mass $m$ and {\em
  extrapolated} local environment overdensity $\ed_{\ell}$,
$dn/(dm\,d\ed_{\ell})$. We then use linear theory and the spherical
evolution model to establish a conversion relation $\delta(\ed,a)$
and calculate $dn/(dm \, d\delta_{\ell})$ as
\begin{equation} \label{convtotrue}
\frac{dn}{dmd\delta_{\ell}}(\delta_{\ell},m,a)dm d\delta_{\ell} 
= \frac{dn}{dmd\ed_{\ell}} \left[\ed_{\ell}(\delta_{\ell},a),m,a\right]dm\frac{\partial
  \ed}{\partial \delta_{\ell}}d\delta_{\ell} \,.
\end{equation}

First of all, we need to define the local environment extrapolated 
overdensity $\ed_{\ell}$ in a precise way. We would like $\ed_{\ell}$ to be a
measure of the density contrast of the
medium in which a collapsed structure is embedded. Clearly, the value
of $\ed_{\ell}$ depends on how far from the structure itself its 
``environment''
extends. We quantify this notion by introducing the 
{\bf clustering scale parameter}, $\beta$, which is defined in the
following way: the ``environment'' of an object of mass $m$ is a
surrounding region in space which encompasses mass $\beta m$
(including the mass of the object). Hence, 
the local environment
extrapolated overdensity $\ed_{\ell}$ is the result of a filtering 
of $\ed(\vec{x},a_0)$ with a filter of scale $\beta m$ centered on the
object. 

Formally, the above definition translates as follows. Consider 
the sharp in $k-$space filtering function 
$\hat{W}_m(k)$ discussed previously (eq. \ref{what}). 
The relation between the cutoff wavenumber and the filter mass,
$k_{\rm c}(m)$, is found by considering the form of the filter function in
configuration space, 
\begin{equation}
W_m(r) = \frac{\sin \left[k_{\rm c}(m)r\right] 
- k_{\rm c}(m) r \cos \left[k_{\rm c}(m)r\right]}{2\pi^2r^3}\,,
\end{equation}
and multiplying by $\rho_{\rm m,0}$ and integrating over all space, which
yields
\cite{lc93}, 
\begin{equation}
k_{\rm c}(m) = \left(\frac{6\pi^2\rho_{\rm m,0}}{m}\right)^{1/3}\,.
\end{equation}
For a collapsed structure at an epoch $a$ which has mass $m$ and is
centered at a point $\vec{x}_0$, we can write
\begin{equation}
\ed(m,\vec{x_0}) = 
\int W_m\left(|\vec{x}_0-\vec{x}|\right)\ed(\vec{x},a_0)d^3\vec{x}
= \ed_{\rm 0,c}(a)\,,
\end{equation}
since the mean extrapolated overdensity of the collapsed structure
itself is always the critical value for collapse, $\ed_{\rm 0,c}(a)$.
For that same object, the {\em local environment extrapolated
  overdensity}, $\ed_{\ell}$, is
\begin{equation} \label{defbet}
\ed_{\ell}(m,\vec{x_0}) = \int W_{\beta m}
\left(|\vec{x}_0-\vec{x}|\right)\ed(\vec{x},a_0)d^3\vec{x}\,.
\end{equation}
Equation (\ref{defbet}) is then the definition of $\ed_{\ell}$ for a given
$\beta$. In our double distribution, $\beta$ is  free parameter, 
which is however constrained to be
between 1 and a few on physical grounds. It cannot be $<1$ since the
mass of the object's environment always includes the mass of the
object itself. In fact, as $\beta$  approaches $1$, the averaging
which produces $\ed_{\ell}$ is taken {\em only} over the collapsed object
itself, and inevitably returns the critical overdensity for
collapse, $\ed_{\rm 0,c}$, for all objects.   
In the other extreme, $\beta \gg 1$, the average
$\ed$ on a scale $\beta m$ is no longer a local quantity with respect to the
central object. When $\beta$ grows without bound, $\ed_{\ell}$ 
approaches $0$ for all collapsed structures, since averaging over the whole
universe identically returns the background matter density,
which corresponds to a vanishing density contrast. In appendix \ref{ap0} we
show that  our double distribution becomes proportional to a 
Dirac delta-function around $\ed_{\ell}=0$ in the limit $\beta
\rightarrow \infty$ and proportional to a Dirac delta-function around
$\ed_{\ell}=\ed_{\rm 0,c}$ in the limit $\beta \rightarrow 1$.

We are now ready to use random walk theory results to 
calculate the double distribution. We first find 
the fraction of points in space which belong to
structures of mass between $m$ and $m+dm$, which in turn are embedded
in a medium of mean linearly extrapolated overdensity between $\ed_{\ell}$
and $\ed_{\ell}+d\ed_{\ell}$, $f(m, \ed_{\ell}, \beta)d\ed_{\ell}\,dm$. The double
distribution then is
\begin{equation}\label{prel1}
\frac{dn}{dm d\ed_{\ell}}= \frac{\rho_{\rm m,0}}{m}f(m,\ed_{\ell},\beta)dm d\ed_{\ell}.
\end{equation}

The quantity $f$ can be written as
\begin{equation}\label{prel2}
f dm \, d\ed_{\ell} = (f_1d\ed_{\ell})(f_2dm)
\end{equation}
where $f_1d\ed_{\ell}$ is the fraction
 of points in space which have an average overdensity between $\ed_{\ell}$
 and $\ed_{\ell}+d\ed_{\ell}$ on a smoothing scale $\beta m$, and $f_2 dm$ is the
 fraction of points satisfying the previous condition which belong to
 collapsed structures of mass between $m$ and $m+dm$.

The first of the two factors above is the fraction of points 
still in the walk which are found between
$\ed$ and $\ed+d\ed$ at a ``time''$\beta m$.
This is the solution of the 
1D random walk problem of $\ed_{\ell}$ as a function of 
$S$, with an absorbing boundary at 
the critical collapse threshold $\ed_{\rm 0,c}$, as given by equation (\ref{wforPS}) 
but for a smoothing scale $\beta m$, 
\begin{equation}
f_1d\ed_{\ell} = 
\frac{
\exp\left[-\frac{\ed_{\ell}^2}{2S(\beta m)}\right]
-\exp \left[\frac{(\ed_{\ell}-2\ed_{\rm 0,c}(a))^2}{2S(\beta m)}\right]}
{\sqrt{2\pi S(\beta m)}}
d\ed\,.
\end{equation}

The second factor ($f_2$) is the {\em conditional probability} that a point in
space originating from $(\beta m, \ed)$ in the mass - overdensity
plane,will reach the ``wall'' for the first time for  
a smoothing scale between $m$ and $m + dm$.
This is then the probability that 
a particular point in space is absorbed by the ``wall''
$\ed_{\rm 0,c}(a)$ at a particular ``time'' $S(m)$, 
provided that the origin of the walk is transferred from $(0,0)$ 
to $(S(\beta m),\ed_{\ell})$. This probablility can then be found if,
in the expression for $dP(>m,a)/dm$ (eq. \ref{dPdm}), we perform the
substitutions $\ed_{\rm 0,c} \rightarrow \ed_{\rm 0,c}-\ed_\ell$ and 
$S(m) \rightarrow S(m)-S(\beta m)$.
Similar conditional probabilities were originally calculated by
\cite{b91} and \cite{lc93}
in the context of rates of mergers between halos. In our case, it is
\begin{equation}
f_2 dm = \frac{\left[\ed_{\rm 0,c}(a)-\ed_{\ell}\right]
\exp\left[-\frac{\left(\ed_{\rm 0,c}(a)-\ed_{\ell}\right)^2}{2\left[S(m)-S(\beta m)\right]}\right]}
{\sqrt{2\pi}\left[S(m)-S(\beta m)\right]^{3/2}}
\left|\frac{dS}{dm}\right|_m dm\,.
\end{equation}

Equations (\ref{prel1}) and (\ref{prel2}) then give
\begin{widetext}
\begin{equation} \label{dd}
\frac{dn}{dmd\ed_{\ell}}(m,\ed_{\ell},\beta,a)=
\frac{\rho_{\rm m,0}}{m} \,\,
\frac{\ed_{\rm 0,c}(a)-\ed_{\ell}}{2\pi }\,\,
\frac{
\exp \left[-\frac{\ed_{\ell}^2}{2S(\beta m)}\right]
- \exp\left[-\frac{\left(\ed_{\ell} - 2 \ed_{\rm 0,c}(a)\right)^2}{2S(\beta
    m)}\right]
}
{[S(\beta m)]^{1/2}
\left[S(m)-S(\beta m)\right]^{3/2}}
\left|\frac{dS}{dm}\right|_m 
\exp\left[-\frac{\left(\ed_{\rm 0,c}(a)-\ed_{\ell}\right)^2}
{2\left[S(m)-S(\beta m)\right]}\right]
\end{equation}
\end{widetext}
with $\ed_{\ell} \le \ed_{\rm 0,c}(a)$ and $\beta > 1$ so $S(m) > S(\beta m)$
\footnote{since $S(m)$ monotonically decreases with $m$ for all
physically interesting power spectra}.
Equation (\ref{dd}) is the double distribution we have sought and is the
central result of this paper. 
Integrating $dn/(dmd\ed_{\ell})$ over $\ed_{\ell}$ yields the Press-Schechter 
mass function, as it should. The result is independent of the value of $\beta$.
We explicitly perform this integration in appendix \ref{ap0}.

Note that the functional form of our
double distribution is similar with that of the integrand used by \cite{mw96}
in their calculation of the cross-correlation between dark halos and
mass using random walk theory, however the second variance of the
field (corresponding to our $S(\beta m)$) in their case refers to a
fixed clustering radius and is independent of object mass.

\subsection{Converting between \boldmath{$\ed_{\ell}$} and \boldmath{$\delta_\ell$}}

In appendices \ref{ap1} and \ref{ap2} we derive exact expressions for 
$\ed_{\ell}(\delta_{\ell},a)$ in the case of the spherical evolution
model, for an $\Omega_{\rm m}=1$ (appendix \ref{ap1}) and an
$\Omega_{\rm m}+\Omega_\Lambda=1$ (appendix \ref{ap2}) universe 
(note however that all of the equations we have presented up to this point are
cosmology-independent, and can therefore be adapted for any cosmological
model). 

An excellent approximation 
to these conversion relations can be derived from
the expression
\begin{equation}\label{magic}
\ed_a \approx \ed_{\rm c}\left[1-(1+\delta_a)^{-1/\ed_{\rm c}}\right]\,.
\end{equation}
Similar approximations were suggested by \cite{b94} and \cite{s98}.
Equation (\ref{magic}) relates the linear overdensity at a time $a$ to
the true overdensity at the same time, and its accuracy is {\em  better than
$2\%$  throughout its domain} for both $\Omega_{\rm m}=1$ and
$\Omega_{\rm m}+\Omega_\Lambda=1$ cosmologies. Its functional form is
much simpler and more intuitive than the more accurate fit of
\cite{mw96}. 
The cosmological model enters only through $\ed_{\rm c}$.
For the Einstein-deSitter universe,
$\ed_{\rm c}$ is given by equation (\ref{matteredc}) for $a_{\rm coll}=1$, while
for the $\Omega_{\rm m}+\Omega_\Lambda=1$ universe it is given by equation 
\ref{edc_lambda} for $a=1$. Note that $\ed_{\rm c}$ is related to the
quantity $\ed_{\rm 0,c}(a)$ (which appears explicitly in our double
distribution expression) through 
\begin{equation}
\ed_{\rm 0,c}(a) = \ed_{\rm c} \frac{D(a_0)}{D(a)}
\end{equation}
where $D(a)$ is the linear growth factor in the relevant cosmology.

The limits of equation (\ref{magic}) are the same as the ones required
for the exact conversion relation. When
$|\delta_a| \ll 1$, $\ed_a \approx \delta_a$. In addition, 
$\ed_a \rightarrow -\infty$ as  $\delta_a \rightarrow -1$, and 
$\delta_a \rightarrow \infty$ as $\ed_a \rightarrow \ed_{\rm c}$.

Using equation (\ref{magic}), 
\begin{equation}\label{apconv}
\ed_{\ell} \approx \frac{D(a_0)}{D(a)}\ed_{\rm
  c}\left[1-(1+\delta_{\ell})^{-1/\ed_{\rm c}}\right]\,,
\end{equation}
where $a$ is the time at which 
we want to evaluate the double distribution.
Note that close to virialization, equation (\ref{apconv}) loses its
applicability (as does the spherical collapse model), and has to be 
replaced by a recipe which does not diverge in $\delta$. We present
such recipes in appendices \ref{ap1} and \ref{ap2}, however the exact
functional form of the conversion relation in this regime cannot
affect any of the physically interesting results as the amplitude of
the double distribution decreases rapidly enough with $\delta$
that the contribution of the high-delta tail 
to the integrated mass function is negligible. We have
verified this fact by comparing the integral of our double distribution
over $\delta$ with the Press-Schechter mass function. When we extended
the integration up to $\delta_{\rm c}$, the results agreed to the accuracy of
the numerical integration. When we extended our integration only up to
$\delta_{\rm v}$ (just below the application of our virialization
recipe), the error relative to the Press-Schechter mass function was
less than $0.02\%$.

\subsection{Clustering Scale Lengths
and Correction for Central Object Contamination}

The definition of $\beta$ and $\ed_{\ell}$ described above was sufficient
for us to derive the double distribution from random walk
theory. However, from a physical point of view, the presence of a
collapsed structure at the center of the ``environment sphere''
contaminates the evaluation of the average ``environmental''
overdensity. If we want the double distribution to describe the
properties of matter {\em surrounding} collapsed objects, we need to
correct for the presence of the objects themselves. 

We will employ a simple, ``top-hat'' physical picture to calculate an
appropriate correction (see also \cite{s98}). Note however that our correction is 
approximate, since the filter we used to smooth the
overdensity field was $k-$sharp rather than top-hat in space. 

Let $\delta_{\rm c}$ be the (true) overdensity of a collapsed object of mass
m and radius $R_{\rm v}$, and $\delta_{\ell}$ be the overdensity of the
``environment sphere'' of radius $R_{\rm e}$. The ``environment sphere''
encompasses a mass $\beta m$, including the central collapsed
object. We want to find the average overdensity $\delta_{\rm ext}$ of
that part of the ``environment sphere'' which is {\em external} to the central object. For the
collapsed object we can write
\begin{equation}\label{cor1}
m = \frac{4}{3} \pi R_{\rm v}^3(1+\delta_{\rm c})\rho_{\rm m}\,,
\end{equation}
where $\rho_{\rm m}$ is the mean matter density of the universe at the
epoch of interest. For the environment sphere, including the central
object, we can write
\begin{equation}\label{cor2}
\beta m = \frac{4}{3} \pi R_{\rm e}^3(1+\delta_{\ell})\rho_{\rm m}\,.
\end{equation}
From equations (\ref{cor1}) and (\ref{cor2}) we get $R_{\rm v}^3 = R_{\rm
  e}^3
(1+\delta_{\ell})/\beta(1+\delta_{\rm c})$. 
It thus 
follows that the length scale $R_{\rm e}$ associated
with the clustering parameter $\beta$ is
\begin{eqnarray}
\label{eq:size}
R_{\rm e} &=& \left( \frac{\beta(1+\delta_{\rm
 c})}{1+\delta_{\ell}} \right)^{1/3}  R_{\rm v} \nonumber \\
 &=& \left( \frac{3 \beta m}{4 \pi (1+\delta_{\ell}) \rho_{\rm m}} \right)^{1/3} \, .
\end{eqnarray}
We see that for a fixed clustering parameter $\beta$,
the length scale associated with an object of mass $m$
is mass-dependent, scaling linearly with the virial
radius but larger by a factor
$\left[\beta(1+\delta_{\rm c})/(1+\delta_{\ell})\right]^{1/3} > 1$.
Thus, we can roughly think of the clustering scale parameter
as a measure of how many virial radii
we include as the local environment around each
structure \footnote{Note however that for fixed $\beta$, the number of
virial radii included in the environment depends on $\delta_{\ell}$
and is larger for underdense environments.}.

Having identified the environmental length scale,
we can now isolate the environmental overdensity
from that of the collapsed object.
The volume of the environment sphere external to the
central object contains a mass
\begin{equation}\label{cor3}
(\beta-1) m = \frac{4}{3} \pi (R_{\rm e}^3-R_{\rm v}^3)
(1+\delta_{\rm ext})\rho_{\rm m}\,.
\end{equation}
Using equation (\ref{eq:size}) to eliminate $R_{\rm v}$, and dividing by
equation (\ref{cor2}) we obtain 
\begin{equation}
\delta_{\rm ext} = \frac{(\beta-1)(1+\delta_{\ell})(1+\delta_{\rm c})}
{\beta(1+\delta_{\rm c})-(1+\delta_{\ell})}-1\,,
\end{equation}
which is the contamination-corrected overdensity for an environment
sphere with uncorrected overdensity 
$\delta_{\ell}$. Then, the contamination-corrected double
distribution will be given by 
\begin{equation}
\frac{dn}{dmd\delta_{\rm ext}}(\delta_{\rm ext},m,a)=
\frac{dn}{dmd\delta_{\ell}} \left[\delta_{\ell}(\delta_{\rm
    ext},a),m,a\right]\frac{d
  \delta_{\ell}}{d\delta_{\rm ext}} \,, 
\end{equation}
where 
\begin{equation}
\delta_{\ell}(\delta_{\rm ext}) = \frac{\beta(1+\delta_{\rm
    ext})(1+\delta_{\rm c})} {(\beta-1)(1+\delta_{\rm c})+(1+\delta_{\rm ext})}-1
\end{equation}
and 
\begin{equation}
\frac{d\delta_{\ell}}{d\delta_{\rm ext}} = 
\frac{\beta (\beta -1)(1+\delta_{\rm c})^2}
{\left[(\beta-1)(1+\delta_{\rm c})+(1+\delta_{\rm ext})\right]^2}\,.
\end{equation}

\subsection{Derivative Quantities}

We now have enough tools to calculate derivative quantities of
interest. The number density of collapsed objects of mass greater than
some minimum $m_{\rm min}$ \footnote{The introduction of a finite
  minimum mass $m_{\rm min}$ is necessary for both physical and
  technical reasons. Physically, the mass of collapsed objects is
  strictly forced to have a lower bound, not only due to the finite
  mass of the dark matter particle, but also due to the existence of a
dark matter Jeans mass, however small this may be. Practically, 
the Press-Schechter $dn/dm$ diverges as $m \rightarrow 0$ and setting
a minimum mass is required to extract interesting information.}
embedded 
in a medium of local overdensity between $\delta_{\rm ext}$ and 
$\delta_{\rm ext}+d \delta_{\rm ext}$ is 
\begin{equation}
\frac{dn}{d\delta_{\rm ext}}(>m_{\rm min})d\delta_{\rm ext} = 
d\delta_{\rm ext} \frac{\partial\ed_{\ell}}{\partial\delta_{\ell}}
\frac{d\delta_{\ell}}{d\delta_{\rm ext}}
\int_{m=m_{\rm min}}^{\infty} 
\!\!\!\frac{dn}{dmd\ed_{\ell}} dm\,,
\end{equation}
while the density of matter in collapsed objects of mass $ > m_{\rm min}$
embedded in a medium of local overdensity between $\delta_{\rm ext}$ 
and $\delta_{\rm ext}
+d\delta_{\rm ext}$ is
\begin{equation}\label{matdelta}
\frac{d\rho}{d\delta_{\rm ext}}(>m_{\rm min})d\delta_{\rm ext} = 
d\delta_{\rm ext} \frac{\partial\ed_{\ell}}{\partial\delta_{\ell}}
\frac{d\delta_{\ell}}{d\delta_{\rm ext}}
\int_{m=m_{\rm min}}^{\infty} \!\!\!m 
\frac{dn}{dmd\ed_{\ell}} dm\,.
\end{equation}

Of all the matter in the universe which belongs to collapsed objects
of mass $>m_{\rm min}$, the fraction by mass which lives in underdense
neighborhoods is
\begin{equation}
f_{\rm \rho,un} = \frac{
\int_{\delta_{\rm ext}=-1}^{0} \frac{d\rho}{d\delta_{\rm
    ext}}(>m_{\rm min})d\delta_{\rm ext}
}
{\int_{\delta_{\rm ext}=-1}^{\delta_{\rm c}} 
\frac{d\rho}{d\delta_{\rm ext}}
(>m_{\rm min})d\delta_{\rm ext}}\,.
\end{equation}
Then, the mass fraction of the matter defined above which 
lives in overdensities will be $f_{\rm \rho, ov} = 1 - f_{\rm \rho, un}$.

Similarly, of all the objects with mass $m>m_{\rm min}$, a fraction
by number which lives inside underdensities is
\begin{equation}
f_{\rm n,un} = \frac{
\int_{\delta_{\rm ext}=-1}^{0} \frac{dn}{d\delta_{\rm ext}}(>m_{\rm
  min})d\delta_{\rm ext}
}
{\int_{\delta_{\rm ext}=-1}^{\delta_{\rm c}} \frac{dn}{d\delta_{\rm
      ext}}(>m_{\rm min})d\delta_{\rm ext}}\,.
\end{equation}
The complementary number fraction of such structures living inside
overdensities will be $f_{\rm n, ov} = 1 - f_{\rm n,un}$.

The number-density--weighted mean $\delta_{\rm ext}$ for structures of mass
$>m_{\rm min}$ is 
\begin{equation}\label{parnmean}
\langle\delta\rangle_{\rm n} = 
\frac{\int_{\delta_{\rm ext}=-1}^{\delta_{\rm c}}\delta_{\rm ext}
  \frac{dn}{d\delta_{\rm ext}}(>m_{\rm
    min})d\delta_{\rm ext}}
{\int_{\delta_{\rm ext}=-1}^{\delta_{\rm c}}\frac{dn}{d\delta_{\rm ext}}(>m_{\rm
    min})d\delta_{\rm ext}}
\end{equation}
with a variance
\begin{equation}\label{parnvar}
\sigma^2_{\rm \delta,n} =  
\frac{\int_{\delta_{\rm ext}=-1}^{\delta_{\rm c}}(\delta_{\rm ext}
  -\langle\delta\rangle_{\rm n})^2 \frac{dn}{d\delta_{\rm ext}}(>m_{\rm
    min})d\delta_{\rm ext}}
{\int_{\delta_{\rm ext}=-1}^{\delta_{\rm c}}\frac{dn}{d\delta_{\rm ext}}(>m_{\rm
    min})d\delta_{\rm ext}}\,.
\end{equation}

Similarly, the matter-density--weighted mean $\delta$ for structures of mass
$>m_{\rm min}$ is 
\begin{equation}
\langle\delta\rangle_{\rm \rho}\label{parrhomean} = 
\frac{\int_{\delta_{\rm ext}=-1}^{\delta_{\rm c}}\delta_{\rm ext}
  \frac{d\rho}{d\delta_{\rm ext}}(>m_{\rm
    min})d\delta_{\rm ext}}
{\int_{\delta_{\rm ext}=-1}^{\delta_{\rm c}}\frac{d\rho}{d\delta_{\rm ext}}(>m_{\rm
    min})d\delta_{\rm ext}}
\end{equation}
with a variance
\begin{equation}\label{parrhovar}
\sigma^2_{\rm \delta,\rho}  =
\frac{\int_{\delta_{\rm ext}=-1}^{\delta_{\rm c}}\left(\delta_{\rm ext} -
  \langle\delta\rangle_{\rho}\right)^2 
\frac{d\rho}{d\delta_{\rm ext}}(>m_{\rm
    min})d\delta_{\rm ext}}
{\int_{\delta_{\rm ext}=-1}^{\delta_{\rm c}}\frac{d\rho}{d\delta_{\rm ext}}(>m_{\rm
    min})d\delta_{\rm ext}}\,.
\end{equation}

\section{Results}\label{res} 

\begin{figure*}
\resizebox{3in}{!}{
\includegraphics{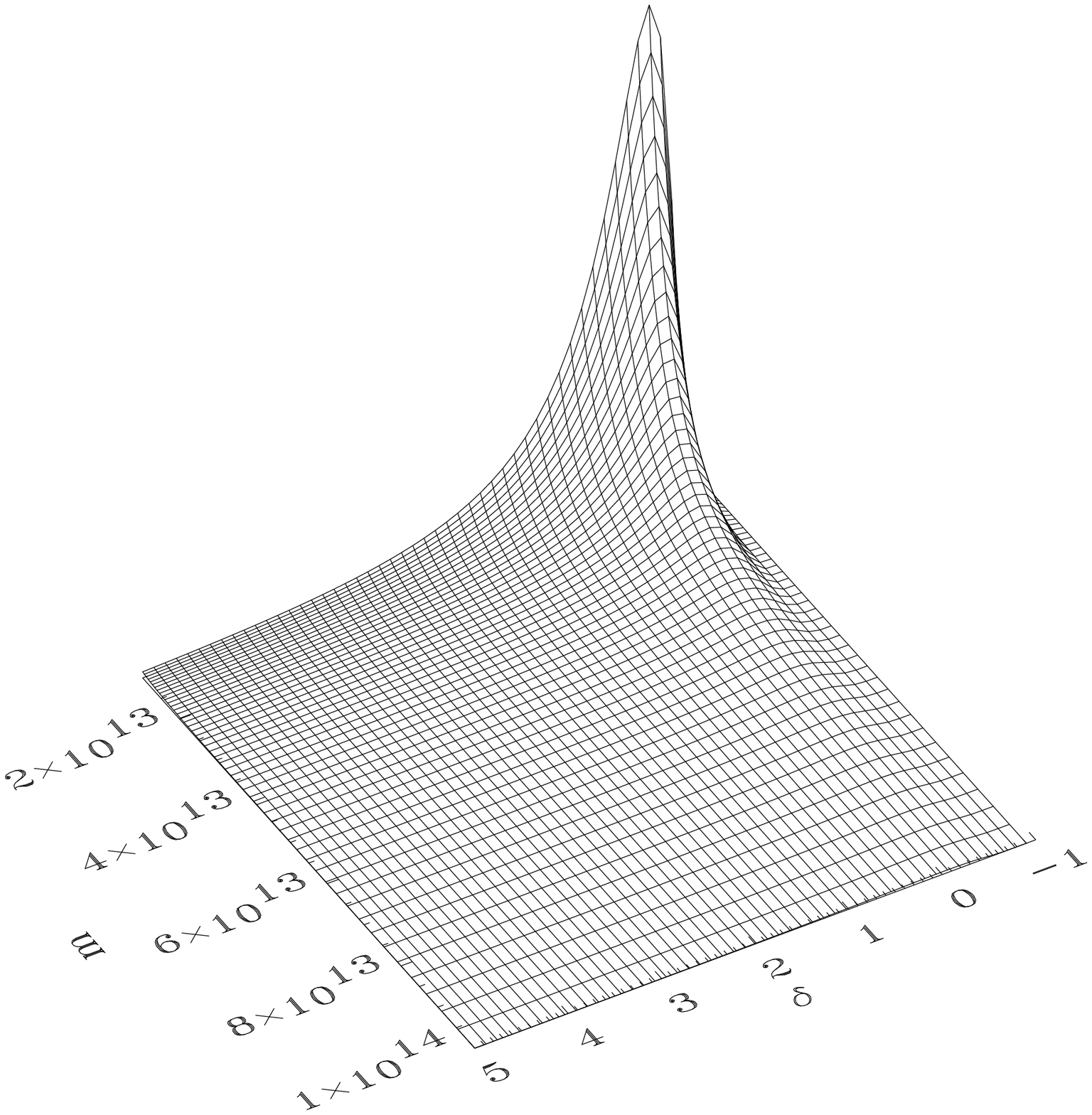}}
\resizebox{3in}{!}{
\includegraphics{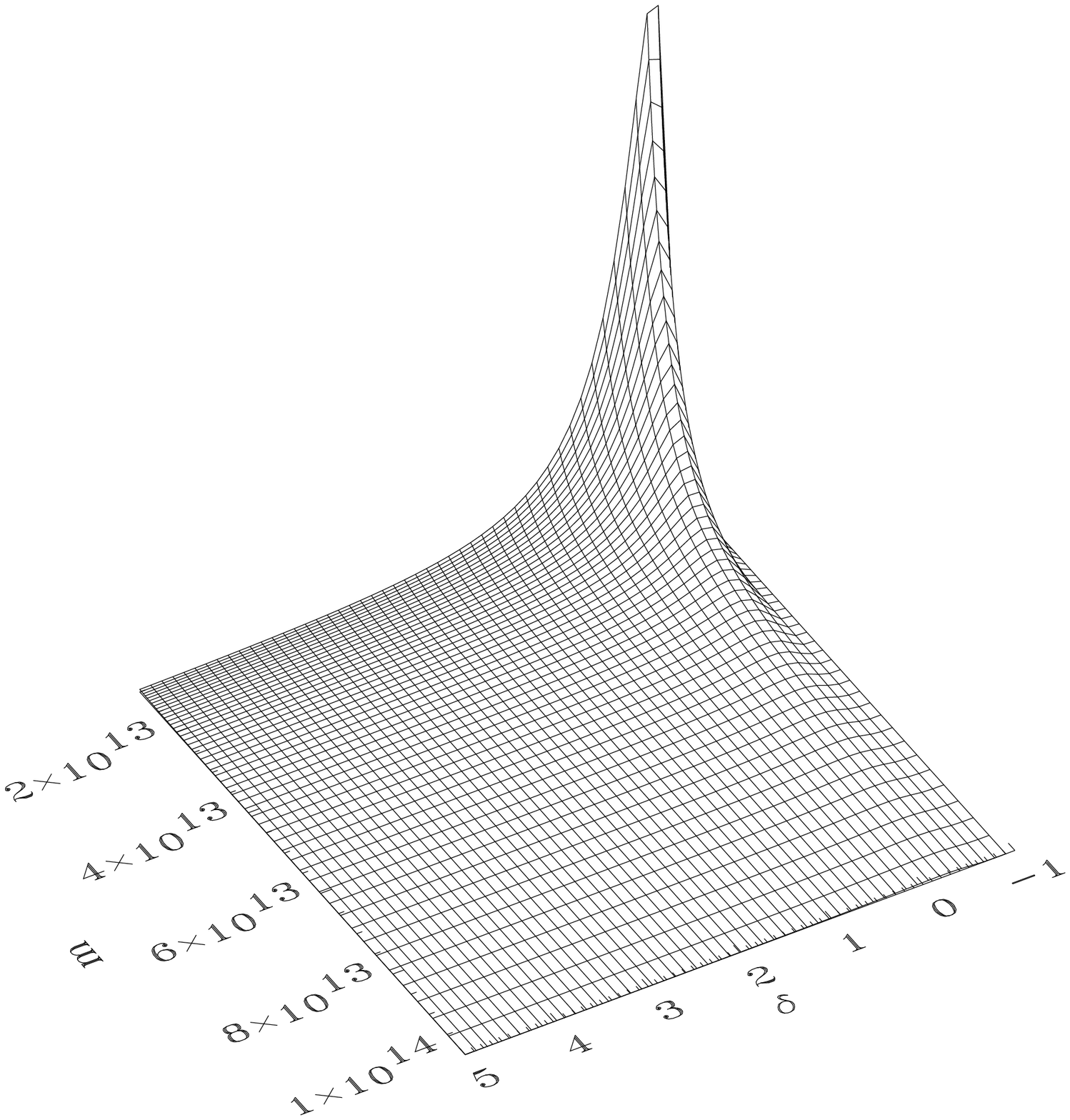}}
\caption{\label{fig:surfaces} Surface plots of the double distribution
for $z=0$ and $\beta=2$ in $\Omega_{\rm m}+\Omega_{\rm \Lambda} =
1$ (left panel) and Einstein-deSitter (right panel) universes. The mass
is measured in ${\rm M_\odot}$. The vertical axis is linear, with the
$m-\delta$ axes level corresponding to $dn/dmd\delta_{\ell} = 0$ and the highest
point corresponding to $dn/dmd\delta_{\ell} = 2.72\times 10^{-2}$ (left
panel)  and
$dn/dmd\delta_{\ell} =1.89\times10^{-1} $ (right panel) objects per 
${\rm Mpc ^3}$ per $10^{15} {\rm M_\odot}$.}
\end{figure*}

In this section we present plots of the double distribution itself
as well as of its various physically interesting derivative
quantities. We compare results derived for a concordance, $\Omega_{\rm
m} +\Omega_\Lambda=1$ universe with WMAP parameters ($\sigma_8=0.84$, 
$h=0.71$, $\Omega_{\rm m}=0.27$, $\Omega_{\rm b} = 0.04$,
\cite{wmap}), and for an Einstein-deSitter ($\Omega_{\rm m}=1$)
universe  with $h=0.71$, $\Omega_{\rm b} = 0.04$, but $\sigma_8=0.45$.
The different power-spectrum normalization in the Einstein-deSitter
case was selected so that the Press-Schechter mass function in this
case coincides with that of the concordance universe on a mass scale 
of $5.5\times 10^{14} {\rm M_\odot}$, which is between the values of
$m_8$ (mass included in a sphere of comoving radius $8h^{-1}$ Mpc) for
the two cosmologies ($m_8=2\times10^{14} {\,\rm M_\odot}$ 
for the concordance universe while $m_8=8\times10^{14} {\,\rm
  M_\odot}$ for the Einstein-de Sitter universe). 
This value of $\sigma_8$ is also consistent with
the fits of \cite{ecf} given the WMAP result for the concordance
universe. 
Finally, we use fitting formulae of \cite{bard86} for the adiabatic cold dark
matter transfer function to calculate the density field variance $S(m)$.
In this section, $\delta$ always refers to $\delta_{\rm ext}$, the
true overdensity of that part of the ''environment sphere'' which is
external to the central object. 

Figure \ref{fig:surfaces} shows a 3-dimensional rendering of our double
distribution as a function of mass and overdensity for fixed $\beta =
2$ and $z=0$.  The left panel corresponds to the concordance universe
while the right panel corresponds to the Einstein-deSitter
universe, and this arrangement is retained throughout this section. 

\begin{figure*}
\resizebox{3in}{!}{
\includegraphics{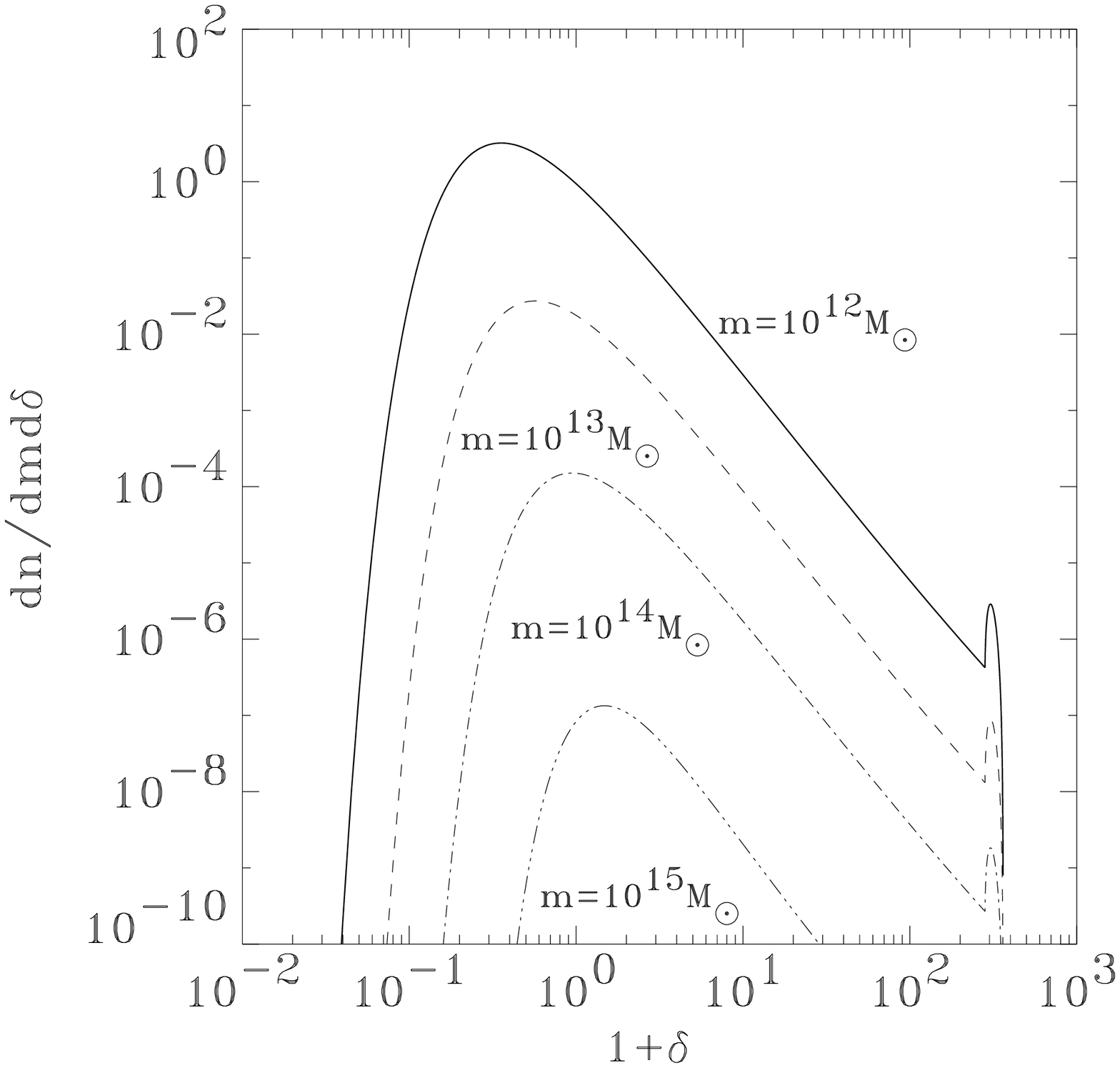}}
\resizebox{3in}{!}{
\includegraphics{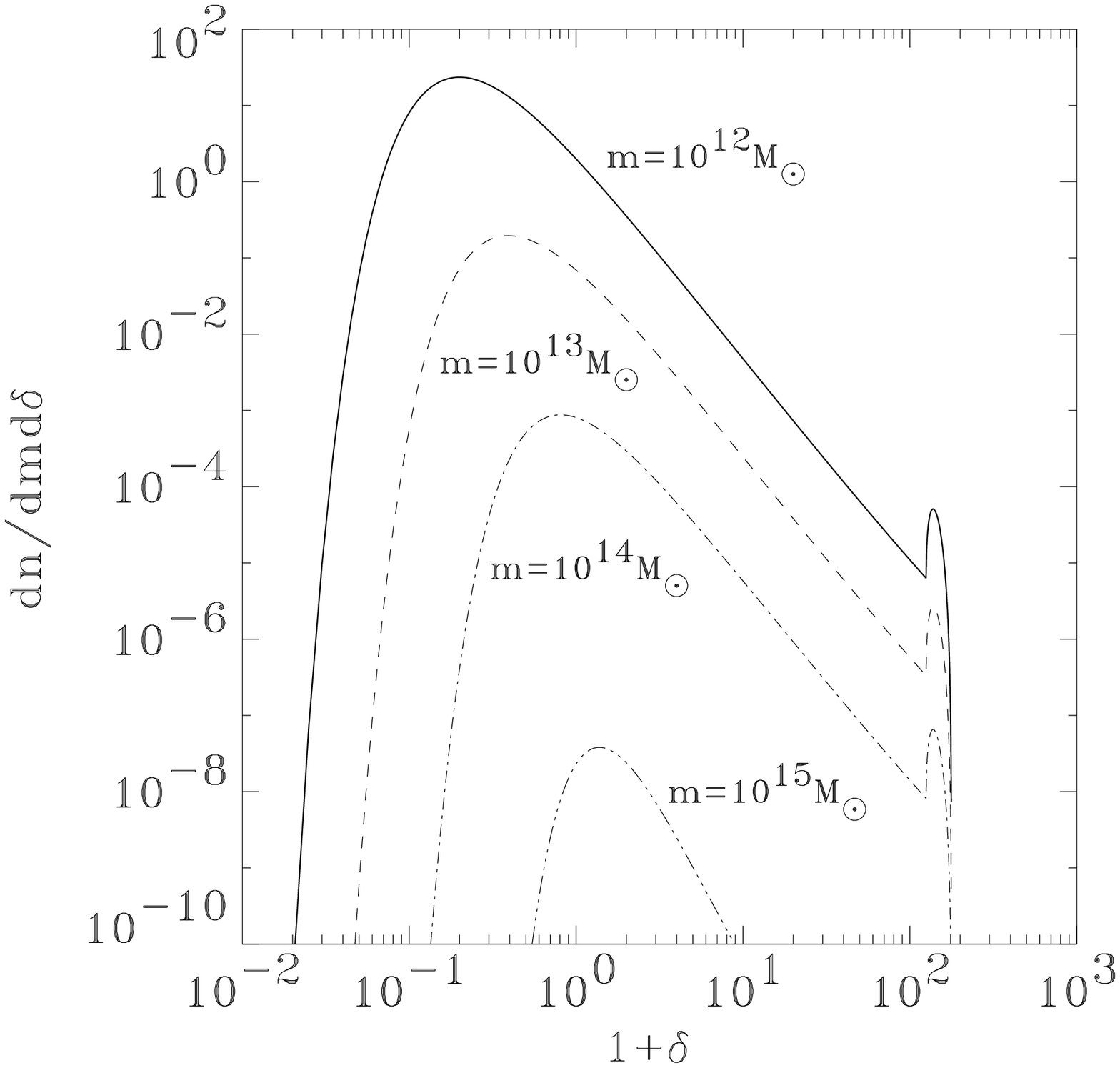}}
\caption{\label{fig:efofm} Slices of the double distribution function
  at various fixed values of the mass for 
$\Omega_{\rm m}+\Omega_{\rm \Lambda} =
1$ (left panel) and  Einstein-deSitter (right panel) universes. The
  units of the double distribution are number of objects per ${\rm Mpc
  ^3}$ per $10^{15} {\rm M_\odot}$.}
\end{figure*}

\begin{figure*}
\resizebox{3in}{!}{
\includegraphics{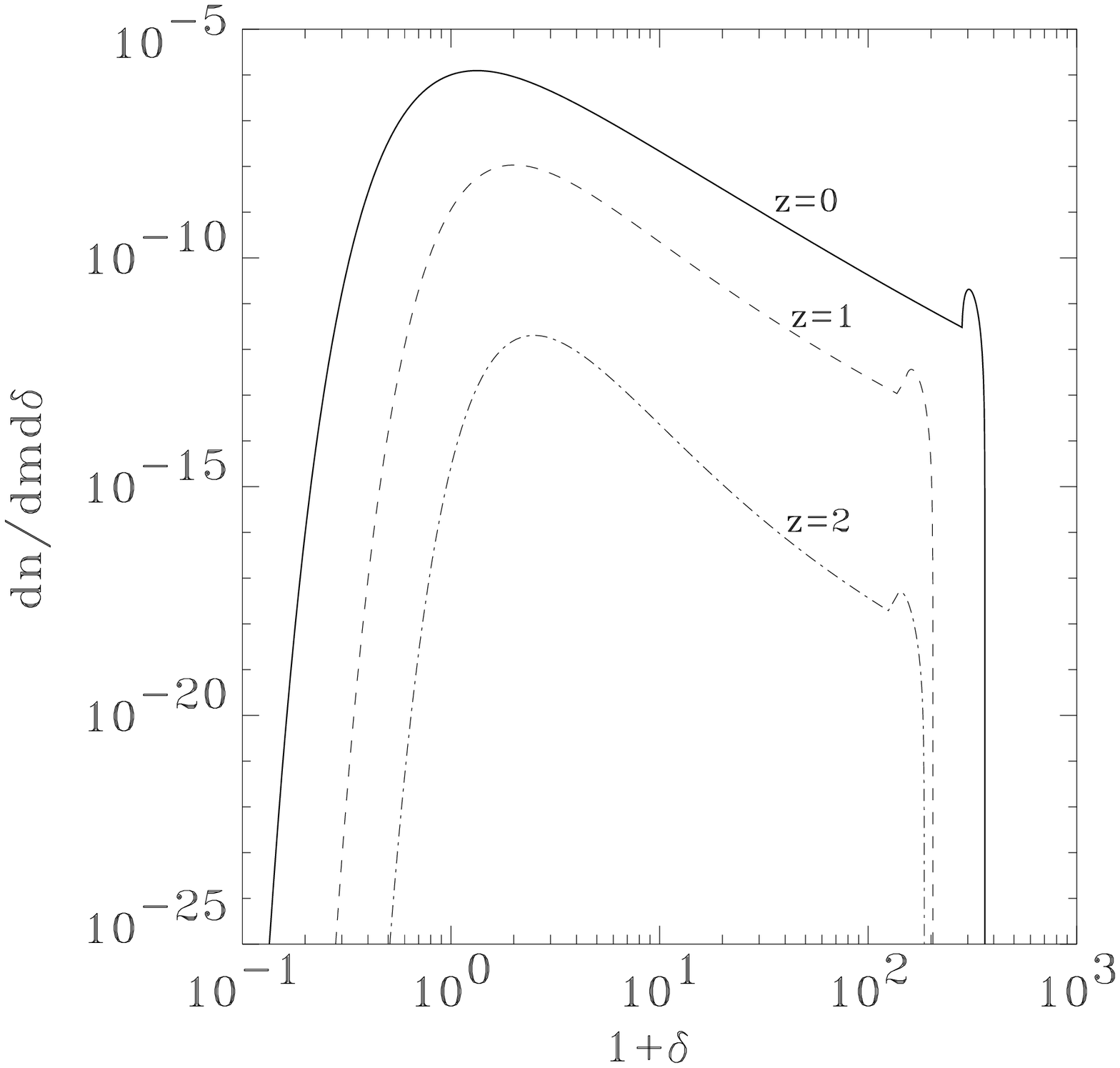}}
\resizebox{3in}{!}{
\includegraphics{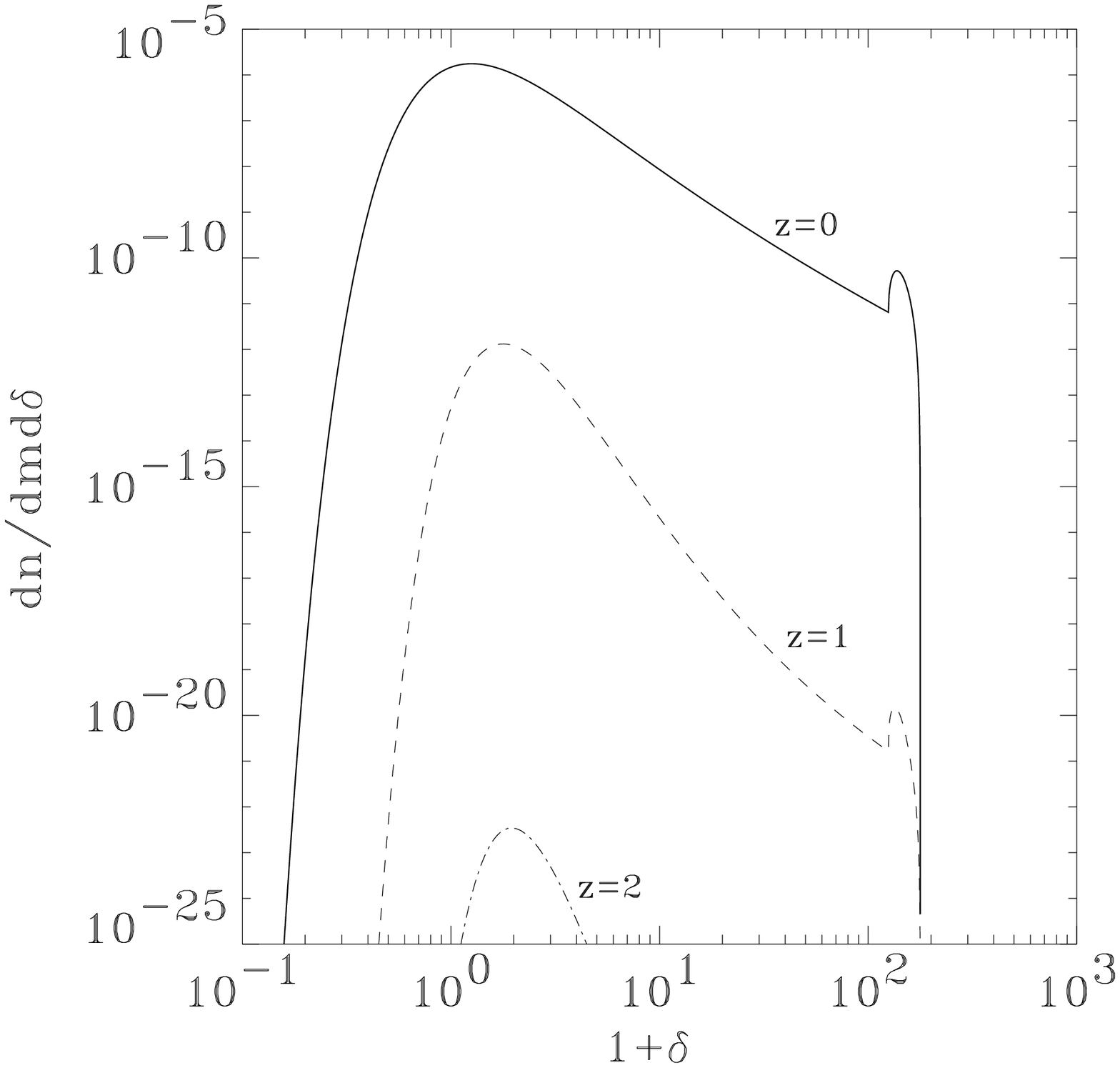}}
\caption{\label{fig:efofz} Slices of the double distribution function
  at $m=5.5\times10^{14}{\rm ,M_\odot}$ and for various values of
  redshift $z$, 
for $\Omega_{\rm m}+\Omega_{\rm \Lambda} =
1$ (left panel) and  Einstein-deSitter (right panel) universes. The
  units of the double distribution are number of objects per ${\rm Mpc
  ^3}$ per $10^{15} {\rm M_\odot}$.}
\end{figure*}

\begin{figure*}
\resizebox{3in}{!}{
\includegraphics{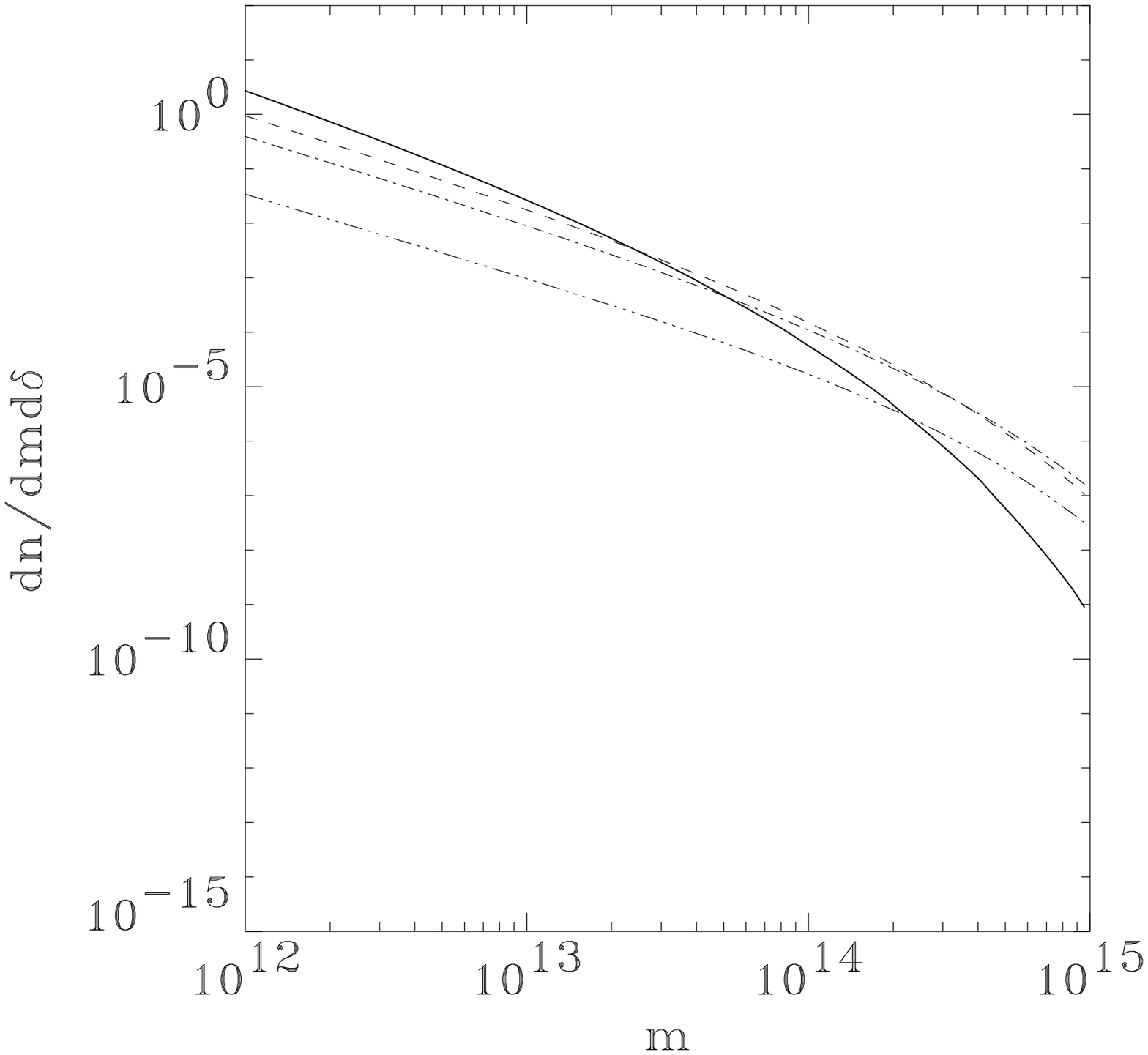}}
\resizebox{3in}{!}{
\includegraphics{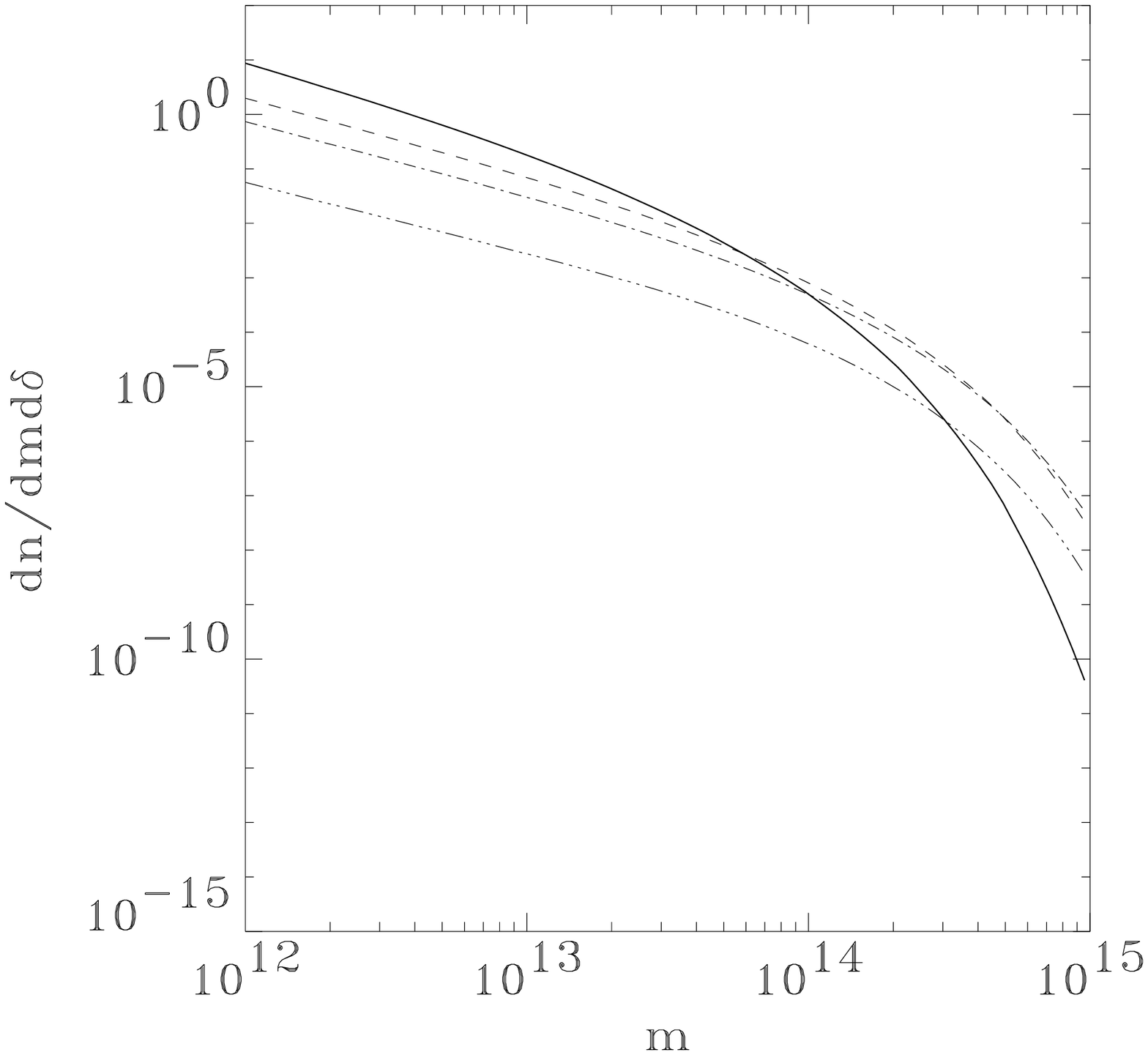}}
\caption{\label{fig:efofd_low} Slices of the double distribution function
  at constant values of $\delta$ 
for $z=0$, $\beta=2$ and for $\Omega_{\rm m}+\Omega_{\rm \Lambda} =
1$ (left panel) and  Einstein-deSitter (right panel) universes. Solid
  line: $\delta = -0.5$; dashed line: $\delta=0$; dot-dashed line:
  $\delta=0.5$; double-dot--dashed line: $\delta=3$. The
  units of the double distribution are number of objects per ${\rm Mpc
  ^3}$ per $10^{15} {\rm M_\odot}$.}
\end{figure*}

\begin{figure*}
\resizebox{3in}{!}{
\includegraphics{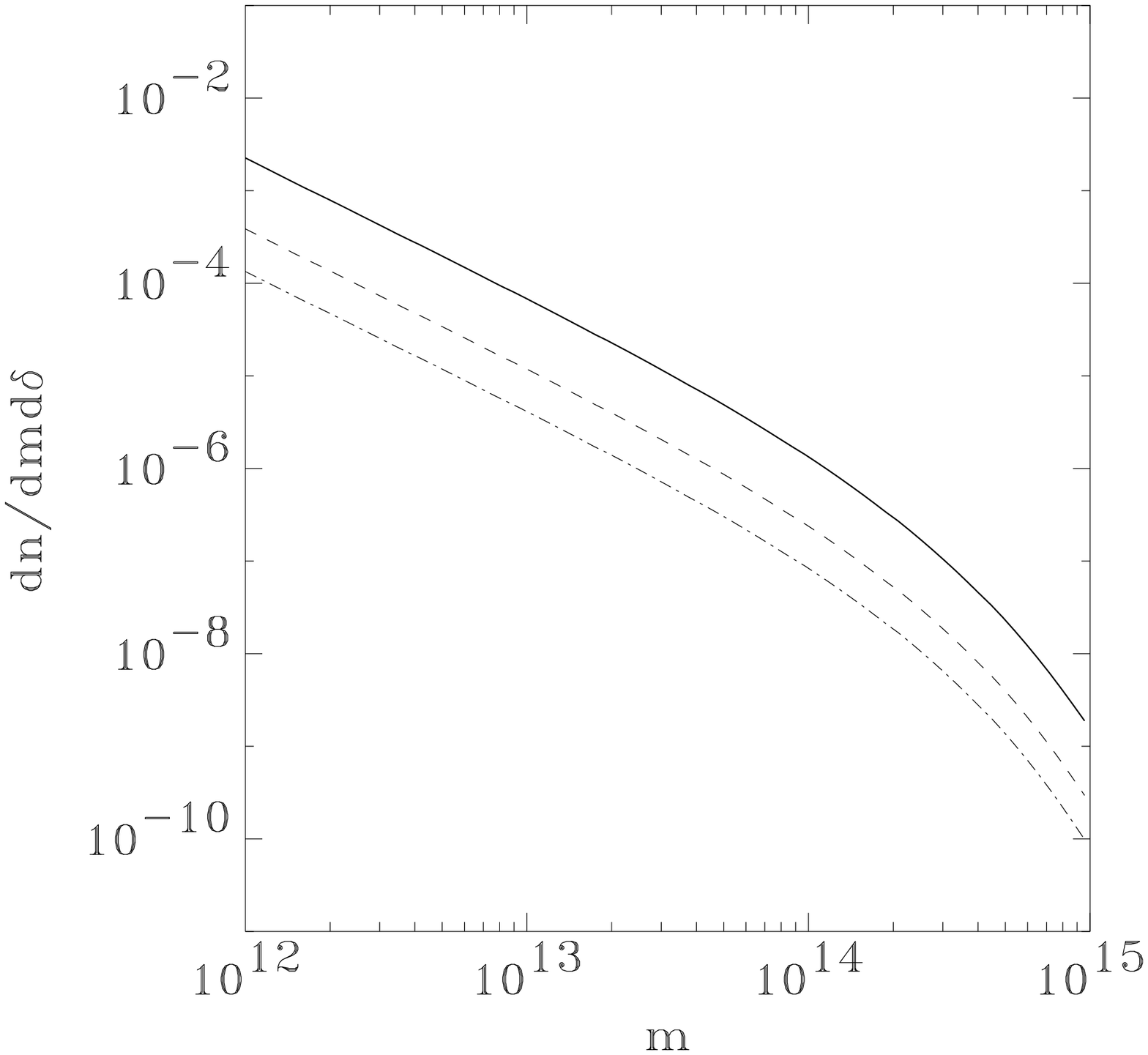}}
\resizebox{3in}{!}{
\includegraphics{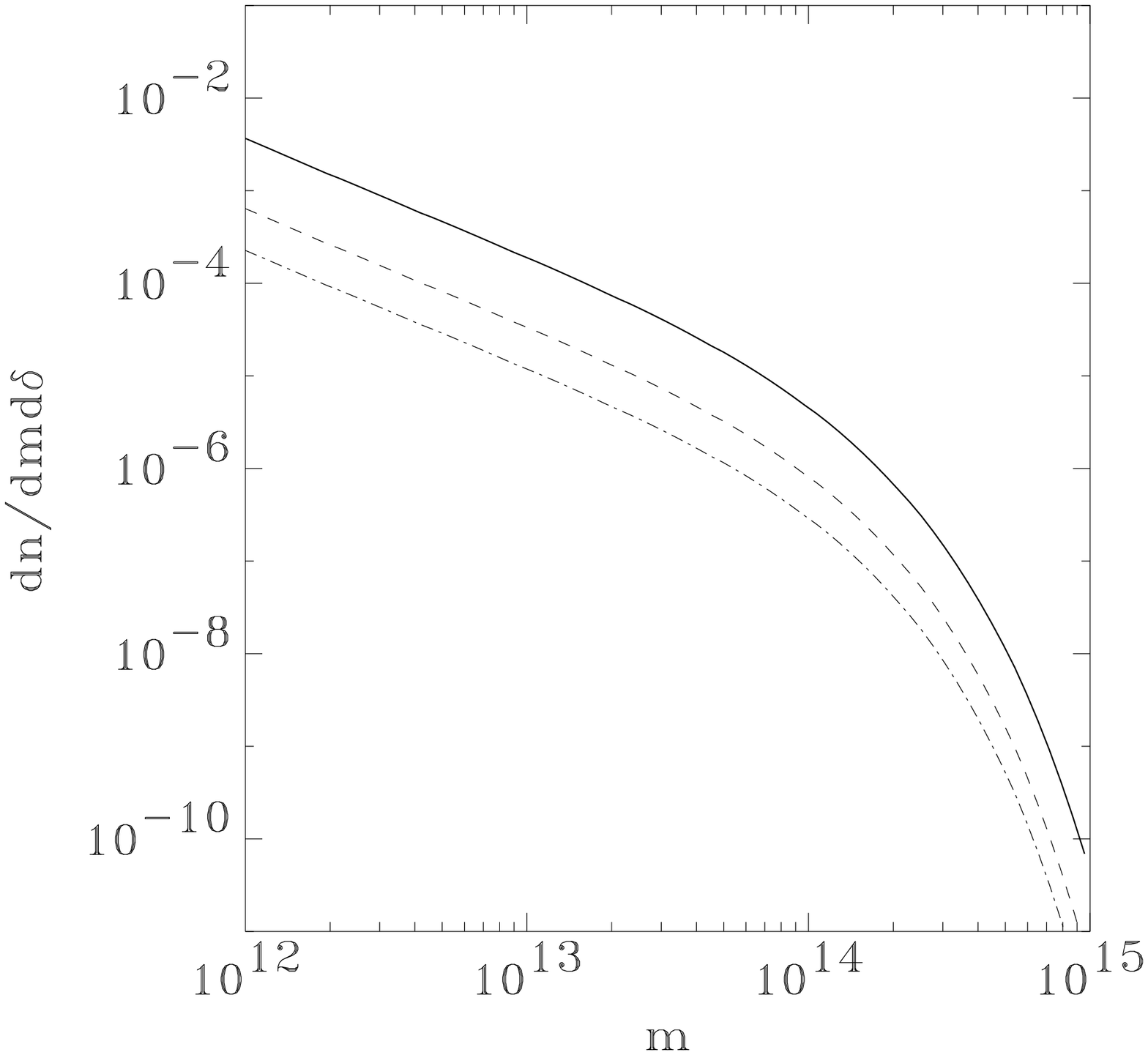}}
\caption{\label{fig:efofd_high} Slices of the double distribution function
  at constant values of $\delta$ 
for $z=0$, $\beta=2$, and for $\Omega_{\rm m}+\Omega_{\rm \Lambda} =
1$ (left panel) and  Einstein-deSitter (right panel) universes. Solid
  line: $\delta = 10$; dashed line: $\delta=20$; dot-dashed line:
  $\delta=30$. The
  units of the double distribution are number of objects per ${\rm Mpc
  ^3}$ per $10^{15} {\rm M_\odot}$.}
\end{figure*}

The features of the double distribution are demonstrated in more quantitative 
detail in Figures \ref{fig:efofm}-\ref{fig:efofd_high}. Figures 
\ref{fig:efofm} and \ref{fig:efofz} show slices of the double
distribution at constant values of mass. In Figure \ref{fig:efofm},
different curves correspond to different values of the central object
mass. In Fugure \ref{fig:efofz}, all curves are for an object mass
$m=5.5 \times 10^{15} {\rm \, M_\odot}$, and different curves
correspond to different redshifts. Their most prominent feature is
the pronounced peak at a relatively low value of $|\delta|$,
indicating that for each given pair of $z$ and $m$, there is a
preferred, ``most probable'' value of the local environment density
contrast. As we can see in Figure \ref{fig:efofm}, the location of
this peak moves to higher values of the density contrast as the mass
of the object increases: small structures are preferentially located 
in relative isolation, while larger structures are more likely to be found in
clustered environments. This result fits well in the picture of
hierarchical structure formation, as smaller structures tend to be
merged into higher-mass objects as time progresses. Lower-mass
objects which are initially part of underdensities are less probable
to undergo mergers, and hence are more likely to survive at late times
than objects which are initially part of overdensities. Conversely,
higher-mass structures are more likely to be parts of overdensities
where they can accumulate mass more easily through mergers with smaller
structures.  

Note, however, that in the hierarchical structure formation picture,
the mass scale where the exponential suppression of collapsed
structures sets in increases with time. Thus, any given mass scale
starts out as being a ``high mass'' at early times and eventually
becomes a ``lower mass'' as it enters the power-law regime of
the Press-Schechter mass function. Hence, according to the argument
we used to explain Figure \ref{fig:efofm}, the double distribution for
any given mass scale should peak at increasing $\delta$ values with
increasing redshift. This is because a particular mass scale used to be closer
to the high-mass end of the halo distribution in the past than it is
today. Indeed, this is the trend seen in Figure
\ref{fig:efofz}. As we would expect, the peak of the distribution moves to
higher $\delta$ values with increasing redshift.  
The significantly
more pronounced suppression of this mass scale in high redshifts in
the Einstein-deSitter universe is due to the different power-spectrum
normalization in the two cosmological models. Because of our choice 
in the power-spectrum normalization,
the exponential suppression in the number density of structures sets in at lower
masses  in the Einstein-de Sitter case than in the concordance
universe. 
Thus, there is a tendency to 
see more structures of higher mass in our concordance results than in
the Einstein-de Sitter case, despite the intuitive expectation that a
higher $\Omega_{\rm m}$ universe should have more massive structures
at late times due to its ability to continue to form structures even
at the present epoch. This would indeed have been the case if 
the power-spectrum had been normalized in the same way. 

That halos of a given mass are more strongly clustered with increasing
redshift was also found by \cite{mw00}, who used $\Delta_8 (m)$ (the rms
overdensity in the number of haloes more massive than some mass scale
after smoothing with a spherical top-hat filter of comoving radius $8
h^{-1}$ Mpc) as a measure for halo clustering. A tendency of higher
mass objects to be found in overdense regions was discussed by
\cite{mw96} and \cite{st02}, who interpreted it by viewing halos today
as progenitors of future larger-scale structures viewed at ``high'' or
``low'' redshift.


\begin{figure*}
\resizebox{3in}{!}{
\includegraphics{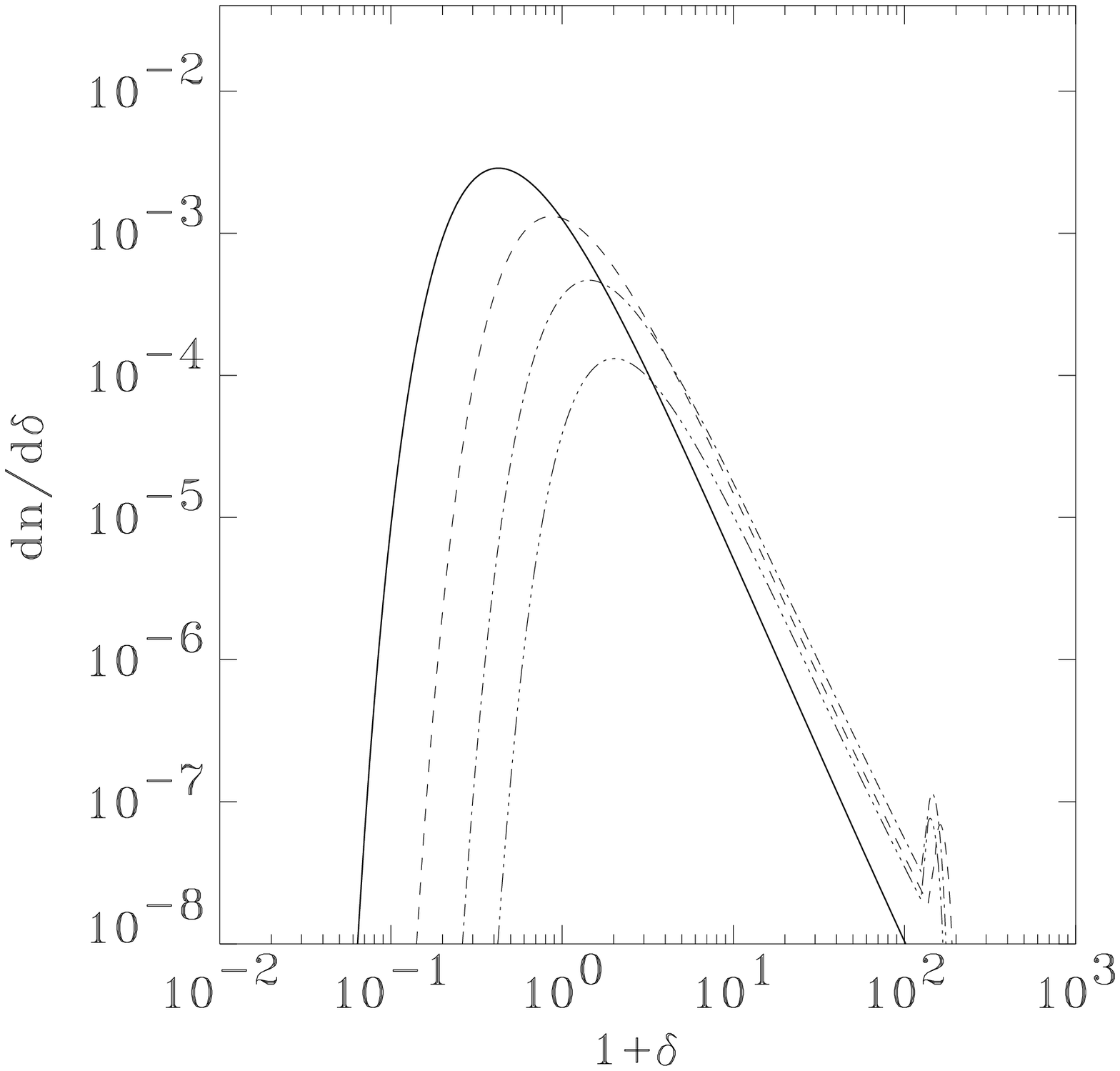}}
\resizebox{3in}{!}{
\includegraphics{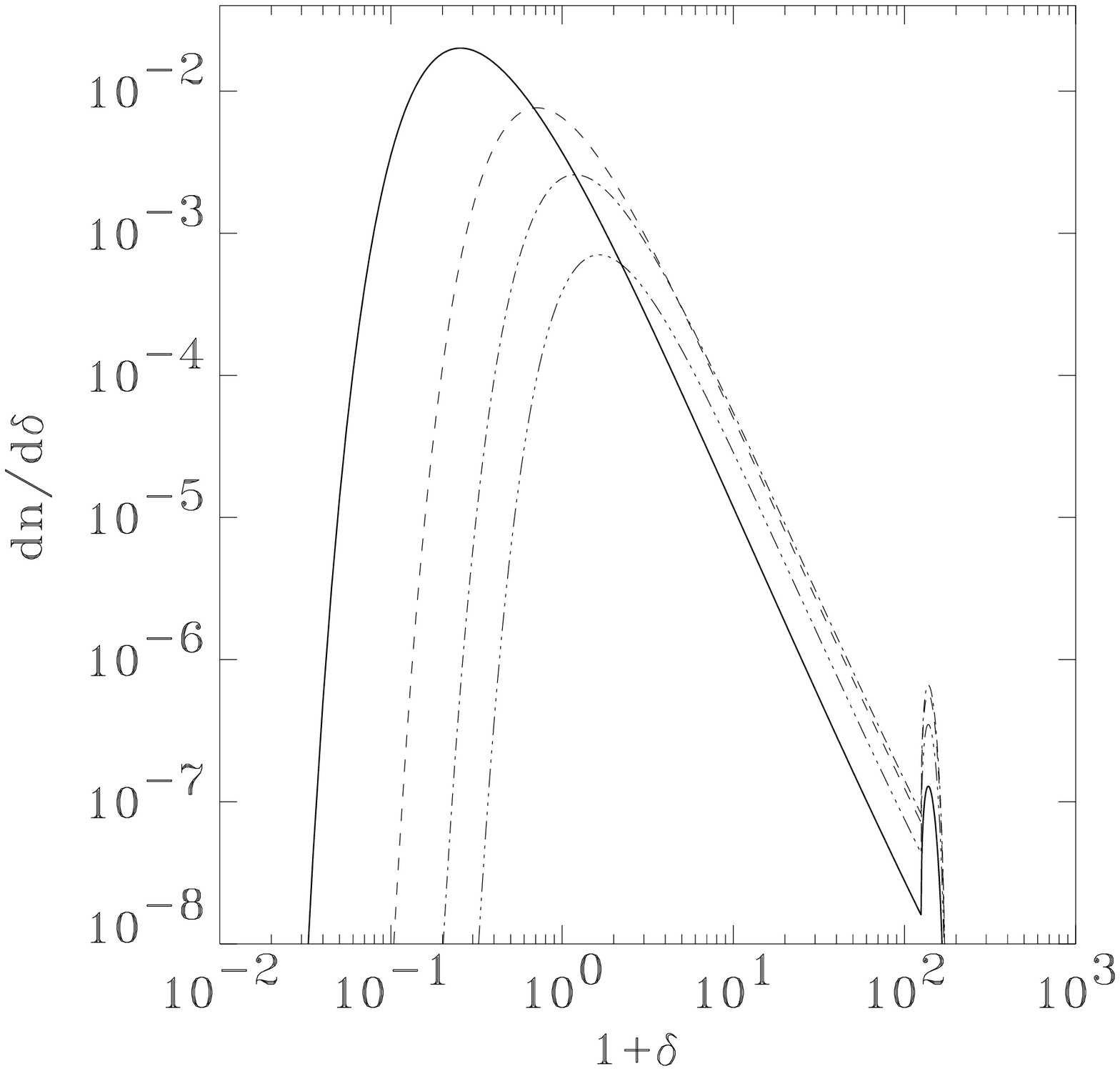}}
\caption{\label{fig:nz} Distribution of structures 
with respect to local density contrast, 
  $dn/d\delta (> 10^{12} {\, \rm M_\odot})$, 
for $\beta = 2$ and for $\Omega_{\rm m}+\Omega_{\rm \Lambda} =
1$ (left panel) and  Einstein-deSitter (right panel) universes. 
  Solid line: $z=0$; dashed line: $z=1$; dot-dashed line: $z=2$;
 double-dot--dashed line: $z=3$. The
  units of  $dn/d\delta$ are number of objects per ${\rm Mpc
  ^3}$.}
\end{figure*}

\begin{figure*}
\resizebox{3in}{!}{
\includegraphics{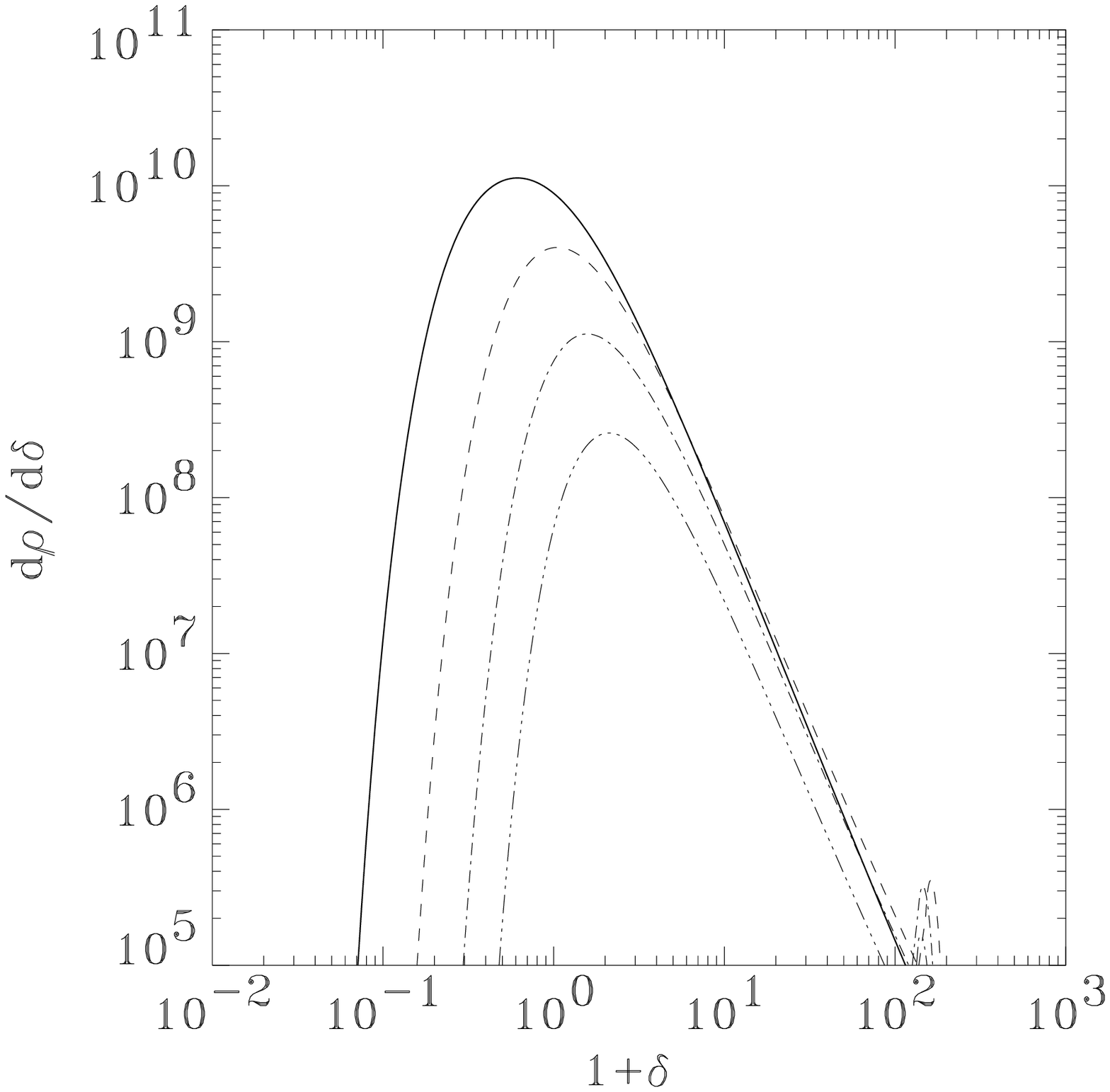}}
\resizebox{3in}{!}{
\includegraphics{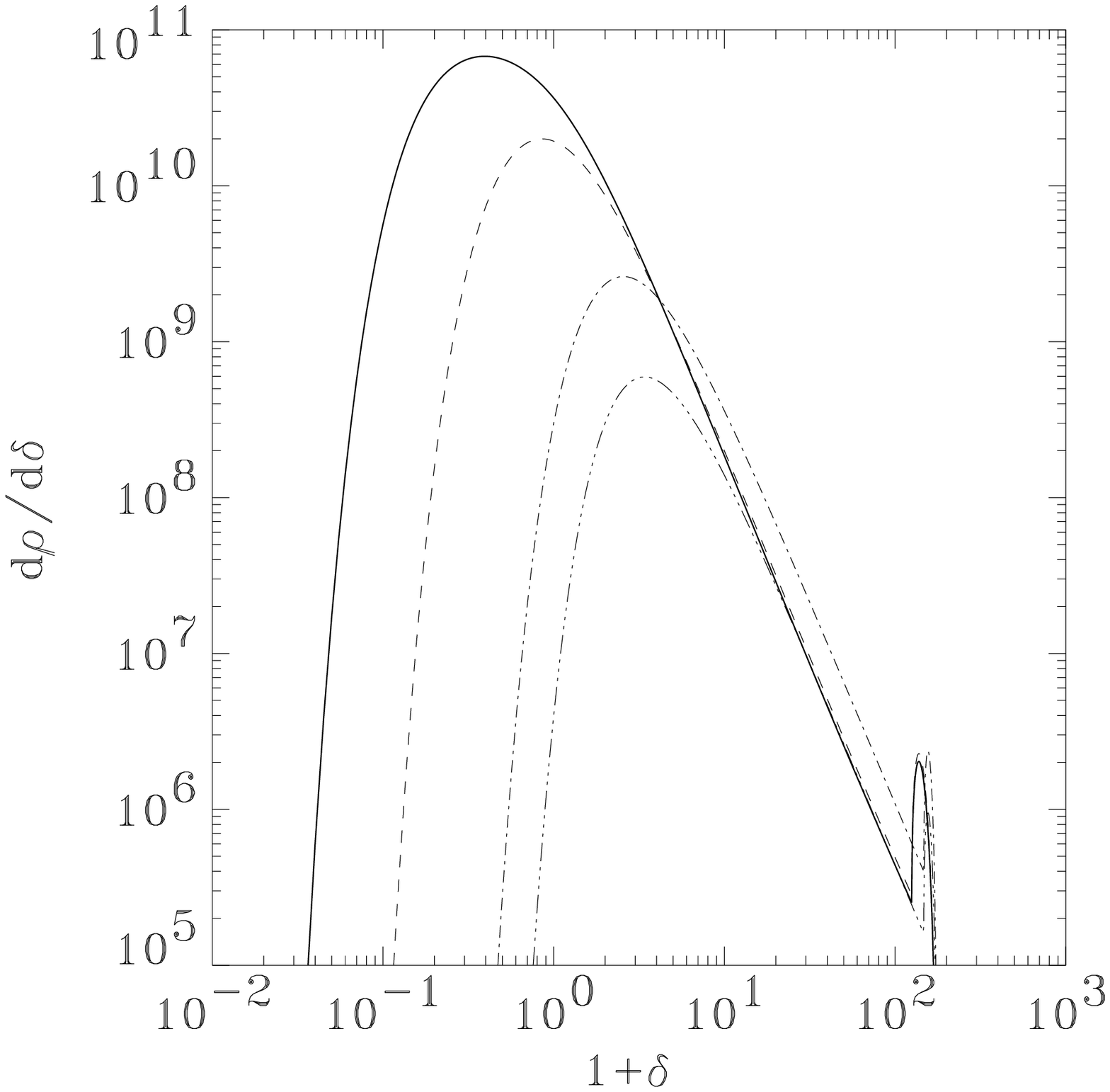}}
\caption{\label{fig:rhoz} Distribution of density of matter inside
  collapsed structures with respect to local density contrast, 
  $d\rho/d\delta (> 10^{12} {\, \rm M_\odot})$, for $\beta =2$ and
for $\Omega_{\rm m}+\Omega_{\rm \Lambda} =1$ 
(left panel) and  Einstein-deSitter (right panel) universes.
  Solid line: $z=0$; dashed line: $z=1$; dot-dashed line: $z=2$;
 double-dot--dashed line: $z=3$.
The units of  $d\rho/d\delta$ are ${\rm M_\odot}$ per ${\rm Mpc
  ^3}$.}
\end{figure*}

\begin{figure*}
\resizebox{3in}{!}{
\includegraphics{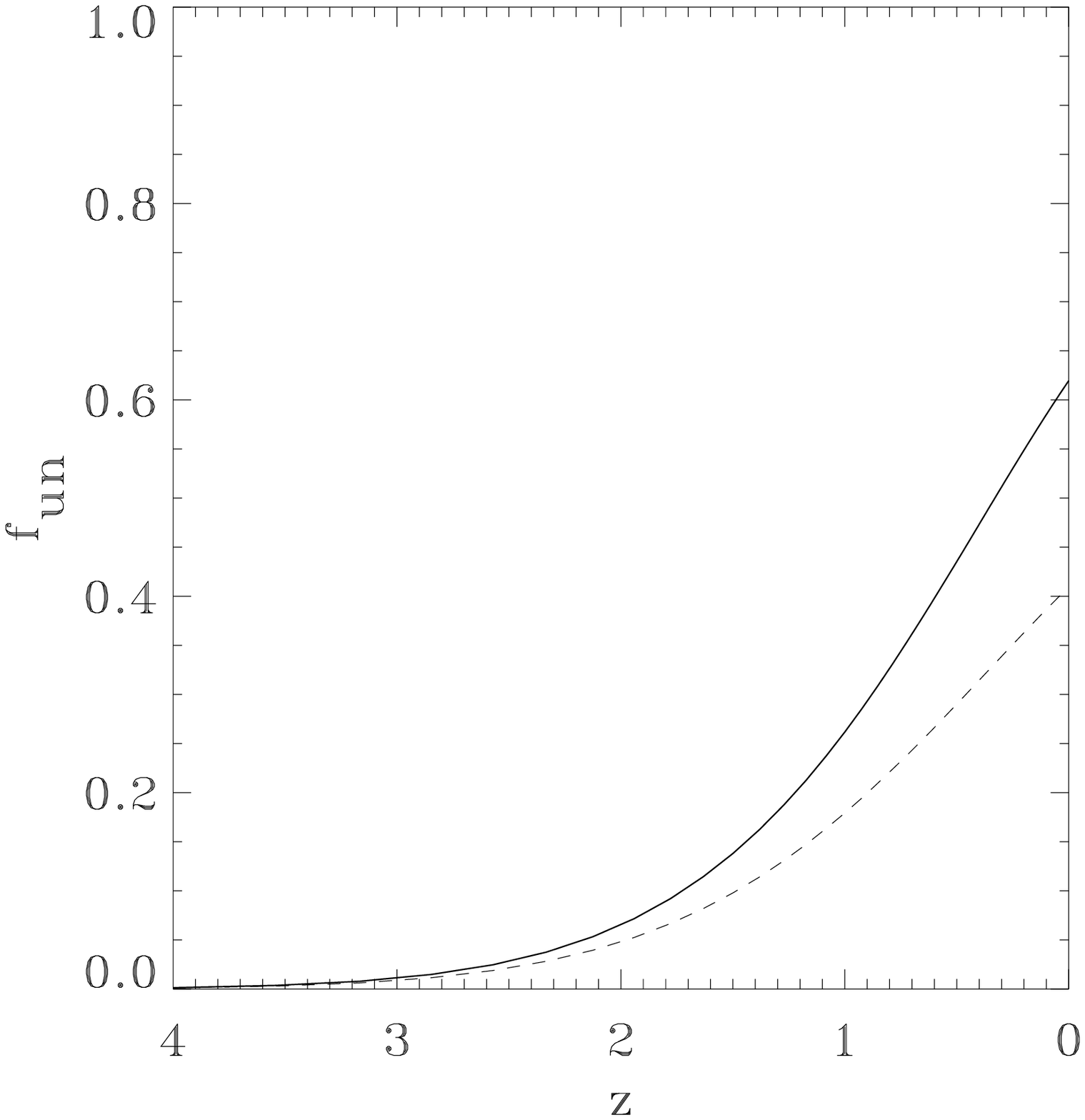}}
\resizebox{3in}{!}{
\includegraphics{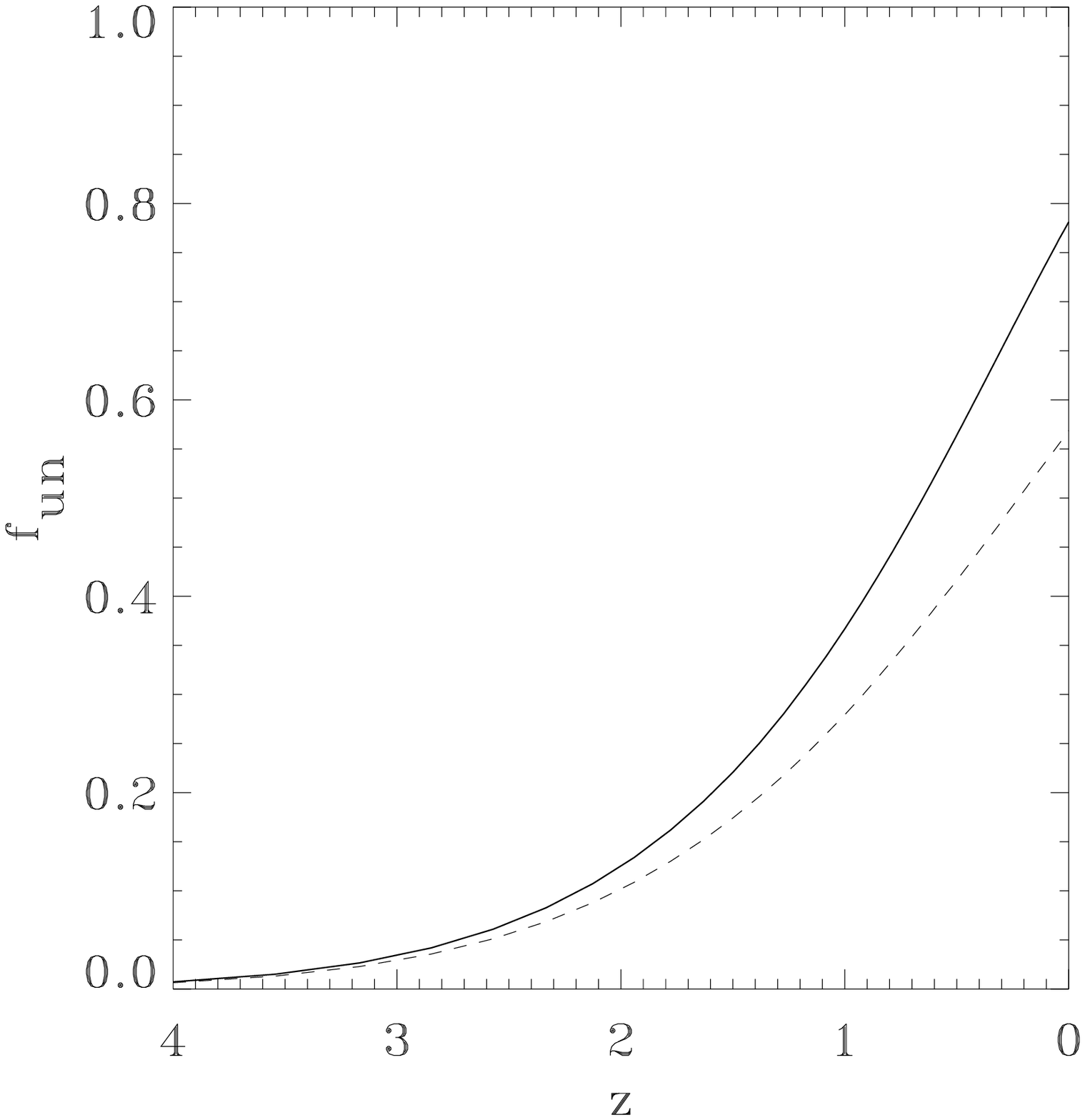}}
\caption{\label{fig:fun} Fraction by number $f_{n,{\rm un}}$
 (solid line) and my mass $f_{\rho,{\rm un}}$ (dashed line)
of objects of mass $>10^{12} {\, \rm M_\odot}$ living in underdense regions, 
as a function of redshift, 
for $\Omega_{\rm m}+\Omega_{\rm \Lambda} =1$ (left panel) and 
Einstein-deSitter (right panel) universes.}
\end{figure*}

In addition to the main peak at low $|\delta|$, an additional, much
lower and sharper peak can be seen right before the critical
overdensity cutoff. This peak is the result of the change of the
functional form of the conversion relation between linearly
extrapolated and true density contrast close to virialization, when
application of the spherical collapse morel would lead $\delta$ to
diverge. The particular shape of the peak is an artifact of the
recipe we adopted for dealing with the virialization regime, and
carries no physical meaning (the shape of the peak is the shape of the
high-$\delta$ end of $d\ed/d\delta$). 
However, since the boundary conditions we use for $\ed(\delta)$ and
its derivative {\em are} physical, we do expect to
have some form of local maximum at the high-$\delta$ end of the
double distribution. Still, as discussed in the previous section, the
effect of the details or even the existence of this local maximum on
the physical quantities of interest is negligible. 

The high-$\delta$ cutoff occurs at higher values of $\delta$ in the
concordance universe than in the Einstein-deSitter universe. This is a
result of the different density contrast achieved at virialization by
structures in the two different cosmologies. In the Einstein-deSitter
case this density contrast is always $18\pi^2$, while in the
concordance universe it is always higher and increases with time.
At high redshifts, before the effect of $\Lambda$ becomes
significant, $\delta_{\rm c}$ is very close to $18\pi^2$ in the concordance
universe as well, as can be seen in Figure \ref{fig:efofz}.

Figures \ref{fig:efofd_low} and \ref{fig:efofd_high} show slices of
the double distribution at various fixed values of $\delta$, with $z=0$
and $\beta=2$. Figure \ref{fig:efofd_low} plots slices corresponding
to relatively low values of $|\delta|$ ($\delta = -0.5, 0, 0.5$ and
$3$, close to the distribution peak in $\delta$). 
At the high-mass end of the distribution, the abundance of
objects increases with increasing $\delta$, while in the low mass end
of the distribution the trend is reversed, and the object abundance
increases with decreasing $\delta$. This is in agreement with the
behavior observed in the constant-$m$ slices.

Figure \ref{fig:efofd_high} plots slices corresponding to high values
of $\delta$ ($\delta = 10, 20$ and $30$), farther from the distribution peack.
In this case, the curves do not cross, and an increase of $\delta$
simply results in an overall suppression of object abundance:
structures of all masses are unlikely to be found overly clustered.
This is because the final stages of collapse proceed rather quickly
compared to the time spent around turnaround. The likelihood of a
region observed in its late stages of collapse but before virialization
is then low because the lifetime of this phase is small. 

In figure \ref{fig:nz} we plot $dn/d\delta(>10^{12} {\rm M_\odot})$ as
a function of $1+\delta$ for different values of redshift. It is
striking that at $z=0$, the distribution peaks at negative $\delta$
values (around $\delta=-0.6$ in the concordance and $-0.7$ in the
Einstein-deSitter universe), indicating that the most probable location for a collapsed
object of mass $>10^{12} {\, \rm M_\odot}$ is an {\em underdense}
environment. For the specific mass range, this trend is reversed by
$z=1$, when the preferred location of these objects is close to the
universe mean ($\delta =0$). This time-evolution pattern is independent of
cosmology, as it is present both in the concordance and the
Einstein-deSitter universes, and appears rather to be a characteristic of the
hierarchical nature of structure formation. Parameters of this
distribution can be calculated using equations (\ref{parnmean}) and
(\ref{parnvar}) which, for the concordance cosmology and $z=0$ give
$\langle \delta\rangle_{\rm n} = 0.43$ and $\sigma_{\rm \delta, n}=4.36$. The
large value of the variance shows that the distribution is
significantly broad. However, the positive value of the mean is an
artifact of the asymmetric boundaries of the distribution and its long
high-$\delta$ tail. This is demonstrated by the
notably different locations of the mean and the median. The value of
the latter is $\delta=-0.22$, therefore more structures in this range
reside inside underdensities.

Figure \ref{fig:rhoz} is the matter-density counterpart of Figure
\ref{fig:nz}, as it plots $d\rho/d\delta(>10^{12} {\rm M_\odot})$
as a function of $1+\delta$ for the same values of redshift as in
Figure \ref{fig:nz}. Again, at the current cosmic epoch, 
the distribution peaks at negative values of $\delta$. Most of the
virialized matter in the universe today appears to reside inside isolated 
objects rather than in clusters (note that decreasing the value of
$m_{\rm min}$ will only enhance this result since the trend towards
isolation is more pronounced for the lower-mass objects).
The trend of the peak with time (towards
larger $\delta$ for higher reshifts) is duplicated here as well.
In particular, note that at present, a significant fraction 
of the mass lies in moderately underdense regions.  Equations (\ref{parrhomean}) and
(\ref{parrhovar}) give for this distribution (in the concordance
cosmology and for $z=0$), $\langle \delta \rangle
_{\rho} = 1.20$ and $\sigma_{\delta,\rho} = 6.23$. The median of this
distribution is at $\delta=0.20$, a positive value.

Finally, Figure \ref{fig:fun} plots the evolution with redshift of the 
fractions by number and by mass, $f_{n,{\rm un}}$ and
$f_{\rho,{\rm un}}$, of objects with $m>10^{12} {\rm \, M_\odot}$, 
living inside underdense regions.
At high redshifts, when the mass of such objects 
is well above the exponential suppression cutoff, 
practically none of them is found inside underdensities. 
This trend is reversed as the redshift 
decreases. In the $\Omega_{\rm}+\Omega_\Lambda=1$ universe, an equal
 number of these structures are 
located inside underdensities by redshift $0.3$ and 
by the current cosmic epoch, about 60\% by number (but only 40\% by
mass)  
of these structures 
are located inside underdensities. 

\begin{figure*}
\resizebox{3in}{!}
{
\includegraphics{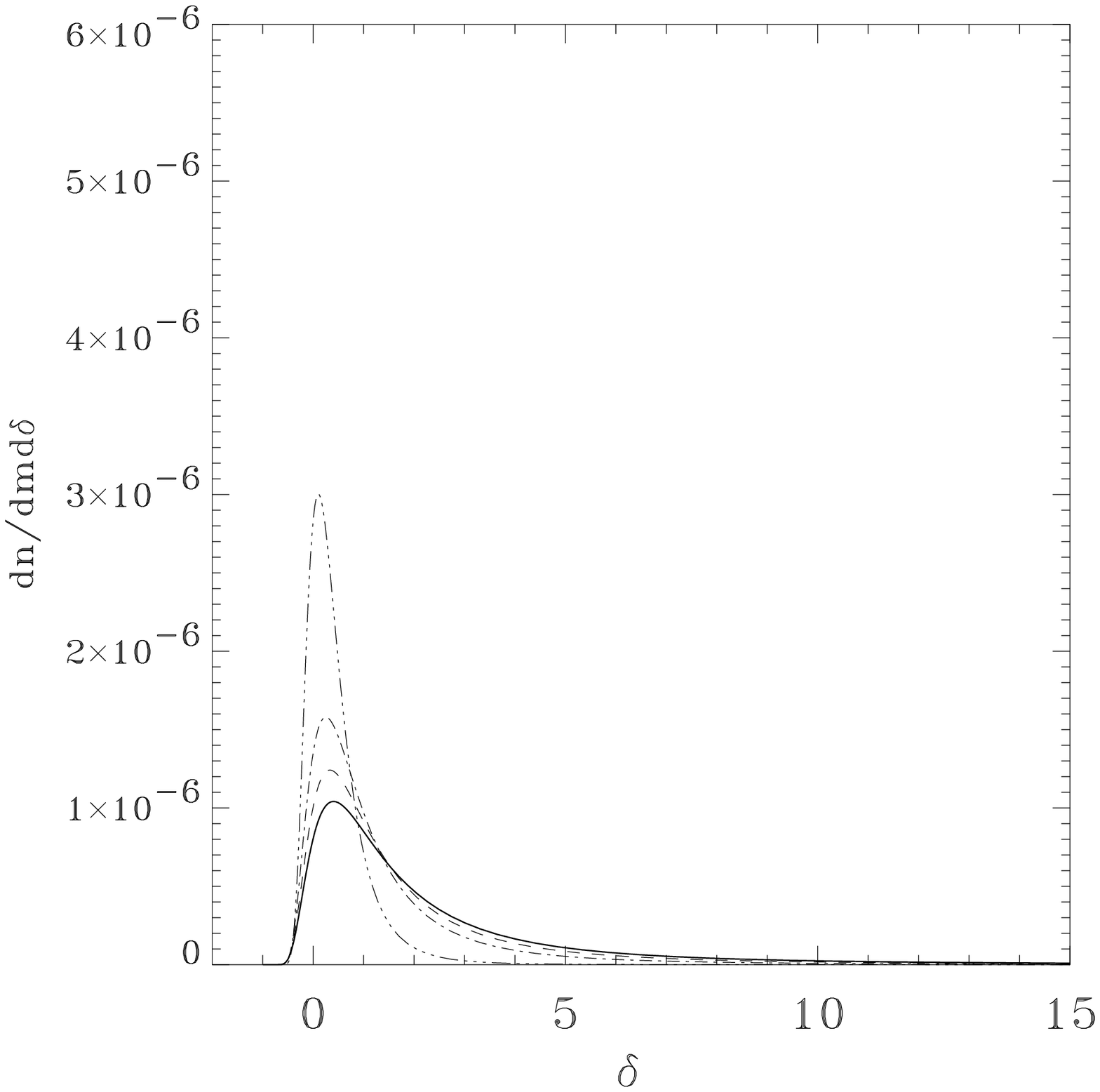}}
\resizebox{3in}{!}{
\includegraphics{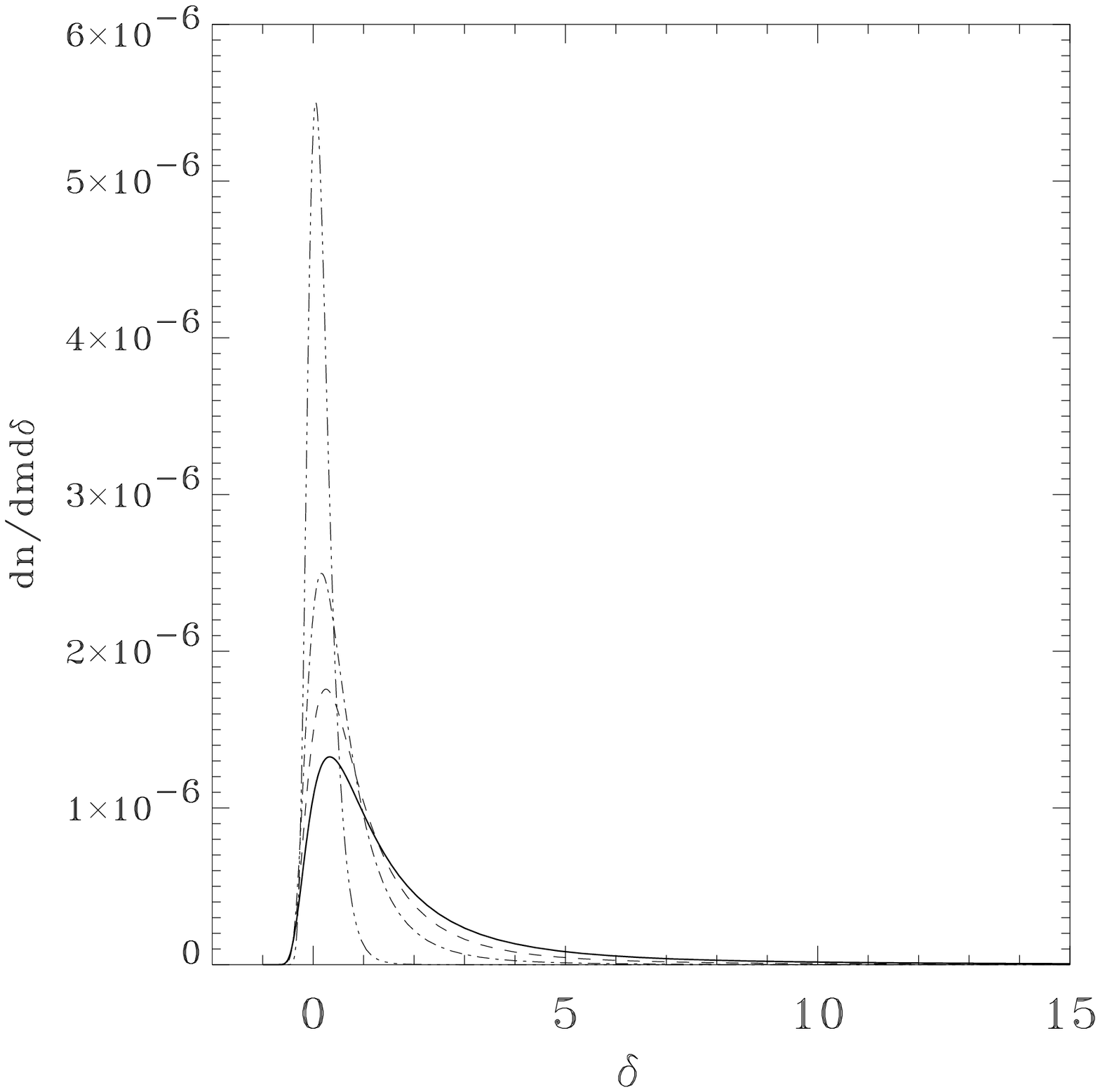}}
\caption{\label{fig:efofbeta} Slices of the double distribution function
  at $m=5.5\times10^{14}{\rm ,M_\odot}$ and for different values of
  the clustering scale parameter $\beta$, 
for $\Omega_{\rm m}+\Omega_{\rm \Lambda} =
1$ (left panel) and  Einstein-deSitter (right panel) universes,
  plotted in linear scale. Solid
  line: $\beta=1.5$; dashed line: $\beta=2$; dot-dashed
  line: $\beta=3$; double-dot--dashed line: $\beta = 10$. The
  units of the double distribution are number of objects per ${\rm Mpc
  ^3}$ per $10^{15} {\rm M_\odot}$.}
\end{figure*}

\begin{figure*}
\resizebox{3in}{!}{
\includegraphics{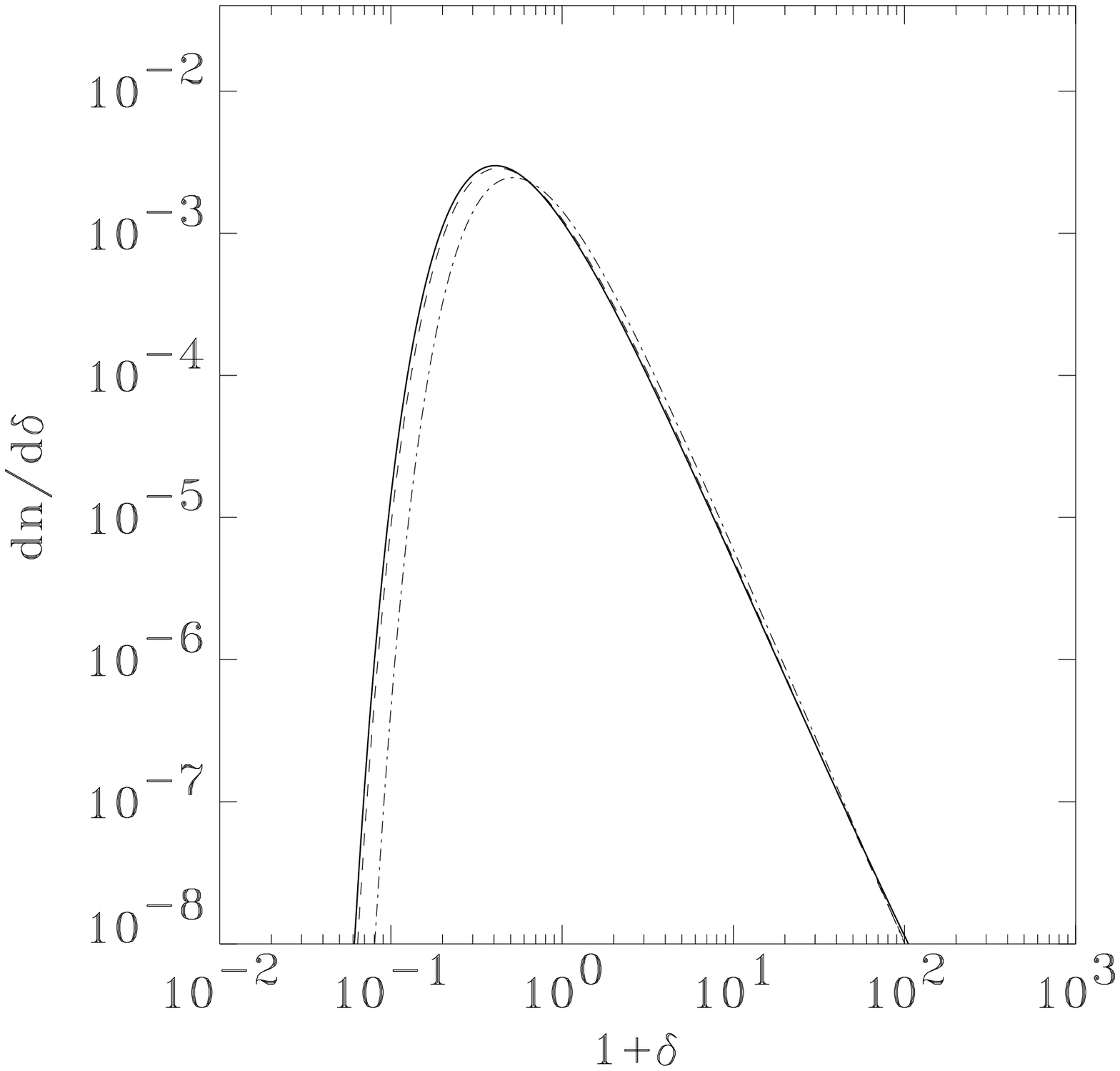}}
\resizebox{3in}{!}{
\includegraphics{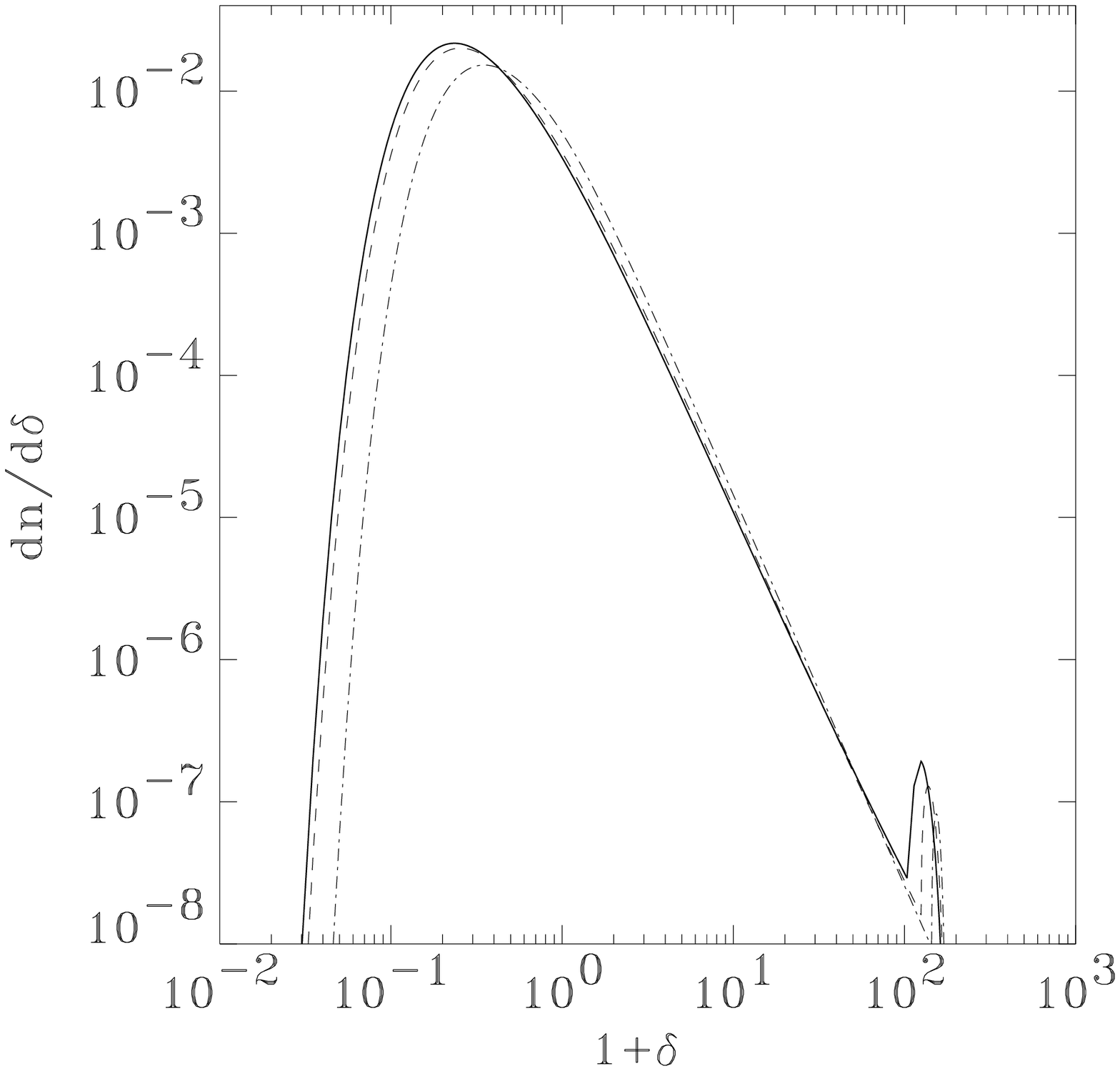}}
\caption{\label{fig:nb} Distribution of structures of mass larger than
  $10^{12} {\, \rm M_\odot}$ with respect to local density contrast, 
  $dn/d\delta$, for $\Omega_{\rm m}+\Omega_{\rm \Lambda} = 
1$ (left panel) and  Einstein-deSitter (right panel) universes,  
at $z=0$, and for
  $\beta=1.5$(solid line), $\beta=2$ (dashed line) and
  $\beta=10$(dot-dashed line).
The  units of  $dn/d\delta$ are number of objects per ${\rm Mpc
  ^3}$.}
\end{figure*}

Figure \ref{fig:efofbeta} demonstrates the effect of changing the
clustering scale parameter on the double distribution. Slices of the
double distribution along $m=5.5\times 10^{14}{\, \rm M_\odot}$ are
plotted (in linear axes)  as a function of $\delta$, and for $\beta =
1.5$ (solid line), $2$ (dashed line), $3$ (dot-dashed line) and $10$
(double-dot--dashed line). The location of the peak appears to be
extremely insensitive to the value of $\beta$ for moderately low
values. It very slowly moves towards $\delta=0$ with increasing
$\beta$, as it should (increasing $\beta$ results in averaging the
overdensity over increasingly large volumes). Note that even as
$\beta$ approaches 1, the peak will {\em not }move towards $\delta_{\rm c}$,
as a result of our correction for the central-object
contamination. This makes our formalism particularly suitable to study
the properties of matter very close but outside a virialized structure
(e.g. the local density of accreted gas). 

The effect of $\beta$ on an integral quantity is shown in Figure
\ref{fig:nb}, which plots  $dn/d\delta(>10^{12} {\rm M_\odot})$
for $\beta = 1.5$ (solid line), $\beta=2$ (dashed line) and $\beta=10$
(dot-dashed line). Again, the results are extremely
insensitive to the value of $\beta$, which gives us confidence about
the robustness of the location of the peak
of our distributions. 

\section{Discussion}\label{disc}

We have presented an extension of the Press-Schechter
mass function, in the form of double distribution of structures
with respect to mass and local
overdensity. We have done so by introducing a clustering
scale parameter $\beta > 1$, which 
we use to associate with each collapsed object of mass
$m$ a larger environment of mass
$\beta m$.  The scale parameter $\beta$ can be expressed as a fucntion of 
the number of virial radii included in the local environment of each structure.
We found that for reasonable
values $\beta \sim 2$,
the shape of the distribution does not depend sensitively on this
parameter. Integration over linearly extrapolated overdensity returns
the original
Press-Schechter mass function, independently of the value of $\beta$.

We present the double distribution in terms of the
true, physical, nonlinear density contrast $\delta$.
However, in calculating the distribution it is 
useful to identify regions using instead
the overdensity obtained via linear analysis,
$\ed$,
extrapolated to the present epoch.
A useful fitting function is given for the
$\delta-\ed$ conversion.

The double distribution
is useful because it allows us to have an explicit analytical
if approximate description of the environment in which collapsed objects
of all masses reside. 
Using the tools we have developed, it can be readily calculated
for any flat cosmology,
and evaluated at any epoch.
Consequently, it offers new insight into the
growth of structure as well as the present
distribution of collapsed objects.

We have evaluated the double distribution and some
of its integral moments for
both a concordance cosmology and an Einstein-de Sitter universe.
Some key results
are that at any redshift, the double distribution
is dominated by a peak
which shifts in mass but is always at
a relatively low value of $|\delta|$.
For each mass,
there is a most probable $\delta$, which
increases with structure mass.
Moreover, at the present epoch in the 
concordance universe, 
the most probable environment
is a modest {\em under}density,
for all objects below about $10^{14} M_\odot$;
thus, underdensities are preferentially
populated by low-mass objects.
Finally, the fraction
of mass in underdensities
increases with time, and in the
concordance cosmology the present
underdense mass fraction in objects of $M>10^{12} {\,\rm M_\odot}$ is about 40\%.
These trends can be understood in terms
of hierarchical clustering
in which overdense regions are the site of
vigorous merging that
clears out low-mass objects,
which then find their last refuge in voids.

These results are consistent with other
analyses in the literature which use 
Press-Schechter-like formalism to
probe the correlations between structures.
For example, \cite{mw96}
extend the Press-Schechter random-walk picture
in a way very similar to ours, but 
identify a local environment in terms of a fixed,
mass-independent radius.  Our results
are thus complementary to these,
because we adopt a mass-dependent environment based on the
virial radius.

Finally, it is well-known that the Press-Schechter mass function
provides an excellent characterization of the
results of numerical $N$-body simulations (e.g., \cite{lc94}).
It would be of great interest to compare the analytic double 
distribution we have presented with
numerical results.
We plan to address this issue in future work.

\acknowledgments{We thank Dimitris Galanakis, Telemachos Mouschovias, 
Kostas Tassis and Ben Wandelt
for enlightening discussions. This work was supported by 
National Science Foundation grant AST-0092939.
The work of VP was partially supported by an Amelia Earhart Fellowship.}

\appendix 

\section{\label{ap0} Analytical Properties of the Double Distribution}

\subsection{Derivation of the Press-Schechter Mass Function From the
  Double Distribution}
 Using $S_1$
to denote $S(m)$ and $S_2$ for $S(\beta m)$ we have:
\begin{widetext}

\begin{eqnarray}
\int_{-\infty}^{\ed_{\rm 0,c}}d\ed \frac{dn}{dmd\ed}&=&
\frac{\rho_{\rm m}}{m}\left|\frac{dS_1}{dm}\right|
\frac{1}{2\pi S_2^{1/2}(S_1-S_2)^{3/2}}
\left\{
\int_{-\infty}^{\ed_{\rm 0,c}}d\ed
(\ed_{\rm 0,c}-\ed)
\exp\left[-\frac{\ed^2}{2S_2}\right]
 \exp\left[-\frac{(\ed_{\rm 0,c}-\ed)^2}{2(S_1-S_2)}\right]-
\right. \nonumber \\
&& \left. \int_{-\infty}^{\ed_{\rm 0,c}}d\ed
(\ed_{\rm 0,c}-\ed)
\exp\left[-\frac{(\ed-2\ed_{\rm 0,c})^2}{2S_2}\right]
 \exp\left[-\frac{(\ed_{\rm 0,c}-\ed)^2}{2(S_1-S_2)}\right]
\right\}\nonumber \\
&=&
\frac{\rho_{\rm m}}{m}
\frac{1}{2\pi S_2^{1/2}(S_1-S_2)^{3/2}}
\left|\frac{dS_1}{dm}\right|
 \int_{-\infty}^{\infty}
\ed'd\ed'
\exp\left[-\frac{(\ed_{\rm 0,c}-\ed')^2}{2S_2}\right]
\exp\left[-\frac{\ed'^2}{2(S_1-S_2)}\right] \nonumber \\
&=& 
\frac{\rho_{\rm m}}{m}
\frac{1}{2\pi S_2^{1/2}(S_1-S_2)^{3/2}}
\left|\frac{dS_1}{dm}\right|
\ed_{\rm 0,c}\left(\frac{S_1-S_2}{S_1}\right)^{3/2}
\sqrt{2\pi S_2}\exp\left[-\frac{\ed_{\rm 0,c}^2}{2S_1}\right]\nonumber \\
&=& \sqrt{\frac{2}{\pi}}
\frac{\rho_{\rm m}}{m^2}\frac{\ed_{\rm 0,c}}{\sqrt{S_1}}
\left|\frac{d\ln \sqrt{S_1}}{d\ln m}\right|
\exp\left[-\frac{\ed_{\rm 0,c}^2}{2S_1}\right]
\end{eqnarray}
\end{widetext}
where we performed a change of variables $\ed'=\ed_{\rm 0,c}-\ed$, 
and we set 
$\ed' \rightarrow -\ed'$ in 
the second integral. The final result is 
the Press-Schechter mass function formula. Note that this result 
is independent of the value of $\beta$.

\subsection{Behavior of the Double Distribution in the limit \boldmath{$\beta
  \rightarrow \infty$}}

In order to find the behavior the double distribution as $\beta
\rightarrow \infty$, we recall that, because $S(m)$ 
decreases monotonically with $m$, its limit in the infinite $\beta$ regime will be
\begin{equation}
\lim _{\beta \rightarrow \infty} S(\beta m) =0 \,.
\end{equation}

Then, using the notation of the previous section, 
\begin{equation}
\lim_{\beta \rightarrow \infty} \frac{dn}{dmd\ed_{\ell}} (m, \ed_{\ell}, \beta,
a) = \lim_{S_2 \rightarrow 0} \frac{dn}{dmd\ed_{\ell}}(S_1, S_2, \ed_{\ell}, a, m)\,.
\end{equation}
The limit of a unit-area Gaussian when its width
vanishes is the Dirac delta-finction $\delta_{\rm D}$, 
\begin{equation}
\lim _{\lambda \rightarrow 0 }\frac{1}{\sqrt{2\pi}\lambda} \exp\left[
-\frac{(x-x_0)^2}{2\lambda^2}\right] = \delta_{\rm D}(x-x_0)\,.
\end{equation}
Using this result, we get
\begin{widetext}
\begin{eqnarray}
\lim_{S_2 \rightarrow 0} \frac{dn}{dmd\ed_{\ell}} &=& 
\frac{\rho_{\rm m}}{m}\left|\frac{dS_1}{dm}\right|
\frac{\ed_{\rm 0,c}-\ed_{\ell}}{2\pi}
\lim_{S_2\rightarrow 0}\left\{
\exp\left[-\frac{(\ed_{\rm 0,c}-\ed_{\ell})^2}{2(S_1-S_2)}\right]
\frac{\exp\left[-\frac{\ed_{\ell}^2}{2S_2}\right]
-\exp\left[-\frac{(\ed_{\ell}-2\ed_{\rm 0,c})^2}{2S_2}\right]
}{S_2^{1/2}(S_1-S_2)^{3/2}} \right\}\nonumber \\
&=& \frac{\rho_{\rm m}}{m}\left|\frac{dS_1}{dm}\right|
\frac{\ed_{\rm 0,c}-\ed_{\ell}}{\sqrt{2\pi}}
\frac{\exp\left[-\frac{(\ed_{\rm 0,c}-\ed_{\ell})^2}{2S_1}\right]}{S_1^{3/2}}
\left\{\lim_{S_2\rightarrow 0}\frac{\exp\left[-\frac{\ed_{\ell}^2}{2S_2}
\right]}{\sqrt{2\pi S_2}}
-\lim_{S_2\rightarrow 0}\frac{\exp\left[-\frac{(\ed_{\ell}-2\ed_{\rm 0,c})^2}{2S_2}
\right]}{\sqrt{2\pi S_2}}
\right\} \nonumber \\
&=& \frac{\rho_{\rm m}}{m}\left|\frac{dS_1}{dm}\right|
\frac{\ed_{\rm 0,c}-\ed_{\ell}}{\sqrt{2\pi}}
\frac{\exp\left[-\frac{(\ed_{\rm 0,c}-\ed_{\ell})^2}{2S_1}\right]}{S_1^{3/2}}
\left[\delta_{\rm D}(\ed_{\ell})-\delta_{\rm D}(\ed_{\ell}-2\ed_{\rm 0,c})
\right]\,.
\end{eqnarray}
\end{widetext}
However, the $\ed_{\ell}-$domain of the double distribution is between
$-\infty$ and $\ed_{\rm 0,c}$, and therefore the value $\ed_{\ell}=2\ed_{\rm 0,c}$ is
outside its domain. Hence the second Dirac delta-function is always zero, and
\begin{equation}
\lim_{\beta \rightarrow \infty} \frac{dn}{dmd\ed_{\ell}} =
\frac{\rho_{\rm m}}{m}\left|\frac{dS_1}{dm}\right|
\frac{\ed_{\rm 0,c}-\ed_{\ell}}{\sqrt{2\pi}}
\frac{\exp\left[-\frac{(\ed_{\rm 0,c}-\ed_{\ell})^2}{2S_1}\right]}{S_1^{3/2}}
\delta_{\rm D}(\ed_{\ell})
\,,
\end{equation}
proportional, as expected, to a Dirac delta-function centered at $\ed_{\ell}=0$.
\subsection{Behavior of the Double Distribution in the limit \boldmath{$\beta
  \rightarrow 1$}}
Denoting $S(\beta m)$ by $S_2$ and letting $\phi = S(m)/S(\beta m)$,
we seek the behavior of the double distribution in the limit $\beta
\rightarrow 1$ or $\phi \rightarrow 1$. Defining 
\begin{equation}
\mathcal{C} = \frac{\rho_{\rm m}}{m}\frac{\ed_{\rm 0,c}-\ed_{\ell}}{\sqrt{2\pi}}
\left|\frac{dS}{dm}\right| \frac{\exp\left[-\frac{\ed_{\ell}^2}{2S_2}\right]
-\exp\left[-\frac{(\ed_{\ell}-2\ed_{\rm 0,c})^2}{2S_2}\right]}{S_2^{1/2}}\,,
\end{equation}
we can write
\begin{eqnarray}
\lim_{\phi \rightarrow 1} \frac{dn}{dmd\ed_{\ell}} &=& 
\mathcal{C} \lim_{\phi \rightarrow 1}\frac{\exp\left[
-\frac{(\ed_{\rm 0,c}-\ed_{\ell})^2}{2S_2(\phi-1)}\right]}
{\sqrt{2\pi}S_2^{3/2}(\phi-1)^{3/2}} \nonumber \\
&=& \mathcal{C}\lim_{\phi \rightarrow 1}
\frac{(\phi-1)^{-3/2}}{\sqrt{2\pi}S_2^{3/2}\exp\left[\frac{(\ed_{\rm 0,c}-\ed_{\ell})^2}
{2S_2(\phi-1)}\right]}\nonumber \\
&\stackrel{\infty/\infty}{=}&\mathcal{C}\lim_{\phi \rightarrow 1}
\frac{-\frac{3}{2}(\phi-1)^{-5/2}}
{\exp\left[\frac{(\ed_{\rm 0,c}-\ed_{\ell})^2}
{2S_2(\phi-1)}\right]\left[-
\frac{\sqrt{2\pi}S_2^{3/2}(\ed_{\rm 0,c}-\ed_{\ell})^2}{2S_2(\phi-1)^{2}}
\right]}\nonumber \\
&=&\mathcal{C}\lim_{\phi \rightarrow 1}\frac{3
\exp\left[-\frac{(\ed_{\rm 0,c}-\ed_{\ell})^2}
{2S_2(\phi-1)}\right]}{\sqrt{2\pi}S_2^{1/2}(\phi-1)^{1/2}(\ed_{\rm 0,c}-\ed_{\ell})^2}
\nonumber \\
&=& \mathcal{C}\frac{3}{(\ed_{\rm 0,c}-\ed_{\ell})^2}\delta_{\rm D}
(\ed_{\rm 0,c}-\ed_{\ell})\,
\end{eqnarray}
proportional, as expected, to a Dirac delta-function around $\ed_{\rm 0,c}$.

\section{\label{ap1} Relating Linearly Extrapolated and True Overdensities in
  an Einstein-de Sitter (\boldmath{$\Omega_{\rm m}=1$}) cosmology}

In this appendix we derive a conversion relation
$\ed_0(a,\delta)$ for an $\Omega_{\rm m}=1$ cosmology (here, $\delta$ is the
density contrast predicted for a density perturbation at cosmic epoch
$a$ by the spherical evolution model and $\ed_0$ is the extrapolation
of the density contrast to the present cosmic epoch using
linear theory). In order to do so, we first calculate $\delta(a)$
from the spherical evolution solution, then calculate $\ed_0$ using
linear theory, and finally require that $\delta(a)$ and $\ed_a$ (the 
linear-theory density contrast at epoch $a$) should
agree at early times.

\subsection{Spherical Evolution Model in an {\boldmath $\Omega_{\rm
      m}=1$ }Universe}

The evolution of a spherically symmetric, overdense perturbation in an
otherwise homogeneous $\Omega_{\rm m}=1$ universe is described by the
parametric equations 
\begin{equation}\label{par2}
a_{\rm p} = \frac{2a_{\rm coll}}{(12\pi)^{2/3}} 
(1-\cos \theta) \, {\rm ,\,\, and \,\,}
a =  a_{\rm coll}\left(\frac{\theta - \sin \theta}{2\pi}\right)^{2/3} \,\, ,
\end{equation}
where $a_{\rm coll}$ is the scale factor of the universe when the
perturbation formally collapses to a point, $a_{\rm p}$ is the scale
factor of the perturbation, and $\theta$ is the development
angle. Note that the perturbation will turn around (reach its maximum
size, $a_{\rm p,max} = 4a_{\rm coll}(12\pi)^{-2/3}$) when $\theta=\pi$, at
a time $a=a_{\rm coll}/2^{2/3}$.

The normalization of equation \ref{par2} is such that the density
contrast $\delta$ can be expressed as 
\begin{equation}\label{gend}
\delta = \left(\frac{a}{a_{\rm p}}\right)^3-1\,.
\end{equation}
Hence, for any density contrast $\delta$, equations \ref{par2} and
\ref{gend} can be combined to give a unique development angle
$\theta(\delta)$ which is the solution to the transcendental equation
\begin{equation}\label{thetaofdelta}
\frac{6^{2/3}(\theta-\sin\theta)^{2/3}}{2(1-\cos \theta)}
-(1+\delta)^{1/3} = 0\,.
\end{equation}

Similarly, the spherical evolution solution for an underdensity is
given by the parametric equations
\begin{equation}\label{upar2}
a_{\rm p}=A_{\rm p}(\cosh \eta -1)  \, {\rm ,\,\, and \,\,} a = A_{\rm
  p}
\frac{6^{2/3}}{2}(\sinh \eta -\eta)^{2/3}\,.
\end{equation}
where $\eta$ is the development angle in this case. Equation
\ref{upar2} together with equation \ref{gend} can be combined as
before to give $\eta(\delta)$ as the solution to the transcendental
equation
\begin{equation}\label{etaofdelta}
\frac{6^{2/3}(\sinh\eta-\eta)^{2/3}}{2(\cosh \eta-1)}
-(1+\delta)^{1/3} = 0\,.
\end{equation}

\subsection{{\boldmath $\ed_0(a,\delta)$} 
according to the spherical evolution model}

The behavior 
of $\delta$ in the linear regime in this cosmology is
\begin{equation}\label{linom1}
\ed = \ed_0 a\,.
\end{equation}
This result should coincide with the linear expansion of the spherical
evolution result at early times. Expanding the 
parametric solution to second nonvanishing order in $\theta$ 
and eliminate $\theta$, we obtain
\begin{equation}\label{oper}
a_{\rm p}(a) = a \left[1-\frac{(12\pi)^{2/3}}{20}
\frac{a}{a_{\rm coll}}\right]\,.
\end{equation}
We then substitute equation \ref{oper} in the definition of
$\delta$ (eq. \ref{gend}) to get
\begin{equation}
\ed = \frac{3(12\pi)^{2/3}}{20 a_{\rm coll}}a
\end{equation}
which, by comparison to equation \ref{linom1} gives
\begin{equation}\label{acollofed}
\ed_0 = \frac{3(12\pi)^{2/3}}{20 a_{\rm coll}}\,.
\end{equation}
Then, the conversion relation we seek is
\begin{equation}\label{otrue}
\ed_0(a,\delta) = \frac{6^{2/3}3}
{20a}\left[\theta(\delta)-\sin \theta(\delta)\right]^{2/3}
\end{equation}
where $\theta(\delta)$ is given by equation \ref{thetaofdelta}.

Equation \ref{otrue} has the undesirable property that it diverges as $\theta$
approaches $2\pi$. This is of course a consequence of the perturbation
formally collapsing to a singularity in the spherical evolution model instead
of reaching virial equilibrium. If we make the usual assumption that at
virialization the radius of the perturbation is $a_{\rm max}/2$ and we
additionally require that 
\begin{itemize}
\item $\ed_0(a,\delta)$ is continuous and smooth at $\theta=3\pi/2$
\item $a_{\rm p}=a_{\rm p,max}$ for all $a\ge a_{\rm coll}$
\end{itemize}
then for $\theta >3\pi/2$ (which corresponds to $\delta>9(3\pi+2)^2/8$)
we can replace equation \ref{otrue} with
\begin{eqnarray}
\ed_0(a, \delta) &=& 
\ed_{\rm 0,v} + \ed_{\rm 0,v}'(\delta - \delta_{\rm v}) \nonumber \\ 
&+&\frac{3(\ed_{\rm0,c}-\ed_{\rm 0,v})-(\delta_{\rm c} - \delta_{\rm v})
(2\ed_{\rm 0,v}'+\ed_{0,c}')}{\left(\delta_{\rm c} - \delta_{\rm
    v}\right)^2}(\delta -\delta_{\rm v})^2 \nonumber\\
&+& \frac{(\ed_{\rm 0,c}'+\ed_{\rm 0,v}')(\delta_{\rm c}- \delta_{\rm
    v})
-2(\ed_{\rm 0,c}-\ed_{\rm 0,v})}{(\delta_{\rm c} -\delta_{\rm v})^3}
(\delta -\delta_{\rm v})^3 \nonumber\\
\label{patch}
\end{eqnarray}
(see appendix \ref{ap2} for a discussion of the reasons for employing
this particular functional form, and \S \ref{form} and \ref{res} for a
discussion on why the effect of such a choice on the double
distribution is negligible). In equation \ref{patch}, 
\begin{eqnarray}
\delta_{\rm v}&=&\left(\frac{a|_{\theta=3\pi/2}}{a_{\rm
    p}|_{\theta=3\pi/2}}\right)^3-1=
 \frac{9(3\pi+2)^2}{8}-1\nonumber \\
\delta_{\rm c}&=& \left(\frac{a|_{\theta=2\pi}}{a_{\rm
    p}|_{\theta=3\pi/2}}\right)^3-1= 18\pi^2-1 \nonumber\\
\ed_{\rm 0,v} &=& \ed_0(a,\delta_{\rm v}) = 
\frac{3^{5/3}}{20a} (3\pi+2)^{2/3} \nonumber \\
\ed_{\rm 0,c}' &=& \left.\frac{\partial \ed_0}{\partial \delta}\right|
_{\delta=\delta_{\rm c}} =
\frac{1}{10a(1+\delta_{\rm c})^{2/3}}
\end{eqnarray}
The last equality coming from the fact that after $a_{\rm coll}$
the radius of a perturbation remains constant and equal to $a_{\rm
  p,max}/2$, while its density contrast $\delta$ changes only due to the
expansion of the background universe, $\delta = (2a/a_{\rm
  p,max})^3-1$ or $\delta = (10 a \ed_0/3)^3-1$. Finally, 
$\ed_{\rm 0,c}$ is given by equation \ref{matteredc} while 
$\ed_{\rm 0,v}'$ is given by equation \ref{oder} for $\delta=\delta_{\rm
    v}$ and $\theta=3\pi/2$.

To get the linear behavior of $\delta$ for an underdensity
we expand the parametric solution \ref{upar2} to
second nonvanishing order in $\eta$ and we eliminate $\eta$ to get
\begin{equation}\label{uper}
a_{\rm p} =  A_{\rm p}\frac{6^{2/3}}{2} \left[
1+\frac{1}{10}\frac{a}{A_{\rm p}}
\right]
\,.
\end{equation}
Substituting equation \ref{uper} in the definition of
$\delta$ (eq. \ref{gend}), we get for the time dependence of 
$\delta$ at early times, 
\begin{equation}\label{ulin}
\ed =  -\frac{3}{10A_{\rm p}}a
\end{equation}
from which, by comparison to equation \ref{linom1}, we get
\begin{equation}
\ed_0 = -\frac{3}{10A_{\rm p}}\,.
\end{equation}
Then, $\ed_0(a,\delta)$ will be 
\begin{equation}\label{utrue}
\ed_0(a,\delta) = -\frac{6^{2/3}3}
{20a}\left[\sinh \eta(\delta) - \eta(\delta)\right]^{2/3}
\end{equation}
where $\eta(\delta)$ is given by equation \ref{etaofdelta}.

Equation \ref{utrue} is valid for all $\eta$ and its limit as $\ed \rightarrow
-\infty$ is $\delta(\ed) \rightarrow -1$. Thus, although the linearly
extrapolated field can become $<-1$, the corresponding value of the actual
$\delta$ is always $\ge -1$, as the physical requirement $\rho_{\rm p} \ge 0$
demands.

\subsection{Critical extrapolated overdensity for collapse, {\boldmath
    $\ed_{\rm 0,c}(a)$}}
The critical extrapolated overdensity for collapse can be found from
equation \ref{acollofed}
\begin{equation}\label{matteredc}
\ed_{\rm 0,c}(a_{\rm coll}) = \frac{3(12\pi)^{2/3}}{20}a_{\rm coll}^{-1}
\approx 1.69 a_{\rm coll}^{-1}\,.
\end{equation}
Note that the above equation has the functional form 
$\ed_{\rm 0,c}(a_{\rm coll})\propto 
1/D(a_{\rm coll})$, where $D(a)$ is the linear growth factor for this
cosmology. This is also true in the $\Omega_{\rm m}+\Omega _\Lambda = 1$ case.

\subsection{{\boldmath$\partial \ed_0 / \partial \delta |_a$}}

In addition to the relation between $\delta$ and $\ed_0$, we will also need the
derivative $\partial \ed_0 / \partial \delta |_a$ in order to convert between
true and extrapolated overdensity differentials in
equation (\ref{convtotrue}). 
In the case of an overdense structure, $\delta>0$, equation
\ref{otrue} gives
\begin{equation}\label{oder}
\left.\frac{\partial \ed_0}{\partial \delta}\right|_{a} =
\frac{6^{2/3}}{10a}\frac{1-\cos \theta (\delta)}
{\left[\theta(\delta)-\sin \theta (\delta)\right]^{1/3}}
\frac{d\theta}{d\delta} \,.
\end{equation}
To evaluate $d\theta/d\delta$ we define the
auxiliary function
\begin{equation}\label{faux}
F_{\rm a}(\theta, \delta) = 6^{2/3}(\theta-\sin \theta)^{2/3}
-2(1-\cos\theta)(1+\delta)^{1/3}\,.
\end{equation}
From equation \ref{thetaofdelta} we get immediately
$F_{\rm a}(\theta,\delta) = 0$, and differentiating we get
$dF_{\rm a} = 0 = \frac{\partial F_{\rm a}}{\partial \theta}d\theta+
\frac{\partial F_{\rm a}}{\partial \delta}d\delta$. Hence, 
\begin{equation}
\frac{d \theta}{d\delta}
= -\frac{\partial F_{\rm a}}{\partial \delta}\left(\frac{\partial
  F_{\rm a}}{\partial \theta}\right)^{-1} \,,
\end{equation}
where
\begin{equation}
\frac{\partial F_{\rm a}}{\partial \delta} = 
-\frac{2}{3}\frac{1-\cos \theta}{(1+\delta)^{2/3}}
\end{equation}
and
\begin{equation}
\frac{\partial F_{\rm a}}{\partial \theta} = 
\frac{6^{2/3}2}{3}\frac{1-\cos \theta}{(\theta-\sin \theta)^{1/3}}
-2(1+\delta)^{1/3}\sin \theta \,.
\end{equation}

Equation \ref{oder} is valid only for $0<\delta<\delta_{\rm v}$. For
$\delta > \delta_{\rm v}$ equation \ref{patch} gives
\begin{eqnarray}
\left.\frac{\partial \ed_0}{\partial \delta}\right|_{a} &=&
\ed_{\rm 0,v}'  \nonumber \\ 
&&\!\!\!\!\! +
2\frac{3(\ed_{\rm0,c}-\ed_{\rm 0,v})-(\delta_{\rm c} - \delta_{\rm v})
(2\ed_{\rm 0,v}'+\ed_{0,c}')}{\left(\delta_{\rm c} - \delta_{\rm
    v}\right)^2}(\delta -\delta_{\rm v}) \nonumber\\
&& \!\!\!\!\! +
3\frac{(\ed_{\rm 0,c}'+\ed_{\rm 0,v}')(\delta_{\rm c}- \delta_{\rm
    v})
-2(\ed_{\rm 0,c}-\ed_{\rm 0,v})}{(\delta_{\rm c} -\delta_{\rm v})^3}
(\delta -\delta_{\rm v})^2\,. \nonumber\\
\end{eqnarray}

In the case of an underdense structure, $\delta<0$, equation
\ref{utrue} gives
\begin{equation}
\left.\frac{\partial \ed_0}{\partial \delta}\right|_{a} =
-\frac{6^{2/3}}{10a}\frac{\cosh \eta (\delta)-1}
{\left[\sinh \eta (\delta)-\eta(\delta)\right]^{1/3}}
\frac{d\eta}{d\delta} \,.
\end{equation}
As before, in order to evaluate $d\eta / d\delta$ we define the
auxiliary function
\begin{equation}\label{gaux}
G_{\rm a}(\eta, \delta) = 
6^{2/3}(\sinh \eta - \eta)^{2/3}
-2(\cosh\eta-1)(1+\delta)^{1/3}\,.
\end{equation}
Equation  \ref{etaofdelta} implies
$G_{\rm a}(\eta,\delta) = 0$ so
\begin{equation}
\frac{d \eta}{d \delta}
= -\frac{\partial G_{\rm a}}{\partial \delta}\left(\frac{\partial
  G_{\rm a}}{\partial \eta}\right)^{-1} \,,
\end{equation}
where
\begin{equation}
\frac{\partial G_{\rm a}}{\partial \delta} = 
-\frac{2}{3}
\frac{\cosh \eta-1}{(1+\delta)^{2/3}}\,,
\end{equation}
and
\begin{equation}
\frac{\partial G_{\rm a}}{\partial \eta} = 
\frac{6^{2/3}2}{3}\frac{\cosh \eta -1}{(\sinh \eta - \eta)^{1/3}}
-2(1+\delta)^{1/3}\sinh \eta\,.
\end{equation}

\section{\label{ap2} Relating Linearly Extrapolated and True Overdensities in
  an \boldmath{$\Omega_{\rm m}+\Omega_\Lambda=1$} cosmology}

In this appendix we will derive a conversion between true and
extrapolated overdensity, $\delta(\ed_0, a)$ for an 
$\Omega_{\rm m}+\Omega_\Lambda = 1$
cosmological model. We will do so by first
calculating the true density contrast $\delta(a)$ of a density
perturbation at cosmic epoch $a$ as predicted by the spherical
evolution model, then calculating $\ed_0$, which is the overdensity of the same
spherical perturbation if extrapolated according to the linear theory 
until the present cosmic epoch, and finally requiring that at early times
linear theory and the linear expansion of the spherical evolution
model should give the same result. 

\subsection{Spherical Evolution Model in an $\Omega_{\rm m} 
+ \Omega_{\rm \Lambda} = 1$ Cosmology}

\subsubsection{The Evolution Equation}

In the spherical evolution model, the spherical density perturbation
under consideration behaves as an independent non-flat sub-universe. Its evolution
is dictated by a Friedmann equation, 
\begin{equation}\label{friedt}
\left(\frac{da_{\rm p}}{dt} \right)^2= H_0^2 \Omega_{\rm m}a_{\rm p}^2
\left(a_{\rm p}^{-3}+\omega -\kappa a_{\rm p}^{-2}\right)
\end{equation}
where $a_{\rm p}$ is the  radius 
of such a spherical density perturbation in an otherwise homogeneous 
universe, $\omega=\Omega_\Lambda/\Omega_{\rm m}=\Omega_{\rm m}^{-1}-1$ (where
$\Omega_{\rm m}$ and $\Omega _{\rm \Lambda}$ are the matter and vacuum
density parameters of the background universe)
and $\kappa$ is a
constant characteristic of the amplitude and sign of the perturbation: the
larger the $|\kappa |$, the larger the deviation from homogeneity at a given
time, while a positive $\kappa$ corresponds to an overdensity and a negative
$\kappa$ to an underdensity. Clearly then in equation \ref{friedt}, the
first term in parentheses on the RHS is the matter term, the second is
the vacuum term and the third is the curvature term, which can have a
positive or negative sign depending on whether we are studying an
``open''(underdensity) or ``closed'' (overdensity) perturbation. 
The normalization of $a_{\rm p}$ 
is such that, had the specific spherical region begun
its evolution with no curvature ($\kappa=0$), $a_{\rm p}$ at the present
cosmic epoch would have been $a_{p(\kappa=0),0}=1$. For this
reason, the density contrast $\delta$ of the perturbation at epoch $a$
is given by equation \ref{gend}

The behavior of the perturbation radius $a_{p}$ as a function of the
universe scale factor $a$ can be found by taking the ratio of the
Friedmann equations of the perturbation and the background universe, 
thus obtaining \cite{peeb84}
\begin{equation}\label{genl}
\left(\frac{da_{\rm p}}{da}\right)^2 = 
\frac{a_{\rm p}^{-1}+\omega a_{\rm p}^2-\kappa}{a^{-1} + \omega a^2}
= \frac{a}{a_{\rm p}}\frac{\omega a_{\rm p}^2-\kappa a_{\rm
    p}+1}{\omega a^3 +1}
\,.
\end{equation}
Equation \ref{genl} implies that the smallest positive perturbation
which will turn around and collapse corresponds to the smallest positive
$\kappa$ for which the equation 
\begin{equation}\label{lf0}
\omega a_{\rm p}^3 -\kappa a_{\rm p} +1=0
\end{equation}
has a real positive solution \cite{ecf}. This gives
\begin{equation}\label{kmin}
\kappa_{\rm min, coll} = 3\omega^{1/3}/2^{2/3}\,.
\end{equation}
Equation \ref{genl} can then be re-written as
\begin{equation}\label{spel}
\frac{da_{\rm p}}{da} = \left\{
\begin{array}{ll}
\left(\frac{a_{\rm p}^{-1}+
\omega a_{\rm p}^2-\kappa}{a^{-1}+\omega a^2}\right)^{1/2},
& \kappa < \kappa_{\rm min,coll} \,\,\,\, {\it {\bf  \,\, or}}\\
& \kappa \ge \kappa_{\rm min,coll}, \,\,a<a_{\rm ta}\\ & \\ & \\
- \left(\frac{a_{\rm p}^{-1}+\omega a_{\rm p}^2-\kappa}{a^{-1}+
\omega a^2}\right)^{1/2},
& \kappa \ge \kappa_{\rm min,coll}, \,\,a>a_{\rm ta}\\
\end{array}
\right.
\end{equation}
where
$a_{\rm ta}$ is the scale factor of the universe when the perturbation
reaches its 
maximum (or {\it turnaround}) radius. The turnaround radius is the
smallest of the two positive solutions of equation  \ref{lf0}, 
\begin{equation}\label{aptaofk}
a_{\rm p,ta} =
\omega^{-1/3}\sqrt{\frac{4}{3}\frac{\kappa}{\omega^{\frac{1}{3}}}}
\cos \frac{1}{3}\left(
\cos ^{-1}\sqrt{\frac{27}{4}\left(\frac{\kappa}{\omega^{\frac{1}{3}}}
\right)^{-3}} +\pi
\right)\,.
\end{equation}
Equation \ref{aptaofk} has an 
asymptotic behavior 
$a_{\rm p,ta} \approx
1/\kappa$ when $\kappa/\omega^{1/3} \gg 1$, 
as expected from equation \ref{lf0}. The maximum possible turnaround
radius, $a_{\rm p,ta,max}$ is achieved for $\kappa=\kappa_{\rm min,coll}$ and
is $a_{\rm p,ta,max}=(2\omega)^{-1/3}$. All other collapsing
overdensities will have $a_{\rm p,ta}<a_{\rm p,ta,max}$.

\subsubsection{Qualitative Description of the Evolution of Structures}

The introduction of the additional vacuum term in the Friedmann
equation considerably complicates the simple classification of density
perturbations to overdensities (all of which turn around and collapse
in an $\Omega_{\rm m}=1$ cosmology) and underdensities (all of which
expand forever). In 
the $\Omega_{\rm m}+\Omega_{\rm \Lambda} =1$
universe there exist overdensities which will continue to expand
forever. The behavior of a perturbation in such a cosmology is
parametrized by the quantity $\kappa/\omega^{1/3}$, and we can
identify the following cases.

{\bf Case I, \boldmath{$\kappa/\omega^{1/3} \leq -1$}: large
  underdensities, expanding forever}. The table below shows the 
  relative magnitude of the three terms in the Friedmann equation
  (matter, curvature and vacuum term) for different values of the
  scale factor of the perturbation. The first line in the table
  indicates the hierarchy of the three terms, from largest to smaller,
  for each range of the scale factor. The second line indicates the
  dominant term in each scale factor range.  The third line shows the
  approximate dependence of the radius of the perturbation, $a_{\rm p}$, on
  time, assuming that only the dominant term contributes to the
  Friedmann equation in each range.
\begin{center}
\begin{tabular}{|c|c|c|c|}
\hline
$a_{\rm p}<\frac{1}{|\kappa|}$ &
$\frac{1}{|\kappa|}<a_{\rm p}< \frac{1}{\sqrt[3]{\omega}} $&
$ \frac{1}{\sqrt[3]{\omega}} < a_{\rm p} < \sqrt{\frac{|\kappa|}{\omega}}$ 
& $a_{\rm p}>\sqrt{\frac{|\kappa|}{\omega}}$\\
\hline
\hline
MCV & CMV& CVM & VCM\\
\hline
matter & curvature & curvature & vacuum \\
\hline
$a_{\rm p} \sim t^{2/3}$ & $a_{\rm p} \sim t$ & $a_{\rm p} \sim t$ &
$a_{\rm p} \sim {\rm e}^t$\\
\hline 
\end{tabular}
\end{center}

{\bf Case II, \boldmath {$-1 < \kappa/\omega^{1/3} \leq 1$}: small
  perturbations, expanding forever}. 
These can be either underdensities ($\kappa <0$) or
  overdensities ($\kappa >0$). In both cases 
  the curvature term never becomes dominant. The following table shows
  their different evolutionary stages (as in Case I).
\begin{center}
\begin{tabular}{|c|c|c|c|}
\hline
$a_{\rm p}<\sqrt{\frac{|\kappa|}{\omega}}$ &
$\sqrt{\frac{|\kappa|}{\omega}}<a_{\rm p}< \frac{1}{\sqrt[3]{\omega}}$&
$ \frac{1}{\sqrt[3]{\omega}} < a_{\rm p} < \frac{1}{|\kappa|}$& 
$a_{\rm p}>\frac{1}{|\kappa|}$\\
\hline
\hline
MCV & MVC& VMC & VCM\\
\hline
matter & matter & vacuum & vacuum \\
\hline
$a_{\rm p} \sim t^{2/3}$ & $a_{\rm p} \sim t^{2/3}$ & $a_{\rm p} \sim
    {\rm e}^t$ &$a_{\rm p} \sim {\rm e}^t$ \\
\hline
\end{tabular}
\end{center}

{\bf Case III, \boldmath{$1 < \kappa/\omega^{1/3} < 3/2^{2/3}$}:
``coasting'' overdensities, expanding forever}.
These overdensities continue to expand forever despite the fact that
they go through a phase in their evolution when the curvature term
becomes dominant and their expansion slows down. 
During this phase, the contributions of the matter 
and vacuum terms, which are the ones driving the expansion, add up to 
a value always higher than the curvature term, although the curvature
term is larger than each one of them. When the
perturbation enters the curvature-dominated phase, the expansion rate 
decreases and the perturbation grows much more mildly than $t^{2/3}$.
The expansion rate reaches a minimum at $a_{\rm p}=(2\omega)^{-1/3}$, 
after which it increases again as the perturbation 
approaches the phase of exponential expansion. This phase between 
the matter-like expansion and the exponential expansion is denoted 
by $(*)$ in the table below.
\begin{center}
\begin{tabular}{|c|c|c|c|}
\hline
$a_{\rm p}<\frac{1}{\kappa}$ &
$\frac{1}{\kappa}<a_{\rm p}< \frac{1}{\sqrt[3]{\omega}} $&
$ \frac{1}{\sqrt[3]{\omega}} < a_{\rm p} < \sqrt{\frac{\kappa}{\omega}}$ 
& $a_{\rm p}>\sqrt{\frac{\kappa}{\omega}}$\\
\hline
\hline
MCV & CMV& CVM & VCM\\
\hline
matter & curvature & curvature & vacuum \\
\hline
$a_{\rm p} \sim t^{2/3}$ &  $(*)$ & $(*)$ &
$a_{\rm p} \sim {\rm e}^t$\\
\hline 
\end{tabular}
\end{center}

Cases I-III are all sub-cases of the Lema\^{i}tre model (\cite{lem1}, \cite{lem2}), which
features an inflection point at $a_{\rm p,e}=(2\omega)^{-1/3}$ where 
$\ddot{a}_{\rm p}=0$ while $\dot{a}_{\rm p}>0$. The rate of expansion
initially decreases to achieve its minimum (positive) value when
$a_{\rm p} = a_{\rm p,e}$, after which point the expansion accelerates again.

{\bf Special Case, \boldmath{$\kappa/\omega^{1/3} = 3/2^{2/3}$}:
Eddington Overdensity}.
This overdensity is the lowest $\kappa$ overdensity which does not 
expand to an infinite radius. However, it does not turn around and 
collapse, but it approaches its (finite) turnaround radius, $a_{\rm
  p,max}=(2\omega)^{-1/3}$ (from eq. \ref{aptaofk}) as $t\rightarrow \infty$.
As seen by an observer inside this overdensity, as $t\rightarrow
\infty$ the part of the universe outside $a_{\rm p,max}$ will accelerate away
and eventually exit the horizon, and the observable universe (``local
Eddington bubble'') 
will asymptotically approach the Einstein static universe (as in the
Eddington model with a cosmological constant).

{\bf Case IV, \boldmath{$\kappa/\omega^{1/3} > 3/2^{2/3}$}:
large overdensities, eventually collapsing}. When such a structure
enters the dominant-curvature-term phase, its expansion rate 
starts to decrease ($a_{\rm p}\sim t^{\epsilon}$
with $\epsilon = \epsilon(t)$ monotonically decreasing from $2/3$ to
0), until the expansion halts, at $a_{\rm p}
= a_{\rm p,ta}$ which occurs at a time $t_{\rm ta}$, given in table
\ref{timescales}. 
After $t_{\rm ta}$ the perturbation turns around and contracts, 
its evolution being symmetrical in time about $t_{\rm ta}$, 
i.e. $a_{\rm p}(t) = a_{\rm p}(2t_{\rm ta}-t)$ for $t > t_{\rm ta}$
(this is a consequence of eq. \ref{friedt} and holds for any
cosmological model as long as the RHS of the Friedmann equation
involves no explicit time-dependence).
Eventually, the perturbation will formally collapse to a singularity 
at time $t_{\rm coll} = 2t_{\rm ta}$.
\begin{center}
\begin{tabular}{|c|c||c|c|}
\hline
$a_{\rm p}<\frac{1}{\kappa}$ &
$\frac{1}{\kappa}<a_{\rm p}< a_{\rm p,ta} $&
$a_{\rm p,ta}>a_{\rm p}>\frac{1}{\kappa}$&
$\frac{1}{\kappa}>a_{\rm p}$
\\
\hline
\hline
MCV & CMV & CMV & MCV\\
\hline
matter & curvature & curvature & matter\\
\hline
$a_{\rm p} \sim t^{2/3}$ & $a_{\rm p} \sim t^\epsilon$ &
$a_{\rm p} \sim (2t_{\rm ta}\!\!-t)^\epsilon$ & 
$a_{\rm p} \!\!\sim \!\!(2t_{\rm ta}\!\!-t)^{2/3}$
\\
\hline 
expansion & expansion & contraction & contraction \\
\hline
\end{tabular}
\end{center}

In all of the cases discussed above, the transitions between different
phases of their evolution occur at characteristic times, those of
matter-vacuum equality $t_{\rm MV}$, matter-curvature
equality $t_{\rm MC}$ and curvature-vacuum equality $t_{\rm CV}$. At
these times (shown in table \ref{timescales}), 
the corresponding terms in the Friedmann equation become
equally important. Note that in the case of the Eddington overdensity 
and of case IV
collapsing overdensities, matter-vacuum equality and curvature-vacuum
equality are never reached, and the vacuum term never dominates over
any of the other terms.

\begin{table}
\caption{Characteristic times of the spherical evolution model in an
  $\Omega_{\rm m}+\Omega_{\rm \Lambda} = 1$
  universe. \label{timescales}}
\begin{tabular}{lll}
\hline \hline
{\bf event} &$\,\,\,$ & {\bf time} \\
$\,$turnaround & &
$t_{\rm ta} = \frac{1}{H_0\sqrt{\Omega_{\rm m}}}
\int _{0}^{a_{\rm p, ta}} da_{\rm p}\sqrt{
\frac{a_{\rm p}}{\omega a_{\rm p}^3-\kappa a_{\rm p}+1}}$\\
\hline 
\begin{tabular}{l}
matter-vacuum \\equality \end{tabular} && 
$t_{\rm MV} = \frac{1}{H_0\sqrt{\Omega_{\rm m}}}
\int _{0}^{\omega^{-1/3}} da_{\rm p}\sqrt{
\frac{a_{\rm p}}{\omega a_{\rm p}^3-\kappa a_{\rm p}+1}}$ \\
\hline 
\begin{tabular}{l}
matter-curvature \\equality \end{tabular} &&
$t_{\rm MC} = \frac{1}{H_0\sqrt{\Omega_{\rm m}}}
\int _{0}^{|\kappa|^{-1}} da_{\rm p}\sqrt{
\frac{a_{\rm p}}{\omega a_{\rm p}^3-\kappa a_{\rm p}+1}}$ \\
\hline
\begin{tabular}{l}
curvature-vacuum \\equality \end{tabular} &&
$t_{\rm CV} = \frac{1}{H_0\sqrt{\Omega_{\rm m}}}
\int _{0}^{\sqrt{\frac{|\kappa|}{\omega}}} da_{\rm p}\sqrt{
\frac{a_{\rm p}}{\omega a_{\rm p}^3-\kappa a_{\rm p}+1}}$\\
\hline \hline
\end{tabular}
\end{table}

In the next section we derive exact solutions for the time-evolution
of $a_{\rm p}$ for perturbations of different curvature. 
However, surprisingly accurate approximate solutions can be derived
using only linear theory and 
equation (\ref{magic}). Solving for the spherical collapse 
density contrast we get 
\begin{equation}
\delta_a \approx  \left(1-\frac{\ed_a}{\ed_{\rm c}}\right)^{-\ed_{\rm c}}-1\,.
\end{equation}
Since $a_{\rm p} = a(1+\delta_a)^{-1/3}$, we can write for collapsing
overdensities 
\begin{eqnarray}\label{apsc}
a_{\rm p} &\approx& a  \left[1-\frac{\ed_{\rm c}
  D(a)/D(a_{\rm c})}{\ed_{\rm c}}\right]^{\ed_{\rm c}/3}\nonumber \\
&=& a\left[1-\frac{D(a)}{D(a_{\rm coll})}\right]^{\ed_{\rm c}/3}\,.
\end{eqnarray}
where the initial conditions (curvature) of the
perturbation are parametrized by its collapse epoch, $a_{\rm coll}$, while the
cosmology enters through the functional form of the linear growth
factor and the linear collapse overdensity, $\ed_{\rm c}$. Similarly, for
perturbations which expand for ever we can write
\begin{eqnarray}\label{apse}
a_{\rm p} &\approx& a  \left[1-\frac{\ed_0
  D(a)/D(a_{0})}{\ed_{\rm c}}\right]^{\ed_{\rm c}/3}\nonumber \\
&=& a\left[1-\frac{\ed_0}{\ed_c}\frac{D(a)}{D(a_0)}\right]^{\ed_{\rm c}/3}\,.
\end{eqnarray}
where the curvature of the perturbation is parametrized by its
extrapolated linear density contrast at the present epoch, $\ed_0$.
Note that for overdensities which expand forever, $\ed_0>0$ and
$a_{\rm p}<a$, while for underdensities $\ed_0<0$ and $a_{\rm p}>a$. Also, because
$D(a)$ asymptotes to a constant value for $a\rightarrow \infty$ (as we
will see in the next sections) , 
$a_{\rm p}$ grows proportionally to $a$ at late
times. This is the exponential expansion phase, described in our
analysis above. 

\subsubsection{Solutions of the Evolution Equation}

For eventually collapsing structures ($\kappa \ge \kappa_{\rm min,coll}$), 
separation of
variables in equation \ref{spel} and integration yields, 
\begin{equation}\label{sepvarint}
\int_0^a \frac{\sqrt{y}dy}{\sqrt{\omega y^3+1}} = 
\left\{
\begin{array}{lr}
\int_{_0}^{^{a_{\rm p}}}\!\!\!\!\!
\frac{\sqrt{x}dx}{\sqrt{\omega x^3 - \kappa x + 1}}
&  a<a_{\rm ta}
\\ \\
2\int_{_0}^{^{a_{\rm p, ta}}}\!\!\!\!\!\!\!\!\!\!\!
\frac{\sqrt{x}dx}{\sqrt{\omega x^3 - \kappa x + 1}}
\!  - \!\! \int_{_0}^{^{a_{\rm p}}} \!\!\!\!\!\!
\frac{\sqrt{x}dx}{\sqrt{\omega x^3 - \kappa x + 1}}
&  a\ge a_{\rm ta}
\end{array}
\right. \,,
\end{equation}
where $a_{\rm ta}$ is the cosmic epoch when $a_{\rm p} = a_{\rm p,ta}$.
Now the integral on the LHS of equation \ref{sepvarint} can be calculated using
\cite{ecf}
\begin{equation}\label{lhsint}
\int \frac{\sqrt{y}dy}{\sqrt{\omega y^3+1}} = 
\frac{2}{3}\omega^{-1/2}\sinh^{-1}\sqrt{\omega y^3}\,\,.
\end{equation}
The integral of the RHS can be re-written as 
\begin{equation}\label{defv1}
\int_0^{a_{\rm p}} \frac{\sqrt{x}dx}{\sqrt{\omega x^3 -
    \kappa x +1}}=
\frac{2}{3}\omega^{-1/2}\mathcal{V}_1(r, \mu)
\end{equation}
where $\mathcal{V}_1$ is the {\em incomplete vacuum integral 
of the first kind}, defined in appendix \ref{vac_ints}, and 
\begin{equation}\label{rmu}
\begin{array}{lcccr}
r = a_{\rm p} / a_{\rm p, ta} \,,
& &&&\mu = (\omega a_{\rm p, ta}^3)^{-1}\\
\end{array} \,\,.
\end{equation}
Note that for this
range of curvature values, $\kappa/\omega^{1/3}$
(which is the quantity which parametrizes the behavior
of the perturbation with time) is a function of $\mu$ alone, with
 \begin{equation}\kappa/\omega^{1/3}=(1+\mu)/\mu^{2/3}\,.\end{equation}

Using equations \ref{lhsint} and \ref{defv1}, equation \ref{sepvarint} 
can be rewritten as, 
\begin{equation} \label{eqfora}
a = \left\{
\begin{array}{lr}
\omega^{-1/3}\left\{\sinh
\left[\mathcal{V}_1(r,\mu)\right]\right\}^{2/3} 
\,, & a\le a_{\rm ta} \\ \\
\omega^{-1/3}\left\{\sinh \left[2 \mathcal{V}_1(1,\mu) - 
\mathcal{V}_1(r,\mu)\right]\right\}^{2/3} \,, & a >a_{\rm ta}
\end{array}
\right. \,.
\end{equation}
Equation \ref{eqfora} is the spherical evolution solution for a collapsing 
perturbation with
amplitude $\kappa$ (parametrized above by $\mu$) 
in an $\Omega_\Lambda + \Omega _{\rm m} =1$,
$\Omega_\Lambda/\Omega_{\rm m}=\omega$ universe, and gives $r$ (and hence
$a_{\rm p}$) as a function of
$a$ for this model. Note that $\mathcal{V}_1(r,\mu)$ is the
development angle for this cosmology.

The scale factor of the universe at turnaround for a given collapsing
overdensity can be found immediately from equation \ref{eqfora}, 
\begin{equation}
a_{\rm ta} = \omega^{-1/3} \left[\sinh \mathcal{V}_1(1,\mu)\right]^{2/3}\,.
\end{equation}
The scale factor of the universe at collapse, $a_{\rm
  coll}$ (when the scale factor of the
perturbation becomes formally zero, $a_{\rm p,c}= r_{\rm c} =0$) is, 
from equation \ref{eqfora} and since $\mathcal{V}_1(0,\mu)=0$, 
\begin{equation} \label{acollapse}
a_{\rm coll} = \omega^{-1/3} \left[\sinh 2
  \mathcal{V}_1(1,\mu)\right]^{2/3} 
\,.
\end{equation}

Equation \ref{eqfora} should not be applied 
"literally" until the final collapse of the perturbation to a
singularity, since the physical picture for the late stages of the
evolution of a perturbation involves virialization at a finite radius.
It has been shown by \cite{lah}
that the analogous arguments which give $a_{p,v}=a_{p,ta}/2$ for the
$\Omega_{\rm m}=1$ cosmology give, for an $\Omega_{\rm m}+\Omega_\Lambda=1$ universe, 
\begin{equation}\label{3vofta}
4\omega a_{\rm p,v}^3 - \frac{2+2\omega a_{p,ta}^3}{a_{p,ta}}a_{p,v} + 1 =0\,.
\end{equation}
The physically meaningful solution of \ref{3vofta} which gives the correct
behavior for $\omega \rightarrow 0$ is (using eq. \ref{rmu})
\begin{eqnarray}\label{apvir}
a_{\rm p,v} 
&=& a_{\rm p,ta} \sqrt{\frac{2\mu+2}{3}} \cos \frac{1}{3}
\left(\cos^{-1}\sqrt{\frac{27\mu^2}{(2\mu+2)^3}}+\pi\right)\,.\nonumber \\
\end{eqnarray}
The scale factor $a_{\rm v}$ of the universe when the scale factor of the 
perturbation {\em past its turnaround} becomes equal to $a_{\rm p,v}$, will be
given by the second branch of equation \ref{eqfora}, for $a_{\rm p} 
= a_{\rm p,v}$. Then, the  
validity range for the second branch of equation \ref{eqfora} is 
$a_{\rm ta}<a<a_{\rm v}$.

For $a>a_{\rm v}$, we can no longer use the spherical evolution
solution to describe the physical picture of interest (virialization). In the
next section we will present a simple recipe we will use to follow
the late stages of evolution of the perturbation which satisfies the
desired boundary conditions ($a_{\rm p} = a_{\rm p,v}$ at $a_{\rm
  coll}$ and constant thereafter).

For perpetually expanding structures ($\kappa < \kappa_{\rm min,coll}$), 
equation \ref{spel} gives 
\begin{equation}\label{sepvarint_nc}
\int_0^a \frac{\sqrt{y}dy}{\sqrt{\omega y^3+1}} = 
\int_0^{a_{\rm p}}\frac{\sqrt{x}dx}{\sqrt{\omega x^3 - \kappa x + 1}}
\,.
\end{equation}
In this case, the integral on the RHS can be rewritten as
\begin{equation}\label{defh1}
\int_0^{a_{\rm p}} \frac{\sqrt{x}dx}{\sqrt{\omega x^3 -
    \kappa x +1}}=
\frac{2}{3}\omega^{-1/2}\mathcal{H}_1(r, \varpi)
\end{equation}
where $\mathcal{H}_1$ is the {\em hyperbolic vacuum integral of the first
kind}, defined in appendix \ref{vac_ints}, and
\begin{equation}\label{rvarpi}
\begin{array}{lcccr}
r = a_{\rm p} / |a_{\rm p, R}| \,,
& &&&\varpi = (\omega |a_{\rm p, R}|^3)^{-1}\\
\end{array} \,\,, 
\end{equation}
where $a_{\rm p,R}$ is the only real (and always negative) root of
equation \ref{lf0} when $\kappa < \kappa_{\rm min,coll}$, 
\begin{equation}
 a_{\rm p,R} = 
\frac{-\omega^{\frac{1}{3}}}
{\!\!\!\!
\sqrt[3]{\left(\frac{1}{2}-\sqrt{\frac{1}{4}-\frac{\kappa^3}{27\omega}}
\right)^2}\!\!\!+\!
{\sqrt[3]{\left(\frac{1}{2}+\sqrt{\frac{1}{4}-\frac{\kappa^3}{27\omega}
}\right)^2}\!-\!\frac{\kappa}{3\omega^{\frac{1}{3}}}
}}
\end{equation}
As in the case of collapsing perturbations,
$\kappa/\omega^{1/3}$ is a function of $\varpi$ alone, with
\begin{equation}\kappa/\omega^{1/3} =
  (1-\varpi)/\varpi^{2/3}\,.\end{equation} Then, $\varpi=1$ is a 
flat subuniverse (not perturbed with respect to the background), and 
perturbations with $\varpi>1$ are underdensities while $1/4<\varpi<1$
correspond to non-collapsing overdensities.

Then, equation \ref{sepvarint_nc} becomes
\begin{equation}\label{eqfora_nc}
a = 
\omega^{-1/3}\left\{\sinh
\left[\mathcal{H}_1(r,\varpi)\right]\right\}^{2/3} 
\end{equation}
which is the spherical evolution solution for a non-collapsing
perturbation with
amplitude $\kappa$ (parametrized above by $\varpi$) 
in an $\Omega_\Lambda + \Omega _{\rm m} =1$,
$\Omega_\Lambda/\Omega_{\rm m}=\omega$ universe.
As for collapsing perturbations, $\mathcal{H}_1(r,\varpi)$ is the
development angle.

\subsection{{\boldmath $\ed_0(\delta)$} 
according to the spherical evolution model}

The linear theory result for a growing-mode perturbation in an
$\Omega_{\rm m}+\Omega_\Lambda=1$ cosmology is \cite{peeb}
\begin{equation}\label{genlin}
\ed = \ed_0\frac{D(a)}{D(a_0)}
\end{equation}  
where $D$, the linear growth factor, is given by $D(a)=A[(2\omega)^{1/3}a]$
with
\begin{equation}
A(x) = \frac{(x^3+2)^{1/2}}{x^{3/2}} \int_0^x 
\left(\frac{u}{u^3+2}\right)^{3/2}du \,.
\end{equation}
To find the relation between $\ed_0$ and $\kappa$, we expand the exact
($\delta = a^3/a_{\rm p}^3-1$) and linear relations for the overdensity to first
order in $a$ and demand that the coefficients be equal. Thus we get
\cite{ecf}
\begin{equation}\label{kofdnot}
\kappa = \frac{(2\omega)^{1/3}}{3A\left[(2\omega)^{1/3}a_0\right]}\ed_0
= \frac{(2\omega)^{1/3}}{3A\left[(2\omega)^{1/3}\right]}\ed_0\,,
\end{equation}
since $a_0=1$ .
This is the value of the constant $\kappa$ for a perturbation which at a
cosmic epoch $a_0=1$ has a linearly extrapolated overdensity $\ed_0$.
Note that since the linear theory result is the same for both underdensities
and overdensities, equation \ref{kofdnot} holds for both cases. For underdensities,
both $\kappa$ and $\ed$ will be negative, while for overdensities both will be
positive. 

At any given epoch $a$, there is a unique perturbation (parametrized
by $\kappa$ or, equivalently, by $\mu$ or $\varpi$) which will have
achieved a true density contrast $\delta$ at that time. Therefore, to
calculate the desired conversion relation $\ed_0(a, \delta)$ we first
calculate $\kappa$ (or, equivalently, $\mu$ or $\varpi$) from the
given $a$ and $\delta$ and the appropriate solution of the evolution
equation (\ref{eqfora} or \ref{eqfora_nc}). Then, we use equation
\ref{kofdnot} to evaluate $\ed_0$. 

\begin{table*}
\caption{Applicability limits for different branches of the spherical
  evolution solution\label{table_limits}}
\begin{tabular}{lllll}
\hline
\hline
{\bf Limit} &$\,\,\,\,\,$ & {\bf Expression} & $\,\,\,\,\,$
&{\bf Auxiliary Relations} \\
\hline
\begin{tabular}{l}Density contrast of \\
Eddington overdensity, $\delta_{\rm Ed}(a)$
\end{tabular}
& & $\delta_{\rm Ed}(a) = 2\omega \left(\frac{a}{r_{\rm
    Ed}(a)}\right)^3 -1$
&&$\sinh ^{-1} \sqrt{\omega a^3} - \mathcal{V}_1(r_{\rm Ed},2) = 0$\\
\hline
\begin{tabular}{l}Density contrast of overdensity\\ turning around at $a$,
    $\delta_{\rm ta}(a)$\end{tabular} &&
$\delta _{\rm ta}(a) = \omega a^3 \mu_{\rm ta}(a)-1$ &&
$\sinh^{-1}\sqrt{\omega a^3} - \mathcal{V}_{1}(1,\mu_{\rm ta}) =0$ \\
\hline
\begin{tabular}{l}Density contrast of overdensity reaching \\
its virial size at $a$, $\delta_{\rm v}(a)$\end{tabular} &&
$\delta_{\rm v}(a) = \left(\frac{a}{a_{\rm p,v}
\left[\mu_{\rm v}(a)\right]}\right)^3 - 1$ &&
$\sinh ^{-1} \sqrt{\omega a^3} - 2\mathcal{V}_1(1,\mu_{\rm v})
+ \mathcal{V}_1\left[
r_{\rm v}(\mu_{\rm v}), \mu_{\rm v}\right] =0$\\
&&&& $r_{\rm v}(\mu) = a_{\rm p,v}(\mu)/a_{\rm p,ta}(\mu)$ \\
\hline 
\begin{tabular}{l}Density contrast of virialized overdensity\\
formally collapsing to a point at $a$, $\delta_{\rm c}(a)$\end{tabular} &&
$\delta_{\rm c}(a) = \left[ \frac{a}{a_{\rm p,v}\left[\mu_{\rm c}(a)\right]}
\right]^3-1$ && 
$\sinh^{-1}\sqrt{\omega a^3}-2\mathcal{V}_{1}(1,\mu_{\rm c}) =0$ \\ 
\hline
\hline
\end{tabular}
\end{table*}
To determine which is the appropriate solution of the evolution
equation we need to use for each $\delta$, we calculate the limits of
applicability of each equation in terms of $\delta$. 
\begin{itemize}
\item Equation \ref{eqfora_nc} is applicable for all forever expanding
  perturbations. For any given epoch $a$, the maximum density contrast
  of such perturbations is achieved by the Eddington perturbation and
  is equal to $\delta_{\rm Ed}(a)$, given in line 1 of Table
  \ref{table_limits}. 
Then, the applicability domain of equation \ref{eqfora_nc} is $-1 <
\delta \le \delta_{\rm Ed}(a)$, and the conversion relation in this
case takes the form shown in the 1st line of Table \ref{mytable1}. 

\item The first branch of equation \ref{eqfora} is applicable for
  eventually collapsing perturbations which, however, have not reached
  their turnaround radius yet. The maximum $\delta$ of all such
  perturbations at a given $a$ is achieved by the perturbation which
  is turning around at $a$, and is equal to $\delta_{\rm ta}(a)$,
  given in line 2 of Table \ref{table_limits}. 
The applicability domain of the first branch of equation \ref{eqfora}
is then $\delta_{\rm Ed}(a) < \delta \le \delta_{\rm ta}(a)$ and the
conversion relation in this case is shown in the 2nd line of table
\ref{mytable1}. 

\item The second branch of equation \ref{eqfora} is applicable for
  eventually collapsing perturbations which are past their turnaround
  but which have not yet reached their virial radius. 
  The maximum $\delta$ of such
  perturbations at a given $a$ is achieved by the perturbation which
  is reaching its virial size at $a$, and is equal to $\delta_{\rm
  v}(a)$, given in line 3 of Table \ref{table_limits}.
The applicability domain of the second branch of equation \ref{eqfora}
is then $\delta_{\rm ta}(a) < \delta \le \delta_{\rm v}(a)$ and the
conversion relation in this case is shown in the 3rd line of table
\ref{mytable1}. 

\item Perturbations which have reached their virial size but have not
  yet reached their designated collapse time, $a_{\rm coll}(\mu)$,
  need to be treated separately, since the spherical collapse model
  fails (does not agree with the physical picture we would like to
  describe, although it is still formally applicable) for radii
  smaller than the virial radius. Since a realistic   
  description of the microphysical dissipation processes which lead to
  virialization is far beyond the scope of this analytical
  calculation, we will adopt a prescription which is driven by
  mathematical simplicity. We will assume that for $\delta_{\rm v}(a)
  < \delta \le \delta_{\rm c}(a)$ the conversion relation
  $\ed_0(a,\delta)$ has the simplest polynomial form which satisfies
  the following physically motivated boundary conditions: 
\begin{itemize}
\item The extrapolated overdensity is continuous and smooth at
  $\delta_{\rm v}$, so $\ed_0(a,\delta_{\rm v}) = \ed_{\rm 0,v}$
and $\left. \partial \ed_0 / \partial \delta \right|_{\delta_{\rm
    v}} = \ed'_{\rm 0,v}$ as given by the appropriate relations of the
  previous branch (3rd line of tables \ref{mytable1} and
  \ref{mytable2} correspondingly).
\item After time $a_{\rm coll}(\mu)$ the radius of the perturbation
  remains constant and equal to the virial radius so changes in the
  (true) overdensity are only due to the increase of the scale factor
  of the background universe. This then implies that
$\ed_0(a,\delta_{\rm c}) = 
\ed_{\rm 0,c}$ given by equation \ref{edc_lambda} and 
$\left. \partial \ed_0 / \partial \delta \right|_{\delta_{\rm
    c}} = (\partial \ed_0 / \partial \mu|_{\mu_{\rm c}})
(\partial \mu / \partial \delta|
_{\delta_{\rm c}})
= \ed'_{\rm 0,c}$ given in Table \ref{mytable1} line 4.
\end{itemize}
The
conversion relation in this case is shown in Table \ref{mytable1} line
4,while its applicability domain is $\delta_{\rm v}(a) < \delta \le 
\delta_{\rm c}(a)$, with $\delta_{\rm c}(a)$ given in Table
\ref{table_limits} line 4.
Past their
collapse time, perturbations are treated as virialized objects
without substructure and are not relevant as ``local environment'' of
other objects for the purposes of our double distribution calculation,
hence it is not necessary to have a conversion relation of $\delta >
\delta_{\rm c}(a)$.
The calculation would be simplified (this last branch would be
unnecessary) if we chose to
regard perturbations as virialized objects after the moment they
reached their virial size after turnaround, at time $a_v$. However,
we will retain the usual assumption that objects virilize at time
$a_{\rm coll}$ for consistency with existing Press-Schechter calculations.
\end{itemize}

\begin{table*}
\caption{\label{mytable1} 
Different branches of conversion relation $\ed_0(a,\delta)$
  for an $\Omega_{\rm m}+\Omega_{\rm \Lambda}=1$ universe.}
\begin{tabular}{lllll}
\hline
\hline
{\bf Branch} & $\,\,\,\,\,\,\,$ 
&{\boldmath $\ed_0(\delta,a)$} & $\,\,\,\,\,\,\,$ 
&{\bf Auxiliary relations} \\ 
\hline 
$-1 < \delta \le \delta_{\rm Ed}$ & &
$\ed_0 (a,\delta) =
\frac{3A\left[(2\omega)^{1/3}\right]}{2^{1/3}}
\frac{1-\varpi(a,\delta)}{\left[\varpi(a,\delta)\right]^{2/3}}$
& & $\sinh^{-1} \sqrt{\omega a^3} - \mathcal{H}_1\left[
a\left(\frac{\varpi \omega}{1+\delta}\right)^{1/3}, \varpi 
\right]=0$\\
\hline 
$\delta_{\rm Ed} < \delta \le \delta_{\rm ta}$ && 
$\ed_0(a,\delta) = 
\frac{3A\left[(2\omega)^{1/3}\right]}{2^{1/3}}
\frac{1+\mu(a,\delta)}{\left[\mu(a,\delta)\right]^{2/3}}$
&& $\sinh^{-1}\sqrt{\omega a^3} -
\mathcal{V}_1\left[a\left(\frac{\mu
    \omega}{1+\delta}\right)^{1/3},\mu\right] =0$\\
\hline
$\delta_{\rm ta} < \delta \le \delta_{\rm v}$ && 
$\ed_0(a,\delta) = 
\frac{3A\left[(2\omega)^{1/3}\right]}{2^{1/3}}
\frac{1+\mu(a,\delta)}{\left[\mu(a,\delta)\right]^{2/3}}$
&&  $  \sinh^{-1} \sqrt{\omega a^3} -2\mathcal{V}_1(1,\mu)
+\mathcal{V}_1\left[a\left(\frac{\mu
    \omega}{1+\delta}\right)^{1/3},\mu\right]=0$\\
\hline
$\delta_{\rm v} < \delta \le \delta_{\rm c}$ &&
$\begin{array}{lll}
\ed_0(a, \delta) &=& 
\ed_{\rm 0,v} + \ed_{\rm 0,v}'(\delta - \delta_{\rm v}) \\ 
&+&\frac{3(\ed_{\rm0,c}-\ed_{\rm 0,v})-(\delta_{\rm c} - \delta_{\rm v})
(2\ed_{\rm 0,v}'+\ed_{0,c}')}{\left(\delta_{\rm c} - \delta_{\rm
    v}\right)^2}(\delta -\delta_{\rm v})^2 \\
&+& \frac{(\ed_{\rm 0,c}'+\ed_{\rm 0,v}')(\delta_{\rm c}- \delta_{\rm
    v})
-2(\ed_{\rm 0,c}-\ed_{\rm 0,v})}{(\delta_{\rm c} -\delta_{\rm v})^3}
(\delta -\delta_{\rm v})^3\\
\end{array}$
&& \begin{tabular}{l}
$\ed_{\rm 0,v} = \ed_0(a,\delta_{\rm v})$, given in this Table line 3\\
$\ed'_{\rm 0,v} = \left. \frac{\partial \ed_0}{\partial \delta}
  \right|_{\delta_{\rm v}} $, given in Table \ref{mytable2} line
  3\\
$ \ed_{\rm 0,c} = \ed_0(a,\delta_{\rm c})$, 
given by equation \ref{edc_lambda} \\
$\ed'_{\rm 0,c} = \left.\frac{\partial \ed_0 }{\partial
  \delta}\right|_{\delta_{\rm c}}  
= - \frac{A\left[(2\omega)^{1/3}\right]
\left(
6\omega^{2/3}a_{\rm p,v}^2\mu_{\rm c}^{2/3} -1 -\mu_{\rm c}\right)}{
\left(2\mu_{\rm c}^{2}\right)^{1/3}(\delta_{\rm c}+1)} $ \\
$\mu_{\rm c}(a)$ given in Table \ref{table_limits} line 4.
\end{tabular}
\\
\hline
\hline 
\end{tabular}
\end{table*}

\subsection{Critical extrapolated overdensity for collapse, {\boldmath
    $\ed_{\rm 0,c}(a)$}}

We need to find the critical $\ed_{\rm 0,c}(a)$ for collapse if the field is linearly
extrapolated to the present epoch, i.e. the value the linearly extrapolated to
the present overdensity must have, for a structure to have collapsed at
universe scale factor $a$. This, from equation \ref{kofdnot}, 
will be 
\begin{equation}\label{edc_lambda}
\ed_{\rm 0,c}(a) = \frac{3
  A\left[(2\omega)^{1/3}\right]}{(2)^{1/3}}\frac{1+\mu_{\rm c}(a)}
{\left[\mu_{\rm c}(a)\right]^{2/3}}
\end{equation}
where again $\mu_{\rm c}(a)$ is given by Table \ref{table_limits} line4.

The dependence of $\ed_{\rm 0,c}$ on $a$ can also be expressed in terms
of the linear growth factor, $D(a)$, as was the case for the
$\Omega_{\rm m}=1$ universe. The conversion relation between
$\delta(a)$ and $\ed_a$ (the linear-theory result for the density
contrast at time $a$) is independent of $a$. In other words, as long
as $\delta$ and $\ed$ both refer to the same time, knowledge of the one
uniquely defines the other, independently of the actual time at which
they are both evaluated. As $\delta(a) \rightarrow \infty$,
$\ed_a\rightarrow \ed_{\rm c}$, the linear-theory density contrast at
the time of collapse (given by eq. \ref{edc_lambda} for $a=1$). 
Therefore $\ed_{\rm c}$ is the same for perturbations of all
curvatures, and, using equation \ref{genlin}, we can write 
\begin{equation}
\ed_{\rm 0,c} (a_{\rm coll })
= \ed_{\rm c} \frac{D(a_0)}{D(a_{\rm coll})}\,.
\end{equation}

\subsection{{\boldmath$\partial \ed_0 / \partial \delta |_a$}}
 
In addition to the relation between $\delta$ and $\ed_0$, we will also
need the 
derivative $\partial \ed_0 / \partial \delta |_a$ in order to convert
between 
true and extrapolated overdensity differentials in
equation (\ref{convtotrue}). The calculation is similar as in the
case of an $\Omega_{\rm m}=1$ universe, and the results are summarized
in table \ref{mytable2}.

\begin{table*}
\caption{\label{mytable2} 
Different branches of derivative $\partial \ed_0/\partial \delta |_a$
  for an $\Omega_{\rm m}+\Omega_{\rm \Lambda}=1$ universe.}
\begin{tabular}{lllllll}
\hline
\hline
{\bf Branch} & $\,\,\,$ & {\bf Auxiliary Function} & $\,\,\,$ 
&{\boldmath $\left.\partial \ed_0 / \partial
  \delta\right|_a(a,\delta)= $} 
&$\,\,\,$& {\bf Definitions of Additional Functions}\\ 
\hline 
$-1 < \delta \le \delta_{\rm Ed}$ && $\Phi_1
= \sinh^{-1}\sqrt{\omega a^3} -
 \mathcal{H}_1(r,\varpi)$ &
&$-\frac{\partial \Phi_1}{\partial \delta}\left(
\frac{\partial \Phi_1}{\partial \ed_0}\right)^{-1}$ &&
$\frac{\partial \Phi_1}{\partial \delta} = 
\frac{a^{3/2}}{2(1+\delta)^{3/2}
\sqrt{\frac{a^3}{(1+\delta)}-\frac{a(\varpi-1)}{\varpi^{2/3}
(1+\delta)^{1/3}}
+ \frac{1}{\omega}}}$\\
&&&&&& $\frac{\partial \Phi_1}{\partial \ed_0}= - \frac{(2\varpi^2)^{1/3} }
{3A[(2\omega)^{1/3}]}
\mathcal{H}_2 \left[
a\left(\frac{\varpi \omega}{1+\delta}\right)^{1/3},\varpi\right]$
\\
&&&&&& $\sinh^{-1} \sqrt{\omega a^3} - \mathcal{H}_1\left[
a\left(\frac{\varpi \omega}{1+\delta}\right)^{1/3}, \varpi 
\right]=0$\\
\hline 
$\delta_{\rm Ed} < \delta \le \delta_{\rm ta}$ && 
$\Phi_2
= \sinh^{-1}\sqrt{\omega a^3} -\mathcal{V}_1(r,\mu)$&&
$-\frac{\partial \Phi_2}{\partial \delta}\left(
\frac{\partial \Phi_2}{\partial \ed_0}\right)^{-1}$&&
$\frac{\partial \Phi_2}{\partial \delta} = 
\frac{a^{3/2}}{2(1+\delta)^{3/2}
\sqrt{\frac{a^3}{(1+\delta)}-\frac{a(\mu+1)}{\mu^{2/3}
(1+\delta)^{1/3}}
+ \frac{1}{\omega}}}$
\\
&&&&&&$\frac{\partial \Phi_2}{\partial \ed_0}=-\frac{(2\mu^2)^{1/3} }
{3A[(2\omega)^{1/3}]}
\mathcal{V}_2 \left[
a\left(\frac{\mu \omega}{1+\delta}\right)^{1/3},\mu\right]$\\
&&&&&& $\sinh^{-1} \sqrt{\omega a^3} - \mathcal{V}_1\left[
a\left(\frac{\mu \omega}{1+\delta}\right)^{1/3}, \mu 
\right]=0$\\
\hline
$\delta_{\rm ta} < \delta \le \delta_{\rm v}$ && 
$\begin{array}{ll}
\Phi_3 =& \sinh^{-1}\sqrt{\omega a^3} \\ 
& -2\mathcal{V}_1(1,\mu) + \mathcal{V}(r,\mu)\end{array}$
&&
$-\frac{\partial \Phi_3}{\partial \delta}\left(
\frac{\partial \Phi_3}{\partial \ed_0}\right)^{-1}$&&
$\frac{\partial \Phi_3}{\partial \delta} = 
- \frac{a^{3/2}}{2(1+\delta)^{3/2}
\sqrt{\frac{a^3}{(1+\delta)}-\frac{a(\mu+1)}{\mu^{2/3}
(1+\delta)^{1/3}}
+ \frac{1}{\omega}}}$
\\
&&&&&&$\frac{\partial \Phi_3}{\partial \ed_0}=\frac{(2\mu^2)^{1/3} }
{3A[(2\omega)^{1/3}]}\left\{\mathcal{V}_2 \left[
a\left(\frac{\mu \omega}{1+\delta}\right)^{1/3},\mu\right]
- \frac{6\mu}{\mu-2}\frac{d\mathcal{V}_1(1,\mu)}{d\mu}
\right\}$\\
&&&&&&  $  \sinh^{-1} \sqrt{\omega a^3} -2\mathcal{V}_1(1,\mu)
+\mathcal{V}_1\left[a\left(\frac{\mu
    \omega}{1+\delta}\right)^{1/3},\mu\right]=0$\\
\\
\hline
$\delta_{\rm v} < \delta \le \delta_{\rm c}$ && ----- & 
\multicolumn{4}{l}{$\begin{array}{l}
\ed_{\rm 0,v}' 
+2\frac{3(\ed_{\rm0,c}-\ed_{\rm 0,v})-(\delta_{\rm c} - \delta_{\rm v})
(2\ed_{\rm 0,v}'+\ed_{0,c}')}{\left(\delta_{\rm c} - \delta_{\rm
    v}\right)^2}(\delta -\delta_{\rm v}) \\
+ 3\frac{(\ed_{\rm 0,c}'+\ed_{\rm 0,v}')(\delta_{\rm c}- \delta_{\rm
    v})
-2(\ed_{\rm 0,c}-\ed_{\rm 0,v})}{(\delta_{\rm c} -\delta_{\rm v})^3}
(\delta -\delta_{\rm v})^2\\
\end{array}$ \begin{tabular}{l} 
 with $\ed_{\rm0,c}$, $\ed_{\rm 0,v}$ 
$\ed_{\rm 0,c}'$, $\ed_{\rm 0,v}'$,\\
as in Table \ref{mytable1} line 4\end{tabular}} 
\\
\hline
\hline 
\end{tabular}
\end{table*}

In table \ref{mytable2}, $\mathcal{H}_2(r, \varpi)$ 
is the {\em hyperbolic  vacuum
integral of the second kind}, and 
$\mathcal{V}_2(r, \mu)$ is the {\em incomplete vacuum
integral of the second kind}, defined in appendix \ref{vac_ints}.

\section{Vacuum Integrals}\label{vac_ints}

 Due to the central importance and frequent
 use of vacuum integrals in this calculation, 
we have generated numerical functions 
based on combinations of tabulated values and asymptotic
 approximations which evaluate each of the vacuum integrals 
in a small fraction of the 
time that would be required for quadrature, and with an  accuracy
 better than $0.5\%$ throughout their domains.
These functions are publicly
 available at \url{http://www.astro.uiuc.edu/~bdfields/DD}.

\subsection{ The incomplete vacuum integral of the first kind 
\boldmath{$\mathcal{V}_1$}}
\subsubsection{Definition} 
We define the incomplete vacuum integral of the first kind as 
\begin{equation}\label{realdefv1}
\mathcal{V}_1(r,\mu) = 
\frac{3}{2}
\int_0^r\frac{\sqrt{x}dx}{\sqrt{(1-x)(-x^2-x+\mu)}}\,,
\end{equation}
with domain $0\le r\le 1$ and $\mu \ge 2$. 
\subsubsection{Properties}
Physically,
$\mathcal{V}_1(r,\mu)$ is proportional to the time required by a
perturbation of normalized curvature parameter
$\kappa/\omega^{1/3}=(\mu+1)/\mu^{2/3}$ to achieve a size $a_{\rm p} =
r a_{\rm p,ta}(\kappa/\omega^{1/3})$ {\em before turnaround}. Its
asymptotic behavior for $r\ll 1$ is 
\begin{equation}
\mathcal{V}_1(r,\mu) \stackrel{r\ll 1}{\approx} \frac{1}{\sqrt{\mu}}
r^{3/2}
\end{equation}
while for $\mu\gg 1$ it is 
\begin{equation}
\mathcal{V}_1(r,\mu) \stackrel{\mu \gg 1}{\approx} 
\frac{1}{\sqrt{2\mu}}\left[\frac{\pi}{2} - \sqrt{r(1-r)}-\sin ^{-1}\sqrt{1-r}
\right]
\end{equation}
In the case of the Eddington perturbation ($\mu =2$), 
we can derive a closed-form expression for $\mathcal{V}_1$: 
\begin{eqnarray}
\mathcal{V}_1(r,2) &=& \frac{3}{2}
\int_0^r\frac{\sqrt{x}dx}{(1-x)\sqrt{x+2}} \nonumber \\
&=& \frac{\sqrt{3}}{2}\left[\ln \frac{1+\sqrt{r}}{1-\sqrt{r}}
 -2\sqrt{3}\sinh^{-1}\sqrt{\frac{r}{2}}
\right. \nonumber \\
&& \left. + \ln \frac{2\sqrt{3}+\sqrt{3r}+3\sqrt{2+r}}
{2\sqrt{3}-\sqrt{3r}+3\sqrt{2+r}}
\right]\,.
\end{eqnarray} 
When $r=1$, the value of $\mathcal{V}_1(1,\mu)$ is the {\em complete} vacuum
integral of the first kind, which is a function of $\mu$ alone.
Physically, the complete vacuum integral of the first kind is
proportional to the time required for a perturbation of curvature
parametrized by $\mu$ to reach turnaround. The derivative of
$\mathcal{V}_1(1,\mu)$ appears in the calculation of the derivative
$\partial \ed_o/\partial \delta$, in the 3rd line of Table
\ref{mytable2}, and it is
\begin{equation}
\frac{d}{d\mu}\mathcal{V}_1(1,\mu) = -\frac{3}{4}\int_0^1
\frac{\sqrt{x}dx}{\sqrt{1-x}(-x^2-x+\mu)^{3/2}}\,.
\end{equation}

\subsection{The hyperbolic vacuum integral of the first kind
\boldmath{$\mathcal{H}_1$}}

\subsubsection{Definition}
We define the hyperbolic vacuum integral of the first kind as 
\begin{equation}
\mathcal{H}_1 (r,\varpi)=\frac{3}{2}
\int_0^r\frac{\sqrt{x}dx}{\sqrt{(1+x)(x^2-x+\varpi)}}\,,
\end{equation}
with domain $0\le r<\infty$ and
$\varpi > 1/4$. 

\subsubsection{Properties}
Physically, $\mathcal{H}_1 (r,\varpi)=$ is
proportional to the time  required by a
perturbation of normalized curvature parameter
$\kappa/\omega^{1/3}=(1-\varpi)/\varpi^{2/3}$ to achieve a size $a_{\rm p} =
r a_{\rm p,R}(\kappa/\omega^{1/3})$. Its asymptotic behavior for $r
\ll 1$ is
\begin{equation}
\mathcal{H}_1 (r,\varpi) \stackrel{r \ll 1}{\approx}
\frac{1}{\sqrt{\varpi}}r^{3/2}
\end{equation}
while for $r \gg 1$ it is 
\begin{equation}
\mathcal{H}_1 (r,\varpi) 
 \stackrel{r \gg 1}{\approx} C(\varpi) + \frac{3}{2}
\ln \left(2\sqrt{r^2-r+\varpi}+2r-1\right)
\end{equation}
where $C(\varpi)$ is a function dependent only on $\varpi$. 
In the case of a flat ($\varpi=1$) perturbation, $\mathcal{H}_1$ can
be integrated immediately to give
\begin{equation}
\mathcal{H}_1 (r,1)=
\frac{3}{2}\int_0^r\frac{\sqrt{x}dx}{\sqrt{x^3+1}}
= \sinh ^{-1}\sqrt{x^3}\,.
\end{equation}

\subsection{The incomplete vacuum integral of the second kind
\boldmath{$\mathcal{V}_2$}} 
\subsubsection{Definition}
We define the incomplete vacuum integral of the second kind as  
\begin{equation}
\mathcal{V}_2 (r,\mu)=
\frac{3}{4}\int_0^r \frac{x^{3/2}dx}{(1-x)^{3/2}(-x^2-x+\mu)^{3/2}}\,, 
\end{equation}
with domain same as for $\mathcal{V}_1(r,\mu)$. 

\subsubsection{Properties}
The incomplete vacuum integral of the second kind is
related to $\mathcal{V}_1(r,\mu)$ through 
\begin{equation}
\frac{\partial}{\partial (\kappa/\omega^{1/3})}\mathcal{V}_1(r,\mu)=
\mu^{2/3}\mathcal{V}_2(r,\mu)\,,
\end{equation}
with 
\begin{equation}
\kappa/\omega^{1/3} = (1+\mu)/\mu^{2/3}\,.
\end{equation}
In the case of the Eddington perturbation ($\mu =2$), we can derive closed-form
expressions for $\mathcal{V}_2$:
\begin{eqnarray}
\mathcal{V}_2(r,2) &=& 
\frac{3}{4}
\int_0^r\frac{x^{3/2}dx}{(1-x)^3(x+2)^{3/2}} \nonumber \\
&=& \frac{\sqrt{3}}{72}\left[
\frac{\sqrt{3r}(2-3r+4r^2)}{\sqrt{2+r}(1-r)^2}\right.\nonumber\\
&&\left.+\log\frac{1-r}{1+2r+\sqrt{3r(2+r)}}
\right]\,.
\end{eqnarray} 

\subsection{
The hyperbolic vacuum integral of the second kind
\boldmath{$\mathcal{H}_2$}}

\subsubsection{Definition}
We define the hyperbolic vacuum integral of the second kind as 
\begin{equation}
\mathcal{H}_2 (r,\varpi)=
\frac{3}{4}\int_0^r
\frac{x^{3/2}dx}{(1+x)^{3/2}(x^2-x+\varpi)^{3/2}}\,, 
\end{equation}
and its domain is that of $\mathcal{H}_1(r,\varpi)$. 

\subsubsection{Properties}
The hyperbolic vacuum integral of the second kind is related to 
$\mathcal{H}_1(r,\varpi)$ through
\begin{equation}
\frac{\partial}{\partial (\kappa/\omega^{1/3})}\mathcal{H}_1(r,\varpi)=
\varpi^{2/3}\mathcal{H}_2(r,\varpi)\,, 
\end{equation}
with 
\begin{equation}
\kappa/\omega^{1/3} = (1+\varpi)/\varpi^{2/3}\,.
\end{equation}
In the case of a flat ($\varpi=1$) perturbation, $\mathcal{H}_2$ takes
the form
\begin{equation}
\mathcal{H}_2 = \frac{3}{2^{4/3}}\int_0^{2^{1/3}r}\left
(\frac{u}{u^3+2}\right)^{3/2}du\,,
\end{equation}
which is the integral entering the linear growth factor in the
$\Omega_{\rm m} + \Omega_\Lambda=1$ universe. Hence, the linear growth
factor function $A(x)$ can be written as
\begin{equation}
A(x) = \frac{2^{4/3}(x^3+2)^{1/2}}{3x^{3/2}}\mathcal{H}_2(2^{-1/3}x,1)\,.
\end{equation}
\end{document}